\documentclass[12pt,twoside,pointlessnumbers,nofootinbib,smallheadings]{revtex4}

\usepackage[CJKbookmarks]{hyperref}
\usepackage{CJK}
\usepackage{amsmath}
\usepackage{graphics}
\usepackage{slashed}
\usepackage{graphicx}
\usepackage{indentfirst}
\usepackage{amssymb}
\usepackage{cancel}
\usepackage{leftidx}
\usepackage{simplewick}
\usepackage[page,title,titletoc,header]{appendix}

\begin{document}

\begin{CJK*}{GBK}{song}

\title{Lightness of Higgs Boson and Spontaneous CP-violation in the Lee Model: An Alternative Scenario}

\author{Ying-nan Mao $^{1,4,}$\footnote{maoyingnan@pku.edu.cn, maoyn@ihep.ac.cn} and Shou-hua
Zhu $^{1,2,3,}$\footnote{shzhu@pku.edu.cn}}

\affiliation{
$ ^1$ Institute of Theoretical Physics $\&$ State Key Laboratory of
Nuclear Physics and Technology, Peking University, Beijing 100871,
China \\
$ ^2$ Collaborative Innovation Center of Quantum Matter, Beijing 100871, China \\
$ ^3$ Center for High Energy Physics, Peking University,
Beijing 100871, China \\
$ ^4$ Center for Future High Energy Physics $\&$ Theoretical Physics Division, Institute of High Energy Physics, Chinese Academy of Sciences, 
Beijing 100049, China
}

\begin{abstract}

Based on the weakly-coupled spontaneous CP-violation two-Higgs-doublet model (named Lee model) and the mechanism to generate the correlation between
smallness of CP-violation and lightness of scalar mass, as we proposed earlier, we predicted a light CP-mixing scalar $\eta$
in which pseudoscalar component is dominant. It is a natural scenario in which $m_{\eta}\sim\mathcal{O}(10\textrm{GeV})\ll v$. It means
new physics might be hidden below the electro-weak scale $v$. Masses of all other scalars ($h$, $H$, $H^\pm$)
should be around the electro-weak scale $v$. Among them, the 125 GeV Higgs boson ($h$) couplings
are standard-model like, and the charged Higgs boson ($H^\pm$) mass should be around the heaviest neutral scalar ($H$) mass.
We discussed all experimental constraints and showed that this scenario is still allowed by data. The strictest constraints 
come from the flavor violation experiments
and the EDM of electron and neutron. We also discussed the future tests for this scenario. It is possible to discover the extra scalars or exclude this
scenario at future colliders, especially at the LHC and $e^+e^-$ colliders with $\mathcal{O}(\textrm{ab}^{-1})$ luminosity. We also pointed out that the
$Z$-mediated Higgs pair production via $e^+e^-\rightarrow h_ih_j$ ($h_i, h_j$ stand for two of the $\eta, h, H$) would be the key observable to confirm or exclude CP-violation in Higgs sector. The sensitivity to test this scenario is worth further studying in detail.

\end{abstract}

\date{\today}

\maketitle

\newpage

\section{Introduction}
The realization of electro-weak symmetry breaking and CP-violation are two important topics both in the standard model (SM)
and beyond the standard model (BSM). It is also attractive to relate them with each other. In our previous work \cite{our},
we proposed the correlation between lightness of Higgs boson and smallness of CP-violation. In this paper, we will continue
to explore an alternative natural scenario and its phenomenology.

In 1964, the Higgs mechanism \cite{Higgs} was proposed. In the Higgs mechanism, a scalar doublet with nontrivial
vacuum expectation value (VEV) was introduced to break the electro-weak gauge symmetry spontaneously. After spontaneous
gauge symmetry breaking in the SM, there exists a scalar named the Higgs boson \footnote{There may exist more
particles in the extension of SM. For example, in the two-Higgs-doublet model (2HDM) \cite{2HDM} in which two scalar
doublets were introduced, there exist five scalars. Two of them are charged and three of them are neutral.}.
In July 2012, both ATLAS \cite{disatlas} and CMS \cite{discms} collaborations
at LHC discovered a new boson with its mass around $125$ GeV \cite{mass}. The subsequent measurements by CMS and
ATLAS \cite{upd1,upd2,upd3} on its signal strengths showed that the scalar behaves similarly with SM Higgs boson.  However there is still spacious
room for the BSM. In some BSM
models, there exist new light particles which may appear in the final states during Higgs decay processes. For example,
in the next-to-minimal super-symmetric standard model (NMSSM) \cite{NMSSM}, the simplest little Higgs model (SLH) \cite{SLH,SLH2,SLH3},
or the left-right-twin-Higgs model (LRTH) \cite{LRTH,LRTH2}, a light scalar $\eta$ with its mass of $\mathcal{O}(10)\textrm{GeV}$ will
naturally appear. For some cases in 2HDM \cite{2HDM0,2HDMexo,2HDMexo2,2HDMexo3,2HDMexo4}, a light scalar $\eta$ is allowed as well, though there are strict
constraints on them. If $m_{\eta}<m_h/2=62.5\textrm{GeV}$, there would be an exotic decay channel $h\rightarrow\eta\eta$; while if $m_{\eta}
<m_h-m_Z=34\textrm{GeV}$, an exotic decay channel $h\rightarrow Z\eta$ should also be open. There is no evidence for exotic Higgs decay
channels at LHC till now, the constraints on the exotic Higgs decay branching ratio is set to be
$\textrm{Br}_{\textrm{exo}}\lesssim(20-30)\%$ \cite{exotic} if the production rate of the Higgs boson is close to that in SM.
The spin-parity property for Higgs boson is expected to be $0^+$ in the SM. Experimentally, a pure pseudoscalar state
($0^-$) is excluded at over $3\sigma$ \cite{par1,par2,par3}. But a mixing state is still allowed, thus the spacious room for
BSM scenarios have not been closed yet.

Theoretically, CP-violation in SM is induced by the Kobayashi and Maskawa (K-M) mechanism \cite{KM} proposed by
Kobayashi and Maskawa in 1973. They proved that a nontrivial phase which leads to CP-violation in quark mixing
matrix (called the CKM matrix \cite{KM,CKM}) would appear if there exist three or more generations of quarks.
The CKM matrix is usually parameterized as the Wolfenstein formalism \cite{wolf}
\begin{equation}
V_{\textrm{CKM}}=\left(\begin{array}{ccc}1-\lambda^2/2&\lambda&A\lambda^3(\rho-i\eta)\\
-\lambda&1-\lambda^2/2&A\lambda^2\\A\lambda^3(1-\rho-i\eta)&-A\lambda^2&1\end{array}\right)+\mathcal{O}(\lambda^4).
\end{equation}
The Jarlskog invariant $J$ \cite{Jar} defined as
\begin{equation}
\det\left(\textrm{i}\left[M_UM_U^{\dag},M_DM_D^{\dag}\right]\right)=
2J\mathop{\prod}_{i<j}\left(m^2_{U_i}-m^2_{U_j}\right)\mathop{\prod}_{i<j}\left(m^2_{D_i}-m^2_{D_j}\right)
\end{equation}
measures the effects of CP-violation where $M_{U(D)}$ is the mass matrix for up (down) type quarks.
$J\approx\lambda^6A^2\eta\approx3\times10^{-5}$ \cite{PDG} means CP-violation in SM is small.
Experimentally, in K- and B-meson systems, several kinds of CP-violation have been discovered \cite{PDG} which represent the
success of K-M mechanism. While it is still attractive to search for new sources of CP-violation, not only to search for
BSM physics, but also to understand the matter-antimatter asymmetry in the universe \cite{PDG,bar1}. SM
itself cannot provide enough baryogenesis effects \cite{PDG,bar1,bar2,bar3}, but in some extensions of SM, for example,
2HDM with CP-violation in Higgs sector, it is possible to generate large enough baryogenesis effect \cite{bar2,bar4}.

Lee model \cite{Lee} is a possible way to connect Higgs mechanism and CP-violation with each other. It was proposed by Lee
in 1973 as the first 2HDM. In Lee model, the lagrangian is required to be CP-conserved, but the VEV of one Higgs doublet
can be complex, thus the CP symmetry is spontaneously broken due to the complex vacuum. In this case, the
neutral scalars are CP-mixing states so that CP-violation effects should occur in the Higgs sector. All the three neutral
scalars should couple to massive gauge bosons with the effective interaction
\begin{equation}
\mathcal{L}_{h_iVV}=\mathop{\sum}_ic_{i,V}h_i\left(\frac{2m^2_W}{v}W^{+\mu}W^-_{\mu}+\frac{m^2_Z}{v}Z^{\mu}Z_{\mu}\right)
\end{equation}
where $c_{i,V}\equiv g_{h_iVV}/g_{hVV,\textrm{SM}}$ is the ratio between the $h_iVV$ coupling strength and that in SM.
$c_{1,V}^2+c_{2,V}^2+c_{3,V}^2=1$ due to the mechanism of spontaneous gauge symmetry breaking. The quantity
\begin{equation}
K\equiv c_{1,V}c_{2,V}c_{3,V}
\end{equation}
measures the CP-violation effects in Higgs sector \cite{2HDM,K} when the masses of the neutral scalars are non-degenerate
\footnote{If at least two of the scalars have the same mass, we can always perform a field rotation between them to keep
at least one $c_{i,V}=0$, thus there would be no CP-violation in Higgs sector.}. In our recent
paper \cite{our}, we proposed the correlation between lightness of Higgs boson and smallness of CP-violation through
small $t_{\beta}s_{\xi}$ in Lee model \footnote{The parameters will be defined next section, or see \cite{our}.}. While in
that paper, we treated the 125 GeV scalar as the lightest one thus it implied a strong-interacted scenario beyond \cite{talk}.
However, another natural scenario with a weakly-interacted scalar in which the heavy scalars have the mass $m_i\sim\mathcal{O}(v)$ is also
possible where $v=246\textrm{GeV}$ is the VEV of the scalar doublet in SM. In this scenario, Lee model would
predict a light scalar $\eta$ with mass $m_{\eta}\ll v$ for the small $t_{\beta}s_{\xi}$ case based on our paper \cite{our}.
In this paper, we will discuss this scenario and its phenomenology.

This paper is organized as follows. In \autoref{model} we introduce the Lee model and its main properties. In \autoref{cons} we discuss the
constraints for this scenario by recent experiments, including data from both high and low energy phenomena. In \autoref{fut} we consider
the predictions and future tests for this scenario. And \autoref{conc} contains our conclusions and discussions.
\section{The Lee Model and a Light Scalar}
\label{model}
In Lee model \cite{Lee}, the lagrangian is required to be CP-conserved in both scalar and
Yukawa sectors. For the scalar sector,
\begin{equation}
\mathcal{L}=(D_{\mu}\phi_1)^{\dag}(D^{\mu}\phi_1)+(D_{\mu}\phi_2)^{\dag}(D^{\mu}\phi_2)-V(\phi_1,\phi_2)
\end{equation}
in which the scalar potential
\begin{eqnarray}
V(\phi_1,\phi_2)&=&\mu_{1}^2R_{11}+\mu_{2}^2R_{22}+\lambda_1R_{11}^2+\lambda_2R_{11}R_{12}\nonumber\\
\label{pot}
&&+\lambda_3R_{11}R_{22}+\lambda_4R_{12}^2+\lambda_5R_{12}R_{22}+\lambda_6R_{22}^2+\lambda_7I_{12}^2.
\end{eqnarray}
Here the scalar doublets
\begin{equation}
\label{vev}
\phi_1=\left(\begin{array}{c}\phi_1^+\\ \frac{v_1+R_1+iI_1}{\sqrt{2}}\end{array}\right),\quad\quad
\phi_2=\left(\begin{array}{c}\phi_2^+\\ \frac{v_2\textrm{e}^{\textrm{i}\xi}+R_2+iI_2}{\sqrt{2}}\end{array}\right)
\end{equation}
and $R(I)_{ij}$ denotes the real (imaginary) part of $\phi_i^{\dag}\phi_j$ \footnote{We can always perform a rotation between
$\phi_1$ and $\phi_2$ to keep the term proportional to $R_{12}$ vanish.}. $\sqrt{v_1^2+v_2^2}=v=246\textrm{GeV}$.
The general Yukawa couplings can be written as
\begin{equation}
\label{Yuk}
\mathcal{L}_y=-\bar{Q}_{Li}((Y_{1d})_{ij}\phi_1+(Y_{2d})_{ij}\phi_2)D_{Rj}
-\bar{Q}_{Li}((Y_{1u})_{ij}\tilde{\phi}_1+(Y_{2u})_{ij}\tilde{\phi}_2)U_{Rj},
\end{equation}
where all coupling constants should be real and $\tilde{\phi}_i\equiv\textrm{i}\sigma_2\phi_i^*$. We choose the Type III \cite{2HDM,type3}
Yukawa couplings because there is no additional discrete symmetry to forbid any term in (\ref{Yuk}). It is possible to generate correct
fermion mass spectrum and CKM matrix from (\ref{Yuk}), for example, see \cite{Yuk1,Yuk2}.

We should minimize the potential (\ref{pot}). For some parameter choices, there is a nonzero $\xi$ which means the spontaneous CP-
violation \footnote{We can always perform a global phase redefinition for $\phi_1$ and $\phi_2$ to keep one of the VEVs real, just like
the case in (\ref{vev}).}. If $v_1,v_2,\xi\neq0$, we have
\begin{eqnarray}
\mu_{1}^2&=&-\lambda_1v_1^2-\frac{\lambda_3+\lambda_7}{2}v_2^2-\frac{\lambda_2}{2}v_1v_2\cos\xi;\\
\mu_{2}^2&=&-\frac{\lambda_3+\lambda_7}{2}v_1^2-\lambda_6v_2^2-\frac{\lambda_5}{2}v_1v_2\cos\xi;\\
\label{cpv}
0&=&\frac{\lambda_2}{2}v_1^2+\frac{\lambda_5}{2}v_2^2+(\lambda_4-\lambda_7)v_1v_2\cos\xi.
\end{eqnarray}
$|\lambda_2v_1^2+\lambda_5v^2_2|<2|\lambda_4-\lambda_7|v_1v_2$ is required to keep $\xi\neq0$. Define
$s_{\alpha}\equiv\sin\alpha,c_{\alpha}\equiv\cos\alpha,t_{\alpha}\equiv\tan\alpha$ in the following
parts of this paper, and $t_{\beta}\equiv v_2/v_1$ is the ratio of the VEVs for scalar doublets. The
vacuum stability conditions can be found in \cite{2HDM} or Appendix. A in \cite{our}.
The Goldstone fields can be written as
\begin{eqnarray}
G^{\pm}&=&c_{\beta}\phi_1^{\pm}+\textrm{e}^{\mp\textrm{i}\xi}s_{\beta}\phi_2^{\pm};\\
G^0&=&c_{\beta}I_1+s_{\beta}c_{\xi}I_2-s_{\beta}s_{\xi}R_2.
\end{eqnarray}
The charged Higgs field is orthogonal to the corresponding charged Goldstone field as
\begin{equation}
H^{\pm}=-\textrm{e}^{\pm\textrm{i}\xi}s_{\beta}\phi_1^{\pm}+c_{\beta}\phi_2^{\pm}
\end{equation}
with the mass square
\begin{equation}
m^2_{\pm}=-\frac{\lambda_7}{2}v^2.
\end{equation}
The symmetric mass matrix $\tilde{m}$ for neutral scalars is written as \cite{our}
\begin{equation}
\left(\begin{array}{ccc}(\lambda_4-\lambda_7)s^2_{\xi}&
-((\lambda_4-\lambda_7)s_{\beta}c_{\xi}+\lambda_2c_{\beta})s_{\xi}&
-((\lambda_4-\lambda_7)c_{\beta}c_{\xi}+\lambda_5s_{\beta})s_{\xi}\\
&&\\
&\begin{array}{c}4\lambda_1c^2_{\beta}+\lambda_2s_{2\beta}c_{\xi}+(\lambda_4-\lambda_7)s^2_{\beta}c^2_{\xi}\end{array}&
\begin{array}{c}((\lambda_3+\lambda_7)+(\lambda_4-\lambda_7)c^2_{\xi}/2)s_{2\beta}\\
+\lambda_2c^2_{\beta}c_{\xi}+\lambda_5s^2_{\beta}c_{\xi}\end{array}\\
&&\\
&&\begin{array}{c}(\lambda_4-\lambda_7)c^2_{\beta}c^2_{\xi}\\+\lambda_5s_{2\beta}c_{\xi}+4\lambda_6s^2_{\beta}\end{array}\end{array}\right)
\end{equation}
in the basis $(-s_{\beta}I_1+c_{\beta}c_{\xi}I_2-c_{\beta}s_{\xi}R_2, R_1,s_{\xi}I_2+c_{\xi}R_2)^T$ in unit of $v^2/2$.
To solve the eigenvalue equation with perturbation method \footnote{For the calculations in details, please see the Appendix.B in our
recent paper \cite{our}, with the same conventions as those in this paper.}, we should expand $\tilde{m}$ in powers of $(t_{\beta}s_{\xi})$
in small $t_{\beta}$ limit as
\begin{equation}
\tilde{m}=\tilde{m}_0+(t_{\beta}s_{\xi})\tilde{m}_1+(t_{\beta}s_{\xi})^2\tilde{m}_2+\ldots
\end{equation}
For the two heavy scalars, we have \cite{our}
\begin{equation}
m^2_{h,H}=\frac{v^2}{2}\left(\left(\tilde{m}_0\right)_{22(33)}+\mathcal{O}(t_{\beta}s_{\xi})\right)
\end{equation}
where
\begin{equation}
(\tilde{m}_0)_{22(33)}=\frac{4\lambda_1+\lambda_4-\lambda_7}{2}\mp
\left(\frac{4\lambda_1-(\lambda_4-\lambda_7)}{2}c_{2\theta}+\lambda_2s_{2\theta}\right).
\end{equation}
Here $\theta=(1/2)\arctan(2\lambda_2/(4\lambda_1-\lambda_4+\lambda_7))$ labels the mixing angle of
the real parts of the two scalar doublets. The scalar fields
\begin{equation}
\left(\begin{array}{c}h\\H\end{array}\right)=\left(\begin{array}{cc}c_{\theta}&s_{\theta}\\
-s_{\theta}&c_{\theta}\end{array}\right)\left(\begin{array}{c}R_1\\R_2\end{array}\right)
+\mathcal{O}(t_{\beta}s_{\xi}).
\end{equation}
We treat the lighter one as $m_h=\sqrt{(\tilde{m}_0)_{22}/2}v=125\textrm{GeV}$. Different from the scenario
in \cite{our}, in this paper, the dominant component for the 125 GeV scalar should be CP-even thus there exists SM limit for its couplings.
While for the lightest scalar $\eta$, to the leading order of $(t_{\beta}s_{\xi})$, we have
\begin{eqnarray}
m^2_{\eta}&=&\frac{v^2t^2_{\beta}s^2_{\xi}}{2}\left((\tilde{m}_2)_{11}-\frac{(\tilde{m}_1)^2_{12}}{(\tilde{m}_0)_{22}}
-\frac{(\tilde{m}_1)^2_{13}}{(\tilde{m}_0)_{33}}\right)\nonumber\\
&=&\frac{v^2t^2_{\beta}s^2_{\xi}}{2}\bigg[4\lambda_6+2\lambda_5(\lambda_3+\lambda_7)s_{2\theta}
\left(\frac{1}{(\tilde{m}_0)_{22}}-\frac{1}{(\tilde{m}_0)_{33}}\right)\nonumber\\
&&-4(\lambda_3+\lambda_7)^2\left(\frac{c^2_{\theta}}{(\tilde{m}_0)_{22}}+
\frac{s^2_{\theta}}{(\tilde{m}_0)_{33}}\right)-\lambda^2_5\bigg(\frac{s^2_{\theta}}{(\tilde{m}_0)_{22}}
+\frac{c^2_{\theta}}{(\tilde{m}_0)_{33}}\bigg)\bigg];\\
\eta&=&I_2-t_{\beta}s_{\xi}\left(\frac{(\tilde{m}_1)_{12}}{(\tilde{m}_0)_{22}}(c_{\theta}R_1+s_{\theta}R_2)
+\frac{(\tilde{m}_1)_{13}}{(\tilde{m}_0)_{33}}(c_{\theta}R_2-s_{\theta}R_1)+\frac{I_1}{t_{\xi}}\right)\nonumber\\
&=&I_2-t_{\beta}s_{\xi}\bigg[\bigg(2(\lambda_3+\lambda_7)\bigg(\frac{c^2_{\theta}}{(\tilde{m}_0)_{22}}+
\frac{s^2_{\theta}}{(\tilde{m}_0)_{33}}\bigg)+\frac{\lambda_5s_{2\theta}}{2}\bigg(\frac{1}{(\tilde{m}_0)_{22}}
-\frac{1}{(\tilde{m}_0)_{33}}\bigg)\bigg)R_1\nonumber\\
&&+\bigg((\lambda_3+\lambda_7)s_{2\theta}\bigg(\frac{1}{(\tilde{m}_0)_{22}}
-\frac{1}{(\tilde{m}_0)_{33}}\bigg)+\lambda_5\bigg(\frac{s^2_{\theta}}{(\tilde{m}_0)_{22}}+
\frac{c^2_{\theta}}{(\tilde{m}_0)_{33}}\bigg)\bigg)R_2+\frac{I_1}{t_{\xi}}\bigg].
\end{eqnarray}
Thus in the limit $t_{\beta}s_{\xi}\rightarrow0$, we have $m_{\eta}\propto t_{\beta}s_{\xi}\rightarrow0$ and
$\eta\rightarrow I_2$, which mean that $\eta$ behaves like a light pseudoscalar but it has small CP-even component.

We can diagonalize the fermion mass matrixes as
\begin{equation}
V_{U,L}M_UV_{U,R}^{\dag}=\left(\begin{array}{ccc}m_u&0&0\\0&m_c&0\\0&0&m_t\end{array}\right),\quad\quad
V_{D,L}M_DV_{D,R}^{\dag}=\left(\begin{array}{ccc}m_d&0&0\\0&m_s&0\\0&0&m_b\end{array}\right)
\end{equation}
in which according to (\ref{Yuk}), the mass matrixes are
\begin{equation}
\left(M_U\right)_{ij}=\frac{v}{\sqrt{2}}\left((Y_{1u})_{ij}c_{\beta}+(Y_{2u})_{ij}s_{\beta}\textrm{e}^{-\textrm{i}\xi}\right),\quad
\left(M_D\right)_{ij}=\frac{v}{\sqrt{2}}\left((Y_{1d})_{ij}c_{\beta}+(Y_{2d})_{ij}s_{\beta}\textrm{e}^{\textrm{i}\xi}\right).
\end{equation}
The CKM matrix $V_{\textrm{CKM}}=V_{U,L}V_{D,L}^{\dag}$ as usual. We can rewrite the Yukawa couplings (\ref{Yuk}) in quark sector
as following adopting the Cheng-Sher ansatz \cite{CS}
\begin{eqnarray}
\mathcal{L}'_{\textrm{Yuk,Q}}&=&-\mathop{\sum}_{f=U_i,D_i}m_f\bar{f}_Lf_R\left
(1+\frac{c_{\beta}R_1+s_{\beta}c_{\xi}R_2+s_{\beta}s_{\xi}I_2}{v}\right)\nonumber\\
&&-\mathop{\sum}_{i,j}\frac{\xi_{ij}^U\sqrt{m^U_im^U_j}}{v}\bar{U}_{i,L}U_{j,R}\left((c_{\beta}R_2-s_{\beta}c_{\xi}R_1+s_{\beta}s_{\xi}I_1)
-\textrm{i}(c_{\beta}I_2-s_{\beta}c_{\xi}I_1-s_{\beta}s_{\xi}R_1)\right)\nonumber\\
&&-\mathop{\sum}_{i,j}\frac{\xi_{ij}^D\sqrt{m^D_im^D_j}}{v}\bar{D}_{i,L}D_{j,R}\left((c_{\beta}R_2-s_{\beta}c_{\xi}R_1+s_{\beta}s_{\xi}I_1)
+\textrm{i}(c_{\beta}I_2-s_{\beta}c_{\xi}I_1-s_{\beta}s_{\xi}R_1)\right)\nonumber\\
&&-\mathop{\sum}_{i,j}\frac{\sqrt{2m^D_im^D_j}}{v}\bar{U}_{i,L}\left(V_{\textrm{CKM}}\cdot\xi^D\right)_{ij}D_{j,R}H^+\nonumber\\
\label{charyuk}
&&-\mathop{\sum}_{i,j}\frac{\sqrt{2m^U_im^U_j}}{v}\bar{D}_{i,L}\left(V_{\textrm{CKM}}^{\dag}\cdot\xi^U\right)_{ij}U_{j,R}H^-+\textrm{h.c.}
\end{eqnarray}
Similarly in the lepton sector
\begin{eqnarray}
\mathcal{L}'_{\textrm{Yuk},\ell}&=&-\mathop{\sum}_{\ell}m_{\ell}\bar{\ell}_L\ell_R\left
(1+\frac{c_{\beta}R_1+s_{\beta}c_{\xi}R_2+s_{\beta}s_{\xi}I_2}{v}\right)\nonumber\\
&&-\mathop{\sum}_{i,j}\frac{\xi_{ij}^{\ell}\sqrt{m^{\ell}_im^{\ell}_j}}{v}\bar{\ell}_{i,L}\ell_{j,R}\left((c_{\beta}R_2-s_{\beta}c_{\xi}R_1+s_{\beta}s_{\xi}I_1)
+\textrm{i}(c_{\beta}I_2-s_{\beta}c_{\xi}I_1-s_{\beta}s_{\xi}R_1)\right)\nonumber\\
&&-\mathop{\sum}_{i,j}\frac{\sqrt{2m^{\ell}_im^{\ell}_j}}{v}\bar{\nu}_{i,L}\left(V_{\textrm{PMNS}}\cdot\xi^{\ell}\right)_{ij}\ell_{j,R}H^++\textrm{h.c.}
\end{eqnarray}
Here $V_{\textrm{PMNS}}$ is the lepton mixing matrix \cite{PMNS} and
\begin{eqnarray}
\xi^U_{ij}&=&(V_{U,L})_{ik}\left(-s_{\beta}\textrm{e}^{\textrm{i}\xi}(Y_{1u})_{kl}+c_{\beta}(Y_{2u})_{kl}\right)(V_{U,R}^{\dag})_{lj};\\
\xi^{D(\ell)}_{ij}&=&(V_{D(\ell),L})_{ik}\left(-s_{\beta}\textrm{e}^{-\textrm{i}\xi}(Y_{1d(\ell)})_{kl}+
c_{\beta}(Y_{2d(\ell)})_{kl}\right)(V_{D(\ell),R}^{\dag})_{lj}.
\end{eqnarray}
The off-diagonal elements of $\xi^{U,D,\ell}_{ij}$ induce the flavor changing processes at tree level. It was proved
in \cite{our} that in the $t_{\beta}s_{\xi}\rightarrow0$ limit, all the four quantities $m_{\eta},c_{\eta,V},K,J\propto
t_{\beta}s_{\xi}$ which means the correlation between the lightest scalar and smallness of CP-violation.

In the scenario we discuss in this paper, there can be exotic Higgs decay channels $h\rightarrow\eta\eta,Z\eta$ induced by
\begin{equation}
\mathcal{L}_{\textrm{exo}}=\frac{c_{h\eta}g}{2c_W}(h\partial_{\mu}\eta-\eta\partial_{\mu}h)Z^{\mu}-\frac{1}{2}g_{h\eta\eta}vh\eta^2.
\end{equation}
It leads to the branching ratios
\begin{eqnarray}
\label{hZeta}
\textrm{Br}(h\rightarrow Z\eta)&=&\frac{g^2c^2_{h\eta}m^3_h}{64\pi m_W^2\Gamma_{h,\textrm{tot}}}\mathcal{F}\left(\frac{m^2_Z}{m^2_h},\frac{m^2_{\eta}}{m^2_h}\right);\\
\label{hetaeta}
\textrm{Br}(h\rightarrow\eta\eta)&=&\frac{g^2_{h\eta\eta}v^2}{32\pi m_h\Gamma_{h,\textrm{tot}}}\sqrt{1-\frac{4m^2_{\eta}}{m^2_h}}
\end{eqnarray}
where $\mathcal{F}(x,y)=(1+x^2+y^2-2x-2y-2xy)^{3/2}$, $g$ is the weak coupling constant and $c_W\equiv m_W/m_Z$.
For the detail couplings, please see \autoref{coup} in appendices, in which all $c_{h,f}$ are defined as the ratio
between Higgs-$f\bar{f}$ couplings and those in SM.
\section{Constraints for this Scenario by Recent Data}
\label{cons}
Besides the 125 GeV Higgs boson ($h$), there are two extra neutral scalars and one of which is expected to be light in this scenario.
For the lightest scalar $\eta$ with its mass $m_{\eta}\sim\mathcal{O}(0.1-1)\textrm{GeV}$, the BESIII \cite{BESIII}, BaBar \cite{ba1,ba2} and CMS
\cite{cmslight} experiments gave strict constraints thus we will focus on the cases $m_{\eta}\sim\mathcal{O}(10)\textrm{GeV}$.
Type II 2HDM including a light scalar with mass $(25-80)\textrm{GeV}$ is excluded \cite{cmslight2} through the search for
$\eta b\bar{b}$ associated production. While for a general
case it is still allowed by collider data, as we will show below. The two extra scalars would face the constraints from the direct
searches at LEP and LHC. In this scenario of Lee model, with a light particle $\eta$, the exotic decay channels
$h\rightarrow\eta\eta,Z\eta$ will modify the total width and signal strengths for the 125 GeV Higgs boson that we should
also consider the constraints from Higgs signal strengths.

In Lee model, there is no additional discrete symmetry to forbid flavor changing processes at tree level, and there are also
new origins for CP-violation. Thus it must face the constraints in flavor physics, including rare decays, meson mixing,
etc. The electric dipole moments (EDM) for electron \cite{eEDM} and neutron \cite{nEDM} would also give strict constraints
in many models with additional CP-violation source \cite{reviewEDM} including Lee model, so we must consider the EDM
constraints here as well.
\subsection{Direct Searches for Extra Scalars}
\label{directsearch}
The LEP experiments \cite{LEP1,LEP2,LEP3} set strict constraints on this scenario through the $e^+e^-\rightarrow Z\eta$ and
$e^+e^-\rightarrow h\eta$ associated production processes. For $\eta$ with its mass $(15-40)\textrm{GeV}$,
\cite{LEP1,LEP2} gave $\sigma_{Z\eta}/\sigma_{\textrm{SM}}\lesssim(1.5-4)\times10^{-2}$ at $95\%$ C.L. which meant
\begin{equation}
c_{\eta,V}\lesssim(0.12-0.2)
\end{equation}
thus $t_{\beta}s_{\xi}\lesssim0.1$ in this scenario. At the same mass region for $\eta$, assuming both $\eta$ and $h$
decay to $b\bar{b}$ final states dominantly, \cite{LEP2,LEP3} gave $c_{h\eta}^2\lesssim(0.2-0.3)$. According to (\ref{rel}),
$c_{H,V}=c_{h\eta}$ thus $c_{H,V}$ should also be small. The results implied that $c_{h,V}\sim1$ thus the couplings of $h$ 
should be SM-like.

\begin{figure}[h]
\caption{$\textrm{Br}(H\rightarrow VV)-m_H$ plots for different parameter choices fixing $c_{\eta,V}=0.1$. The green, yellow,
blue, and red lines stand for $c_{H,V}=0.2,0.3,0.4,0.5$ respectively in each figure. The upper figures are for $m_{\eta}=20\textrm{GeV}$
while the lower figures are for $m_{\eta}=40\textrm{GeV}$. In each line, from left to right, we take $g_{H\eta\eta}=0,0.5,1$.}\label{HVVplot}
\includegraphics[scale=0.4]{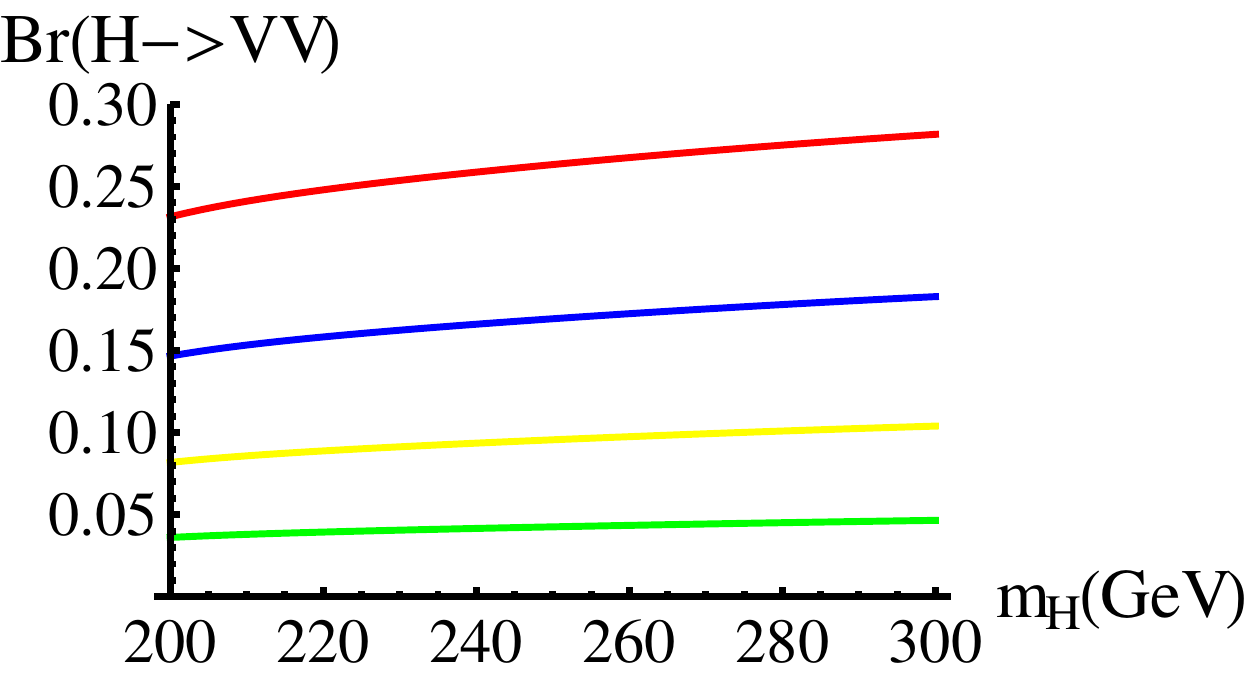}\quad\includegraphics[scale=0.4]{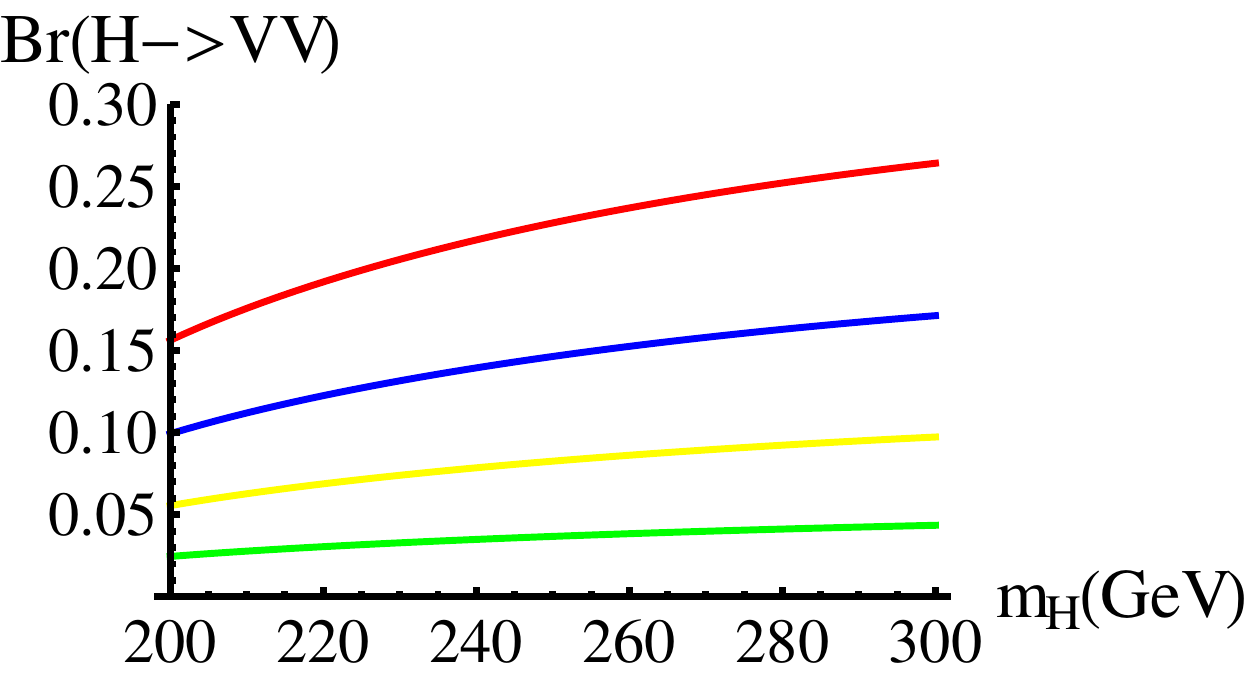}\quad\includegraphics[scale=0.4]{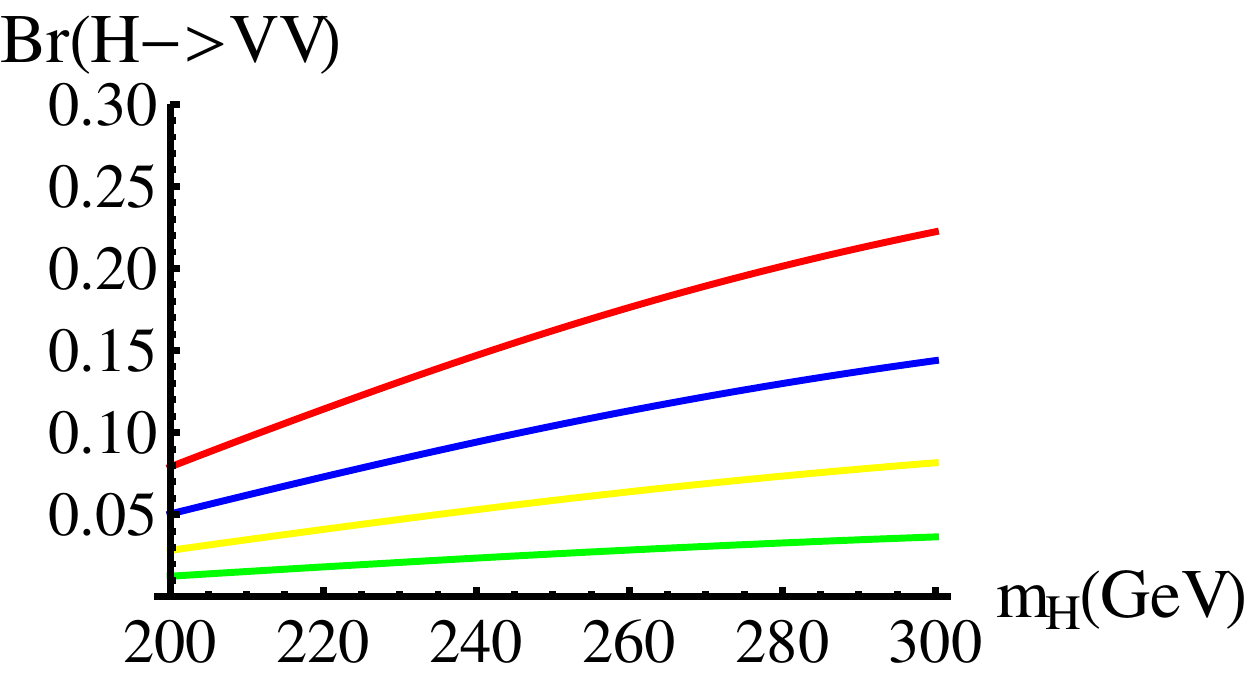}
\includegraphics[scale=0.4]{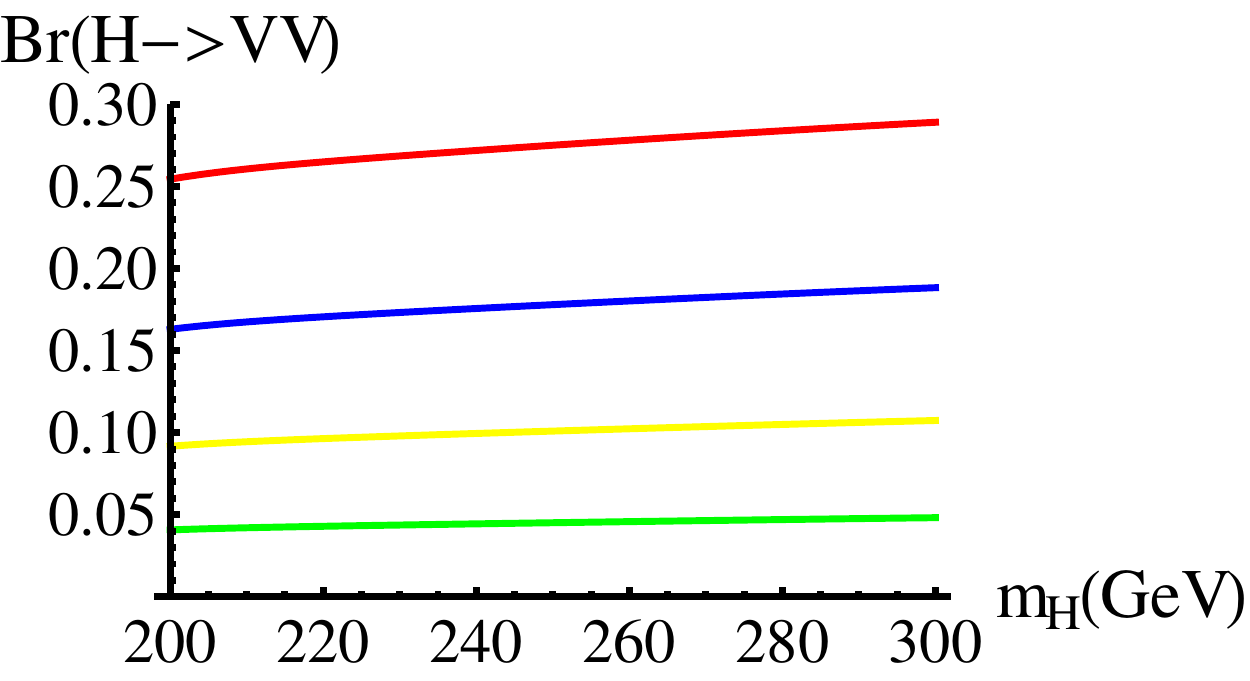}\quad\includegraphics[scale=0.4]{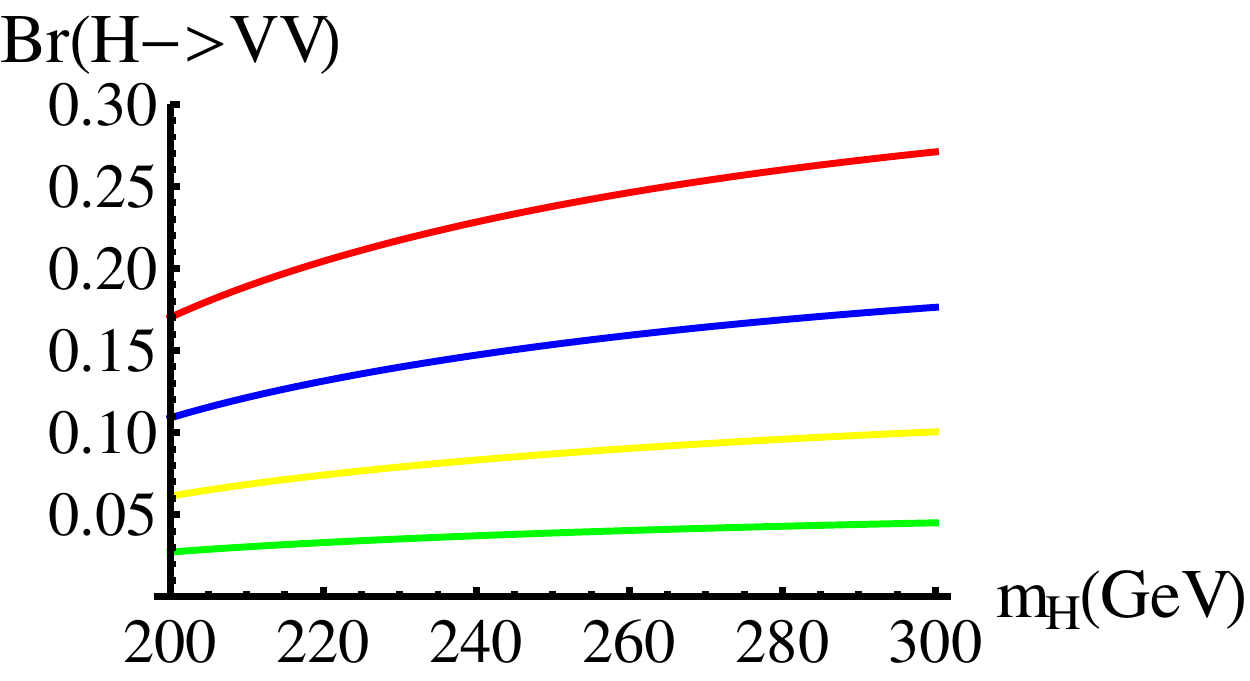}\quad\includegraphics[scale=0.4]{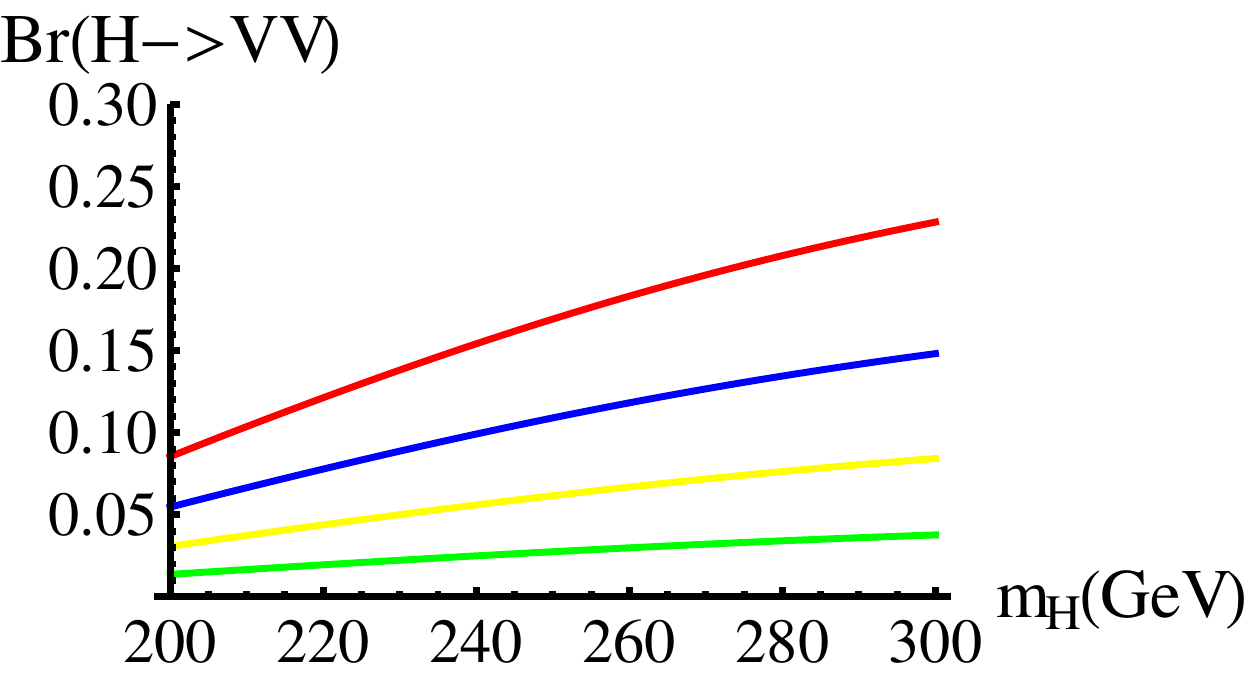}
\end{figure}

The direct searches for a heavy Higgs boson at LHC \cite{cmsheavy,atlasheavy} excluded a SM Higgs boson in the mass region
$(145-1000)\textrm{GeV}$ at $95\%$ C.L. A SM Higgs boson with its mass around $v$ would decay to $WW$ and $ZZ$
final states dominantly with $\textrm{Br}(H_{\textrm{SM}}\rightarrow VV)\approx1$ \cite{smhiggs},
while in 2HDM it can be modified because of a suppressed $HVV$ coupling and the existence of other decay channels like
$H\rightarrow Z\eta,\eta\eta,h\eta,$ and $Zh$ (if $m_H>m_Z+m_h=216\textrm{GeV}$), $hh$ (if $m_H>2m_h=250\textrm{GeV}$),
$H^+H^-$ (if $m_H>2m_{\pm}$). For a heavy scalar $H$, analytically the partial widths should be
\begin{eqnarray}
\Gamma_H(VV)&\approx&c_{H,V}^2\Gamma_{H,\textrm{SM}};\\
\Gamma_H(\eta\eta)&=&\frac{g^2_{H\eta\eta}v^2}{32\pi m_H}\sqrt{1-\frac{4m^2_{\eta}}{m^2_H}};\\
\Gamma_H(Z\eta)&=&\frac{c_{h,V}^2m^3_H}{8\pi v^2}\mathcal{F}\left(\frac{m^2_{\eta}}{m^2_H},\frac{m^2_Z}{m^2_H}\right).
\end{eqnarray}
The suppression in $\Gamma_H(VV)$ comes from small $c_{H,V}$ while $\Gamma_H(Z\eta)\propto c^2_{h,V}$ is not suppressed because
$h$ is SM-like and $c_{h,V}\sim1$. According to CMS results \cite{cmsheavy} which gave the most strict constraint, for $m_H\sim
(200-300)\textrm{GeV}$, the $95\%$ C.L. upper limit for the signal strength is \footnote{For a heavy Higgs boson,
$\textrm{Br}_{\textrm{SM}}(H\rightarrow VV)\sim1$ according to \cite{smhiggs}.}
\begin{equation}
\mu_H\equiv\frac{\sigma_H}{\sigma_{H,\textrm{SM}}}\cdot\frac{\textrm{Br}(H\rightarrow VV)}{\textrm{Br}_{\textrm{SM}}(H\rightarrow VV)}\lesssim(0.1-0.2).
\end{equation}
Numerically, we show the $\textrm{Br}(H\rightarrow VV)-m_H$ plots for different parameter choices fixing $c_{\eta,V}=0.1$ in \autoref{HVVplot}.
From the figures, we can see that if the production cross section $\sigma_H\sim\sigma_{H,\textrm{SM}}$, $c_{H,V}\lesssim0.3$ would be allowed;
while if $\sigma_H\sim0.5\sigma_{H,\textrm{SM}}$, $c_{H,V}\lesssim0.4$ would also be allowed. It is not sensitive to $m_{\eta}$. We did
not consider the $H\rightarrow hh$ channel for $m_H>2m_h=250\textrm{GeV}$ in the discussions above. Numerically, for $g_{Hhh}\sim1$, we
have $\textrm{Br}(H\rightarrow hh)\lesssim0.1$ which leads to $\sigma(pp\rightarrow H\rightarrow hh)\lesssim0.4\textrm{pb}$ \cite{smhiggs}.
For this case, the direct search for $H\rightarrow hh$ channel by CMS \cite{Hhh} cannot give further constraint.
We don't consider the case $m_H\gg v$ here because of the weakly-coupled hypothesis.

No significant evidence for a charged Higgs boson had been found at colliders. Recently the ATLAS searches through
$gb\rightarrow tH^-(\rightarrow\bar{t}b)$ process gave constraint on the $tbH^{\pm}$ vertex as
\cite{chargh}
\begin{equation}
\label{xitt}
|\xi_{tt}|\lesssim(1.5-3)
\end{equation}
for a charged Higgs boson with mass $m_{\pm}$ in the region $(200-600)\textrm{GeV}$. In these searches, some hints for a
charged Higgs signal with about $2.4\sigma$ significance were also found in this mass region. As can be seen below,
it is consistent with this scenario.
\subsection{Global-fits for Higgs Signal Strengths}
\begin{figure}[h]
\caption{Feynman diagrams for $h\rightarrow\gamma\gamma$ decay in this model.}\label{hgaga}
\includegraphics[scale=0.85]{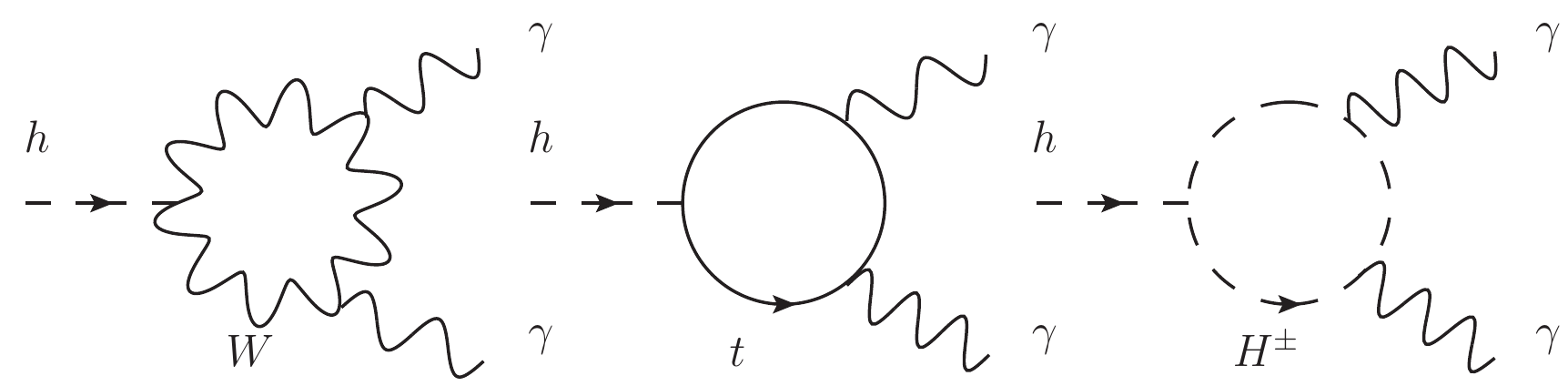}
\end{figure}

The Higgs signal strength for a channel which exists in SM is defined as
\begin{equation}
\label{mu}
\mu_{i,f}\equiv\frac{\sigma\cdot\textrm{Br}}{(\sigma\cdot\textrm{Br})_{\textrm{SM}}}=
\frac{\sigma_i}{\sigma_{i,\textrm{SM}}}\cdot\frac{\Gamma_h(f)}{\Gamma_{h,\textrm{SM}}(f)}
\cdot\frac{\Gamma_{h,\textrm{tot},\textrm{SM}}}{\Gamma_{h,\textrm{tot}}}.
\end{equation}
The SM Higgs boson with its mass $m_h=125\textrm{GeV}$ has a total width $\Gamma_{h,\textrm{tot},\textrm{SM}}=4.1\textrm{MeV}$ \cite{smhiggs}.
In this scenario, $\Gamma_{h,\textrm{tot}}$ is also modified by the exotic decay channels $h\rightarrow Z\eta,\eta\eta$. Here for VBF
or $Vh$ associated production channel, $\sigma/\sigma_{\textrm{SM}}=c^2_{h,V}$; while for gluon fusion production,
\begin{equation}
\frac{\sigma}{\sigma_{\textrm{SM}}}=\left|\textrm{Re}(c_{h,t})+\textrm{i}\frac{\mathcal{B}_{1/2}(x_t)}{\mathcal{A}_{1/2}(x_t)}\textrm{Im}(c_{h,t})\right|^2.
\end{equation}
For the decay channels $h\rightarrow WW^*$ and $ZZ^*$, we have $\Gamma_h(VV)/\Gamma_{h,\textrm{SM}}(VV)=c^2_{h,V}$; for
$h\rightarrow b\bar{b},c\bar{c}$ and $\tau^+\tau^-$, $\Gamma_h(f)/\Gamma_{h,\textrm{SM}}(f)=|c_{h,f}|^2$; while for the
loop induced decay processes,
\begin{equation}
\label{gg}
\frac{\Gamma_h(gg)}{\Gamma_{h,\textrm{SM}}(gg)}=
\left|\textrm{Re}(c_{h,t})+\textrm{i}\frac{\mathcal{B}_{1/2}(x_t)}{\mathcal{A}_{1/2}(x_t)}\textrm{Im}(c_{h,t})\right|^2;
\end{equation}
\begin{equation}
\label{gaga}
\frac{\Gamma_h(\gamma\gamma)}{\Gamma_{h,\textrm{SM}}(\gamma\gamma)}=
\left|\frac{c_{h,V}\mathcal{A}_1(x_W)+\frac{4}{3}\textrm{Re}(c_{h,t})\mathcal{A}_{1/2}(x_t)+\left(\frac{g_{h,\pm}v^2}{2m^2_{\pm}}\right)\mathcal{A}_0(x_{\pm})
+\frac{4}{3}\textrm{i}\textrm{Im}(c_{h,t})\mathcal{B}_{1/2}(x_t)}{\mathcal{A}_1(x_W)+\frac{4}{3}\mathcal{A}_{1/2}(x_t)}\right|^2.
\end{equation}
Here $x_i\equiv m^2_h/4m_i^2$ where $i$ denotes the particles $t,W$ or $H^{\pm}$ in loops. The index $j$ in $\mathcal{A}(\mathcal{B})_j$
denotes the spin of the particle in loops, see the Feynman diagrams in \autoref{hgaga}.
The analytical loop integration functions given by \cite{higgsI,higgsII}
are listed in \autoref{loop} as (\ref{gggaga1})-(\ref{gggaga5}). According to \cite{smhiggs},
\begin{eqnarray}
\Gamma_{h,\textrm{tot}}&=&\Gamma_{h,\textrm{tot,SM}}\left(0.58|c_{h,b}|^2+0.24c^2_{h,V}+0.06|c_{h,\tau}|^2+0.03|c_{h,c}|^2\right.\nonumber\\
&&\left.+0.09|c_{h,t}|^2(1+1.31\sin^2\alpha_t)\right)+\Gamma_{h,\textrm{exo}}
\end{eqnarray}
where $\Gamma_{h,\textrm{tot,SM}}=4.1\textrm{MeV}$ for $m_h=125\textrm{GeV}$. $\alpha_t\equiv\arg(c_{h,t})$ and $\Gamma_{h,\textrm{exo}}$
is the exotic decay width. Define
\begin{equation}
\chi^2\equiv\mathop{\sum}_{i,f}\left(\frac{\mu_{i,f,\textrm{obs}}-\mu_{i,f,\textrm{pre}}}{\sigma_{i,f}}\right)^2
\end{equation}
ignoring the correlations between different channels. $\mu_{i,f,\textrm{obs}(\textrm{pre})}$ means the observed (predicted) signal strength for
production channel $i$ and decay final state $f$ and $\sigma_{i,f}$ means the standard deviation of the signal
strength measurement for the corresponding channel. Numerically, the fitting results are not sensitive to the charged Higgs contribution in
$h\rightarrow\gamma\gamma$ channel.

According to (\ref{chf}), in this scenario, $c_{h,f}\sim1$ holds for all fermions since $h$ contains large component of $R_1$. Thus 
for all $c_{h,f}$, the modifications from $1$ are suppressed
by $t_{\beta}$. We also have $c_{h,V}\sim1$ in the text above. Thus for any channel, according to (\ref{mu})
\begin{equation}
\mu_{i,f,\textrm{pre}}=
\frac{\sigma_i}{\sigma_{i,\textrm{SM}}}\cdot\frac{\Gamma_h(f)}{\Gamma_{h,\textrm{SM}}(f)}
\cdot\frac{\Gamma_{h,\textrm{tot},\textrm{SM}}}{\Gamma_{h,\textrm{tot}}}
\sim\frac{\Gamma_{h,\textrm{tot},\textrm{SM}}}{\Gamma_{h,\textrm{tot}}},
\end{equation}
which means the signal strengths are mainly modified by the exotic decay width $\Gamma_{\textrm{exo}}$.
Numerically $\Gamma_{\textrm{exo}}\lesssim(1-2)\textrm{MeV}$ is still allowed for other couplings close to those in SM. For
$m_{\eta}<m_h/2$, $h\rightarrow\eta\eta$ channel is available. And according to (\ref{hetaeta}), we have
\begin{equation}
g_{h\eta\eta}\lesssim\mathcal{O}(10^{-2})
\end{equation}
which means a strong correlation among $\lambda_i$ in Higgs potential. To the leading order ,
\begin{equation}
g_{h\eta\eta}=(\lambda_3+\lambda_7)c_{\theta}+\frac{1}{2}\lambda_5s_{\theta}+\mathcal{O}(t_{\beta}s_{\xi}),
\end{equation}
which gives $\lambda_3+\lambda_7\simeq-\lambda_5t_{\theta}/2+\mathcal{O}(t_{\beta}s_{\xi})$. While for
$m_{\eta}<m_h-m_Z$, $h\rightarrow Z\eta$ channel is open, (\ref{hZeta}) gave
\begin{equation}
c_{H,V}=c_{h\eta}\lesssim\mathcal{O}(10^{-2}-10^{-1}).
\end{equation}
For $m_{\eta}\sim(15-30)\textrm{GeV}$, $c_{h\eta}=0.05$ is till allowed.

According to the direct searches for heavy neutral Higgs boson $H$, we can see that $c_{H,V}=0.3$ is in the allowed region
for almost all cases. While according to the bounds from Higgs signal strengths, we can see for $m_{\eta}<m_h-m_Z=34
\textrm{GeV}$, there would be further constraint on $c_{H,V}$ from $h\rightarrow Z\eta$ rare decay channel. In this case,
$c_{H,V}=0.05$ would be allowed. Thus we have two groups of typical benchmark points as listed in \autoref{bench}. We choose
$m_{\eta}=20\textrm{GeV}$ and $40\textrm{GeV}$ as the two typical cases.
\begin{table}[h]
\caption{Benchmark points in scalar sector for the following parts of this paper. The first line is a typical choice
for the case $h\rightarrow Z\eta$ decay allowed; while the second line is a typical choice for the case $h\rightarrow Z\eta$
decay forbidden.}\label{bench}
\begin{tabular}{|c|c|c|c|c|c|c|c|}
\hline
Case & $ m_{\eta}$ & $m_H$ & $c_{h,f}$ & $c_{\eta,V}$ & $c_{H,V}$ & $c_{h,V}$ & $t_{\beta}s_{\xi}$ \\
\hline
I & $20$ GeV & $\sim v$ & $\sim1$ & $0.1$ & $0.05$ & $0.994$ & $\sim0.1$ \\
\hline
II & $40$ GeV & $\sim v$ & $\sim1$ & $0.1$ & $0.3$ & $0.95$ & $\sim0.1$ \\
\hline
\end{tabular}
\end{table}
\subsection{Constraints from Oblique Parameters}
The GFitter group gave updated electro-weak fitting results \cite{obl} for oblique parameters \cite{oblth} as
\begin{equation}
\begin{array}{ccc}
S=0.05\pm0.11,&T=0.09\pm0.13,&U=0.01\pm0.11,\\
R_{ST}=+0.90,&R_{SU}=-0.59,&R_{TU}=-0.83;
\end{array}
\end{equation}
where $R$ means the correlation between two variables. Here $U$ is also treated as a free variable and the reference points
are taken as $m_{h,\textrm{ref}}=125\textrm{GeV}$, $m_{t,\textrm{ref}}=173\textrm{GeV}$. In 2HDM, $U$ is expected to be
ignorable thus we can fix $U=0$ and get \cite{obl}
\begin{equation}
S=0.06\pm0.09,\quad T=0.10\pm0.07,\quad R=+0.91.
\end{equation}
In 2HDM, the contribution to $\delta S$ and $\delta T$ \cite{2HDM,DSDT,DSDT2} are
\begin{eqnarray}
\delta S&=&\frac{1}{24\pi}\Bigg[(1-2s^2_W)^2G(z_{\pm},z_{\pm})+c_1^2G(z_2,z_3)+c_2^2G(z_3,z_1)+c_3^2G(z_1,z_2)\nonumber\\
&&+\mathop{\sum}_{i=1}^3\left(c^2_iH(z_i)+\ln\left(\frac{m^2_i}{m^2_{H^{\pm}}}\right)\right)
\label{DS}
-H\left(\frac{m^2_{h,\textrm{ref}}}{m^2_Z}\right)-\ln\left(\frac{m^2_{h,\textrm{ref}}}{m^2_{H^{\pm}}}\right)\Bigg];\\
\delta T&=&\frac{1}{16\pi s^2_Wm^2_W}\bigg[\mathop{\sum}_{i=1}^3(1-c_i^2)F(m^2_{H^{\pm}},m^2_i)-c_1^2F(m^2_2,m_3^2)
-c_2^2F(m^3_3,m_1^2)-c_3^2F(m_1^2,m_2^2)\nonumber\\
\label{DT}
&&+3\mathop{\sum}_{i=1}^3c^2_i(F(m^2_Z,m^2_i)-F(m^2_W,m^2_i))-3(F(m^2_Z,m^2_{h,\textrm{ref}})-F(m^2_W,m^2_{h,\textrm{ref}}))\bigg].
\end{eqnarray}
The arguments above are defined as $z_i\equiv(m_i/m_Z)^2$ and $z_{\pm}\equiv(m_{\pm}/m_Z)^2$.
The analytical loop integration functions given by \cite{2HDM,DSDT,DSDT2} are listed in \autoref{loop} as (\ref{ST1})-(\ref{ST4}).

\begin{figure}[h]
\caption{Oblique parameter constraints for the scenario we discussed in this paper. The green region is $68\%$ C.L.
allowed and the yellow region is $95\%$ C.L. allowed. The left figure is for Case I while the right figure is for
Case II in \autoref{bench}. We plot the curves with a parameter $m_{\pm}$. In each curve, we begin with $m_{\pm}=80
\textrm{GeV}$. In both figures, the curves from left to right are for
$m_H=(200,250,300)\textrm{GeV}$ respectively. For the allowed regions in the curves, we made them thick and black,
please see the allowed regions in \autoref{mpm} in details.}\label{STobl}
\includegraphics[scale=0.6]{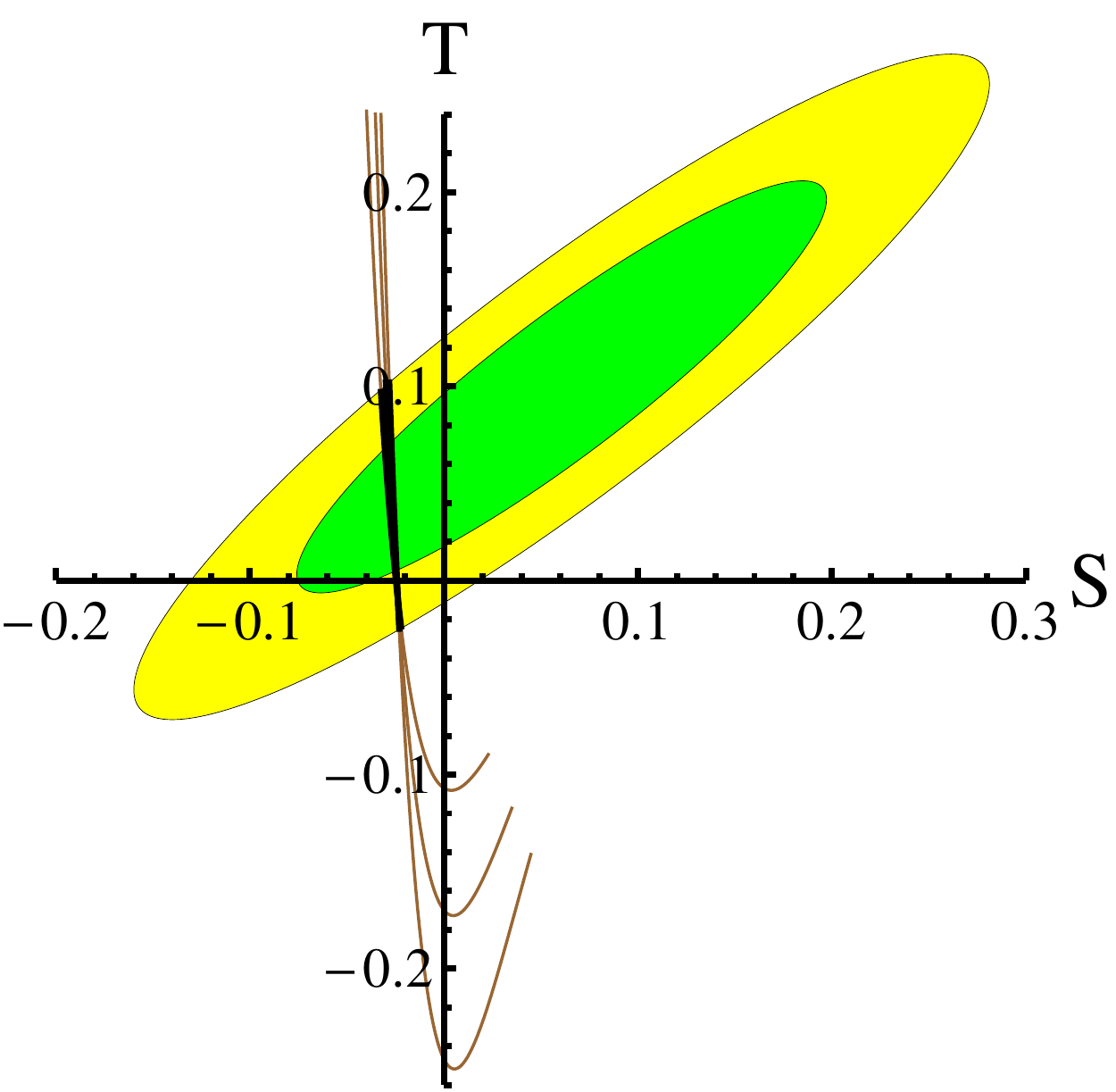}\quad\includegraphics[scale=0.6]{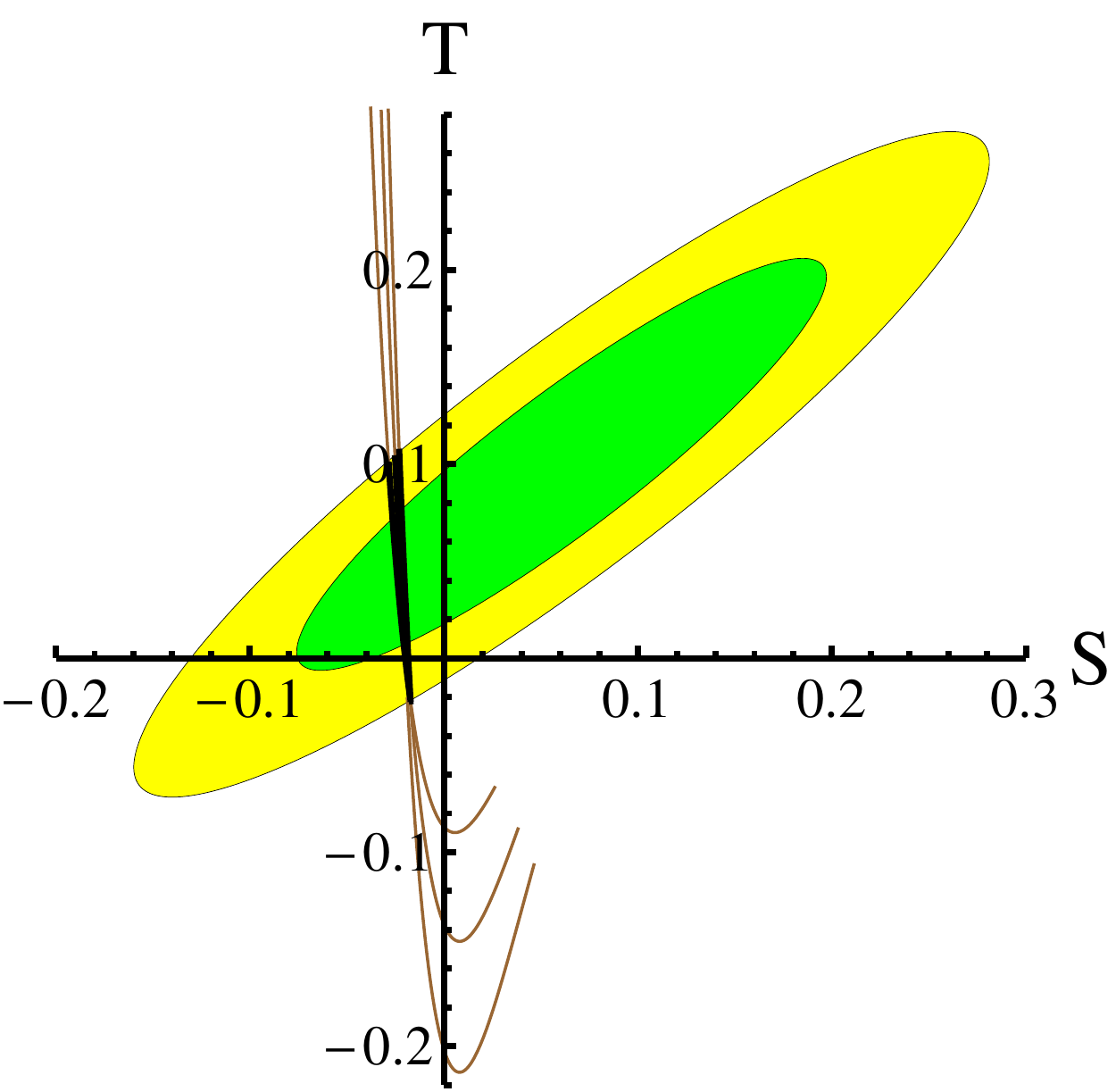}
\end{figure}

We perform the fitting process based on the mathematica code \cite{math} \footnote{The second $\chi^2$ (for $95\%$ C.L.)
should be $6.0$ according to \cite{PDG}.} with the benchmark points in \autoref{bench}. We plot the curves using the
charged Higgs mass $m_{\pm}$ as a parameter in \autoref{STobl}. Direct searches by LEP gave
constraints on charged Higgs boson mass as $m_{\pm}>78.6\textrm{GeV}$ \cite{LEP4}
at $95\%$ C.L. so that we begin from $m_{\pm}=80\textrm{GeV}$. The thick regions in the curves
stands for allowed regions by oblique parameter constrains for both benchmark points.
For both cases in \autoref{bench}, we list the allowed $m_{\pm}$ in \autoref{mpm}.
\begin{table}[h]
\caption{Allowed regions for $m_{\pm}$ for each case above.}\label{mpm}
\begin{tabular}{|c|c|c|c|}
\hline
$m_H$ (GeV) & $200$ & $250$ & $300$ \\
\hline
Allowed $m_{\pm}$ for Case I (GeV) & $190-231$ & $242-277$ & $293-323$ \\
\hline
Allowed $m_{\pm}$ for Case II (GeV) & $185-228$ & $232-269$ & $279-311$ \\
\hline
\end{tabular}
\end{table}
For all the cases, allowed $m_{\pm}$ are around the heavy neutral Higgs mass $m_H$, as the scenario discussed by
\cite{2HDM0,2HDMcus}. For $m_H$ around the electro-weak scale $v$, a charged Higgs boson should also have
its mass around that scale. A light charged Higgs boson (with its mass
$m_{\pm}<m_t$) is disfavored here thus we don't consider the constraints from the rare decay process $t\rightarrow H^+b$.
\subsection{Constraints from Meson Mixing Data}
\label{mixcons}
The neutral mesons $K^0$, $D^0$, $B^0$, and $B^0_s$ should mix with their anti-particles through $W^{\pm}$
mediated box diagrams in the SM. Thus a nontrivial contribution to $\langle\bar{M}^0|\mathcal{H}|M^0\rangle$
leads to the mass splitting effect between different CP eigenstates for meson \footnote{In fact in the real
world, CP is not a good symmetry thus a mass eigenstate is modified a little from a CP eigenstate. See the
details for this formalism in \autoref{mix}.}. Here we list the experimental data \cite{PDG,HFAG} and
SM predictions \cite{mixpre1,mixpre2,mixpre3,mixpre4} \footnote{No SM prediction results for $\Delta m_D$ appears
because the dominant contribution comes from long-distance interactions thus it is difficult to calculate.}
for meson mixing in \autoref{mixing} where the decay constants and bag parameters are from lattice data \cite{lattice}.
\begin{table}[h]
\caption{Experimental data and SM predictions for mass splitting effects in meson mixing.}\label{mixing}
\begin{tabular}{|c|c|c|}
\hline
Meson&$\Delta m_{\textrm{exp}}$ (GeV)&$\Delta m_{\textrm{SM}}$ (GeV)\\
\hline
$K^0(d\bar{s})$&$(3.483\pm0.006)\times10^{-15}$&$(3.30\pm0.34)\times10^{-15}$\\
\hline
$D^0(c\bar{u})$&$(5.9\pm2.6)\times10^{-15}$&$-$\\
\hline
$B^0_d(d\bar{b})$&$(3.36\pm0.02)\times10^{-13}$&$(3.57\pm0.60)\times10^{-13}$\\
\hline
$B^0_s(s\bar{b})$&$(1.1686\pm0.0014)\times10^{-11}$&$(1.14\pm0.17)\times10^{-11}$\\
\hline
\end{tabular}
\end{table}

\begin{figure}[h]
\caption{Additional Feynman diagrams contributed to $\mathbf{m}_{12,M}$ in Lee model.}\label{addi}
\includegraphics[scale=0.9]{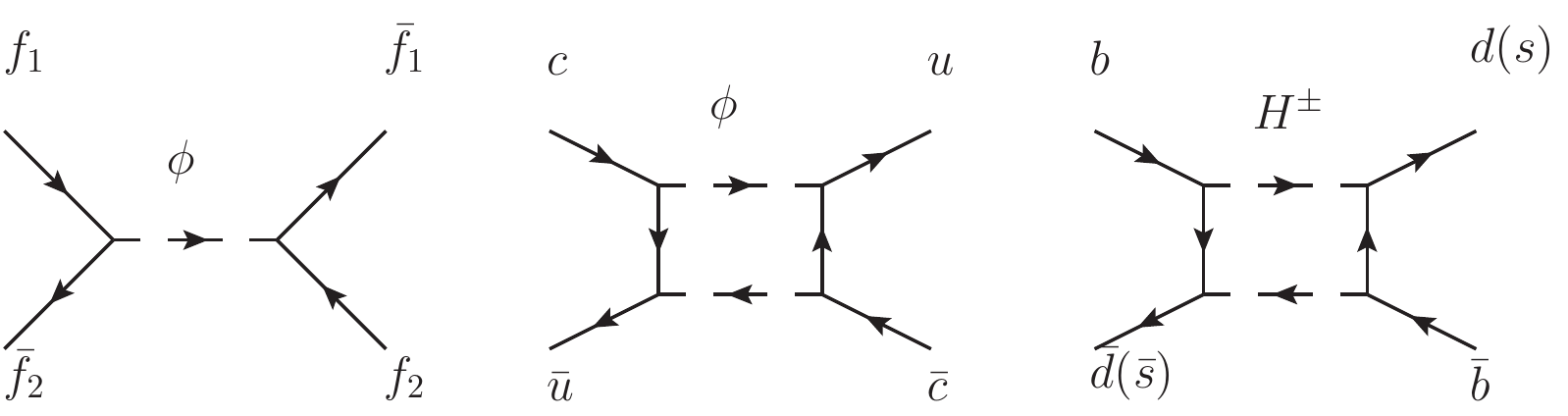}
\end{figure}

In general, we can parameterize the off-diagonal element in mass matrix as \cite{offdiag,offdiag2}
\begin{equation}
\label{mixpar}
\mathbf{m}_{12,M}\equiv\frac{1}{2m_M}\langle\bar{M}^0|\mathcal{H}|M^0\rangle=\mathbf{m}_{12,M}^{\textrm{SM}}(1+\Delta_M\textrm{e}^{2\textrm{i}\delta_M})
\end{equation}
where the factor $(2m_M)^{-1}$ comes from the normalization condition.
In SM we must have $\Delta_M=\delta_M=0$. In $B^0(B^0_s)$ system, $\Delta m_{B(B_s)}=2|\mathbf{m}_{12,B(B_s)}|$; while in $K^0$
system, $\Delta m_{K}=2\textrm{Re}\left(\mathbf{m}_{12,K}\right)$. A nonzero $\delta_M$ would also modify the CP-violation effects
from those in SM. In Lee model, the additional contributions to $\mathbf{m}_{12,M}$ are shown in \autoref{addi}.
The neutral scalars $\phi=\eta,h,H$ in the diagrams.

First, consider the left diagram in \autoref{addi} which induce the mixing directly at tree level. It can contribute to
the mixing of all the four kinds of mesons. The dominant contribution must come from $\eta$ because it is light and its
flavor changing couplings are not suppressed by $t_{\beta}s_{\xi}$ or $s_{\theta}$. The tree level $\eta$ induced
contribution for $M^0(f_i\bar{f}_j)-\bar{M}^0(f_j\bar{f}_i)$ mixing is \cite{mix1,mix2}
\begin{equation}
\mathbf{m}_{12,M}^{\eta,\textrm{tree}}=\frac{f^2_MB_Mm_Mm_im_j}{12m^2_{\eta}v^2}
\left[\left(1+\frac{6m_M^2}{(m_i+m_j)^2}\right)c_{\eta,ij}c^*_{\eta,ji}-\frac{5m^2_M}{2(m_i+m_j)^2}\left(c^2_{\eta,ij}
+c^{*2}_{\eta,ji}\right)\right].
\end{equation}
Here $f_M$ and $B_M$ are the decay constant and bag parameter for meson $M^0$ separately. According to (\ref{etaij}),
$c_{\eta,ij}=\pm\xi_{ij}\left(1+\mathcal{O}(t_{\beta}s_{\xi})\right)$. With the experimental constraints in
\cite{offdiag2,Bmix}, for different $\delta_{B(B_s)}$, $\Delta_{B(B_s)}\lesssim(0.1-0.4)$ at $95\%$ C.L. Assuming
$|\xi_{ij}|\sim|\xi_{ji}|$, numerically for $m_{\eta}\sim(20-40)\textrm{GeV}$, we have 
\begin{equation}
|\xi_{bd(db)}|\lesssim(0.7-3)\times10^{-2},\quad\quad|\xi_{bs(sb)}|\lesssim(0.9-2.5)\times10^{-2}.
\end{equation}
Similarly, $|\xi_{sd(ds)}|\lesssim(0.8-1.7)\times10^{-2}$ for $K^0-\bar{K}^0$ mixing from \cite{offdiag2}. While for
$D^0-\bar{D}^0$ mixing, we have $|\xi_{cu(uc)}|\lesssim(1.7-3.4)\times10^{-2}$. For all the four types of mixing,
the constraints on $\xi_{ij}$ are of $\mathcal{O}(10^{-2})$.

Next, consider the middle diagram in \autoref{addi} which can induce a $D^0-\bar{D}^0$ mixing through top quark and
a scalar mediated in the box. Assuming $|\xi_{tu(c)}|\sim|\xi_{u(c)t}|$, its contribution to $\Delta m_D$ is \cite{etabox}
\begin{equation}
\label{Dbox}
\Delta m_D^{\eta,\textrm{box}}\approx\frac{m_um_c|\xi_{tu}\xi_{tc}|^2}{24\pi^2v^4}f_D^2m_DB_Dr
\mathcal{F}_0\left(\frac{m^2_t}{m^2_{\eta}}\right)
\end{equation}
where $r=(\alpha_s(m_t)/\alpha_s(m_b))^{6/23}(\alpha_s(m_b)/\alpha_s(m_c))^{6/25}=0.8$ describes the QCD effects
and loop function $\mathcal{F}_0(x)$ \cite{etabox} is listed as (\ref{box1}) in \autoref{loop} in the appendices.
Assuming its contribution is less than the complete $\Delta m_D$, numerically we have
\begin{equation}
\label{tutc1}
|\xi_{tu}\xi_{tc}|\lesssim6
\end{equation}
for a $\eta$ with its mass $(20-40)\textrm{GeV}$.

Last, consider the right diagram in \autoref{addi} which induce $B^0(B^0_s)-\bar{B}^0(\bar{B}^0_s)$ mixing
through the box diagram in which one or two $W^{\pm}$ should be replaced by $H^{\pm}$ comparing with the case
in SM. This kind of diagrams are highly suppressed in $K^0-\bar{K}^0$
mixing. In neutral B sector, the contributions from $W^{\pm}-H^{\pm}$ box and $H^{\pm}$ box can be estimated
as \cite{pmbox}
\begin{equation}
\label{Bbox}
\Delta_{B(B_s)}\textrm{e}^{\textrm{i}\delta_{B(B_s)}}=\xi_{tt}^2\cdot\frac{\mathcal{F}_1(m^2_t/m^2_W,m^2_t/m^2_{\pm},m^2_{\pm}/m^2_W)
+\xi_{tt}^2\mathcal{F}_0(m^2_t/m^2_{\pm})}{\mathcal{F}_2(m^2_t/m^2_W)}.
\end{equation}
The loop functions $\mathcal{F}_i$ \cite{pmbox} are listed as (\ref{box1})-(\ref{box3}) in \autoref{loop} in the appendices,
and $\mathcal{F}_0$ is the same as that in the box diagram for $D^0-\bar{D}^0$ mixing in (\ref{Dbox}). It is sensitive
only to $\xi_{tt}$ because the other terms are suppressed by the mass of down type quarks. The S-T parameter fits favor
a charged Higgs boson with mass $m_{\pm}\sim m_H\sim v$ (see also \autoref{mpm}), so numerically we have
\begin{equation}
\label{tt}
|\xi_{tt}|\lesssim(0.6-0.9)
\end{equation}
using the $B^0(B^0_s)-\bar{B}^0(\bar{B}^0_s)$ mixing constraints \cite{offdiag2,Bmix}. This bound is stricter
than that from the direct searches for a charged Higgs boson in (\ref{xitt}).
\subsection{LHC Constraints on Top Quark Flavor Violation}
The $\phi tq$ (where $q=c,u$ and $\phi=\eta,h$) direct interactions in (\ref{etaij}) and (\ref{hij}) would induce
$t\rightarrow\phi q$ rare decay processes. The partial widths can be given by
\begin{equation}
\Gamma(t\rightarrow\phi q)=\frac{m^2_tm_q(|c_{\phi,tq}|^2+|c_{\phi,qt}|^2)}{32\pi v^2}\left(1-\frac{m^2_{\phi}}{m^2_t}\right)^2.
\end{equation}
For $\phi=\eta$, we have $c_{\eta,ij}=\textrm{i}\xi_{ij}+\mathcal{O}(t_{\beta}s_{\xi})\sim\textrm{i}\xi_{ij}$; while for
$\phi=h$, if $m_{\eta}<34\textrm{GeV}$, $c_{h,ij}\sim-\textrm{i}t_{\beta}s_{\xi}\xi_{ij}\sim-0.1\textrm{i}\xi_{ij}$ with
$t_{\beta}s_{\xi}\sim0.1$; else $c_{h,ij}\sim(-0.1\textrm{i}+\mathcal{O}(0.1))\xi_{ij}$. For the latter case,
\begin{equation}
|c_{h,ij}|\sim(0.1-0.3)|\xi_{ij}|.
\end{equation}
All the numerical estimations above are based on (\ref{etaij}) and (\ref{hij}) etc. in \autoref{coup}. The combined
experimental result by ATLAS \cite{htq} gave
\begin{equation}
\textrm{Br}(t\rightarrow hc)<0.46\%\quad\textrm{and}\quad\textrm{Br}(t\rightarrow hu)<0.45\%
\end{equation}
respectively at $95\%$ C.L. Assuming $|\xi_{ij}|\sim|\xi_{ji}|$ as usual, we have
\begin{equation}
\label{tutc2}
|\xi_{tu}|\lesssim(1-3)\times10^2\quad\textrm{and}\quad|\xi_{tc}|\lesssim(5-14)
\end{equation}
using the SM predicted top quark total width $\Gamma_{t,\textrm{tot}}\approx1.3\textrm{GeV}$ \cite{PDG,ttot}.

It is difficult to search for $t\rightarrow\eta q$ rare decay since $\eta$ decays to jets dominantly, but we
can obtain the constraints through the exotic decay branching ratio of top quark. The $t\bar{t}$ production cross section
measurements at LHC with $\sqrt{s}=8\textrm{TeV}$ gave $\sigma_{t\bar{t}}=(237\pm13)\textrm{pb}$ \cite{ttbarex}
assuming $m_t=173\textrm{GeV}$ and $\textrm{Br}(t\rightarrow Wb)=1$, which is consistent with the SM prediction
$\sigma_{t\bar{t},\textrm{SM}}=(246^{+9}_{-11})\textrm{pb}$ \cite{ttbarsm}. Thus we have for the top exotic decay
channels that $\Gamma(t\rightarrow\textrm{exotic})/\Gamma(t\rightarrow Wb)<8\%$ at $95\%$ C.L.
In this scenario, $\textrm{Br}(t\rightarrow\eta q)/
\textrm{Br}(t\rightarrow hq)\sim\mathcal{O}(10-10^2)$, thus $t\rightarrow hq$ is ignorable in this paragraph.
With these data, we have
\begin{equation}
\label{tutc3}
2\times10^{-4}|\xi_{tu}|^2+0.1|\xi_{tc}|^2\lesssim1
\end{equation}
for $m_{\eta}\sim(20-40)\textrm{GeV}$.

The last constraint comes from same sign top production. The $95\%$ C.L. upper limit given by CMS \cite{ttex}
is $\sigma_{tt}<0.37\textrm{pb}$. Theoretically, $\eta$ mediated $uu\rightarrow tt$ process would be the
dominant production channel in this scenario. The cross section can be expressed as
\begin{equation}
\sigma(uu\rightarrow tt)=\int dx_1dx_2f_u(x_1)f_u(x_2)\sigma(s_0)
\end{equation}
where $f_u(x)$ is the parton distribution function (PDF) for up quark and
\begin{eqnarray}
\sigma(s_0)=\frac{m^2_um^2_t\beta_t(|\xi_{tu}|^2+|\xi_{ut}|^2)^2}{64\pi s_0v^4}\int^1_{-1}dc_{\theta}\left[
\left(\frac{1-\beta_tc_{\theta}}{1+\beta^2_t+4m^2_{\eta}/s_0-2\beta_tc_{\theta}}\right)^2\right.\nonumber\\
\left.\left(\frac{1+\beta_tc_{\theta}}{1+\beta^2_t+4m^2_{\eta}/s_0+2\beta_tc_{\theta}}\right)^2
-\frac{1+\beta_t^2(c^2_{\theta}-2)}{(1+\beta^2_t+4m^2_{\eta}/s_0)^2-4\beta^2_tc^2_{\theta}}\right].
\end{eqnarray}
Here $s_0\equiv x_1x_2s_{\textrm{LHC}}$ is the square of energy in the moment center frame of two partons;
$\beta_t\equiv\sqrt{1-4m^2_t/s_0}$ is the velocity of top quark and $\theta$ is the azimuth angle of top quark
in respect to the beam line. Numerically, for $m_{\eta}\sim(20-40)\textrm{GeV}$, assuming $|\xi_{tu}|\sim|\xi_{ut}|$
and using the MSTW2008 PDF \cite{PDF}, we have
\begin{equation}
\label{tutc4}
|\xi_{tu}|\lesssim10^2.
\end{equation}

\begin{figure}[h]
\caption{Allowed region for top flavor changing couplings. Notice in the right figure we used double-log
coordinates to show a very large region.}\label{tutc}
\includegraphics[scale=0.6]{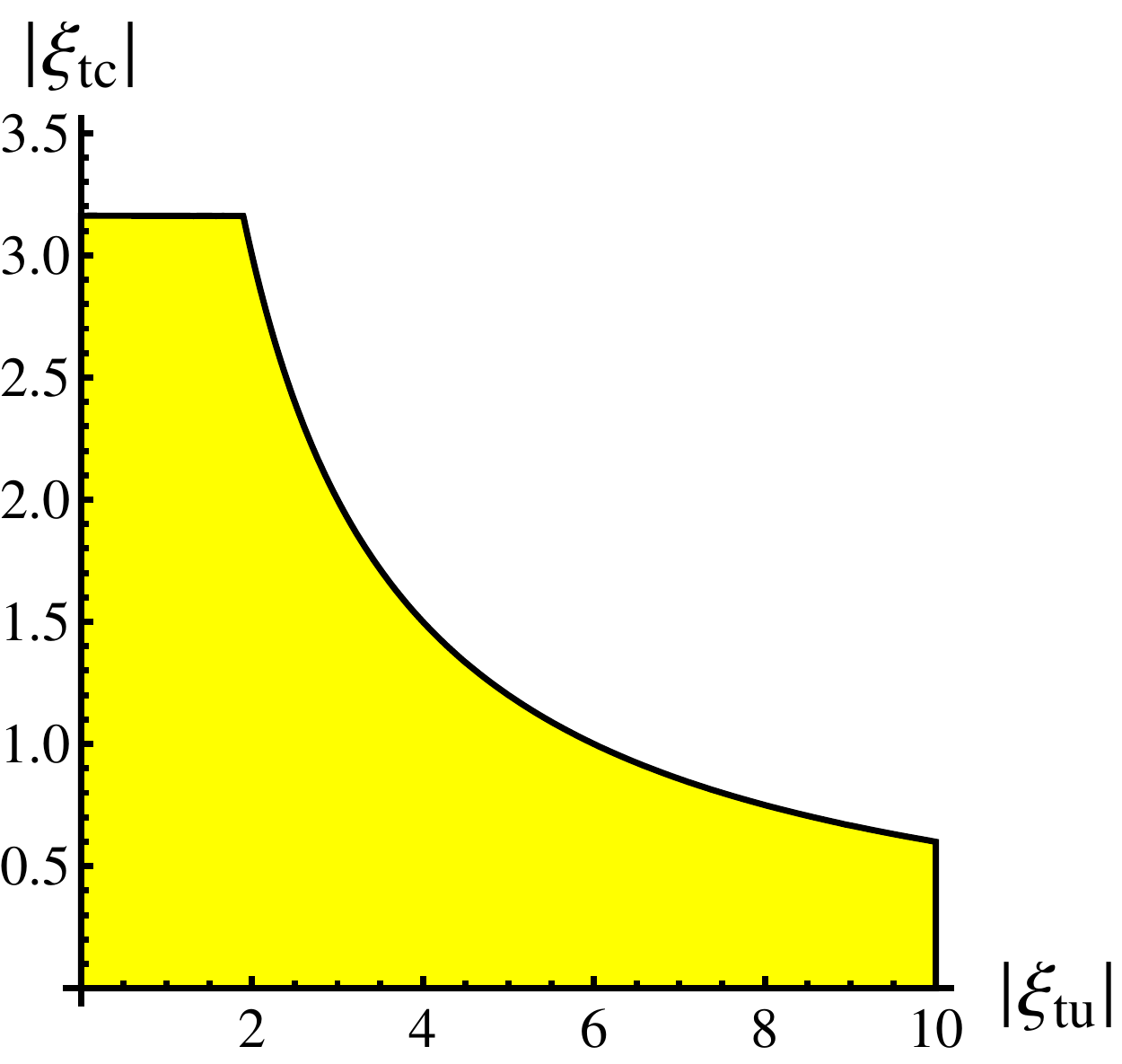}\quad\includegraphics[scale=0.6]{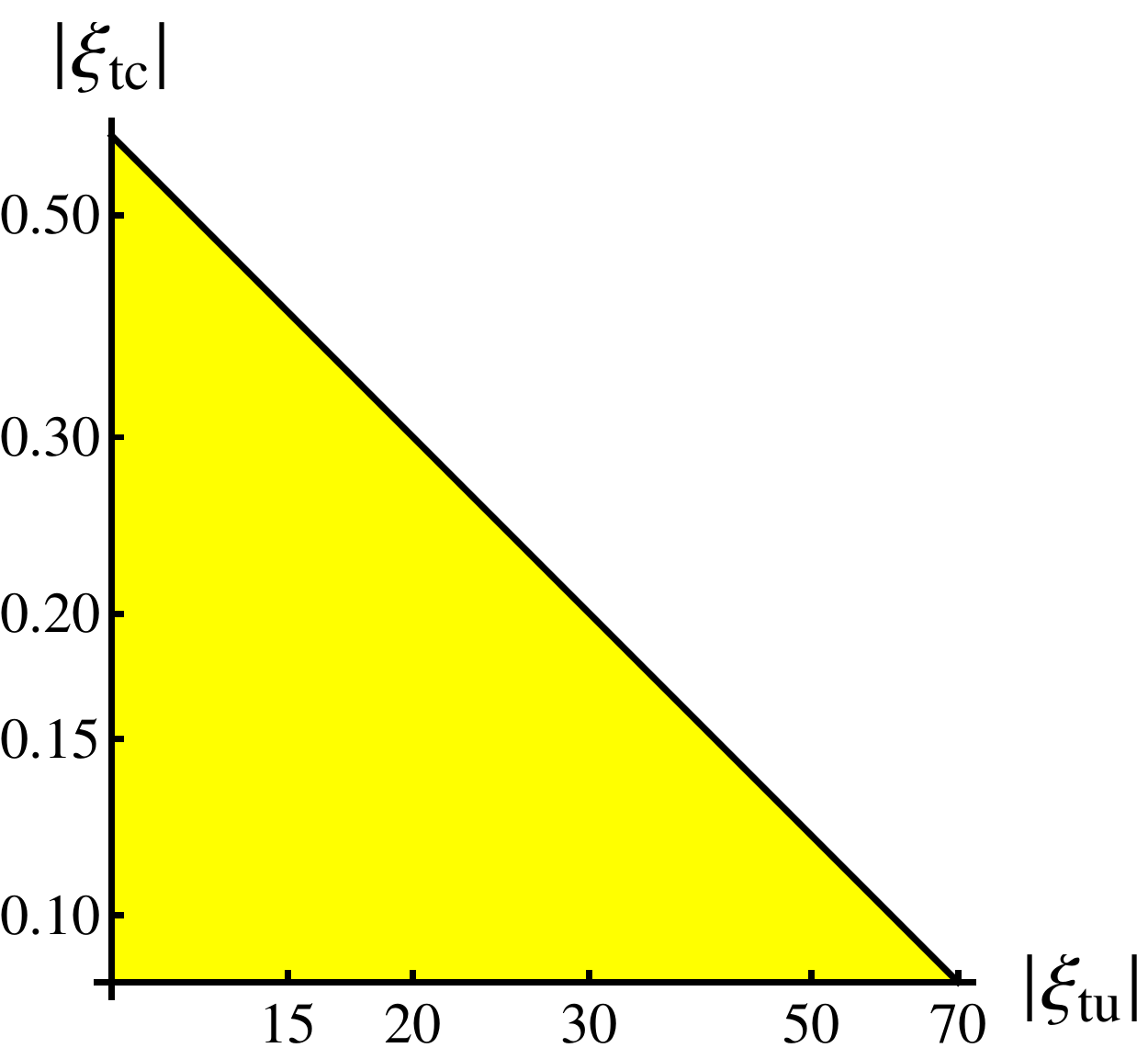}
\end{figure}

Combining the equations (\ref{tutc1}), (\ref{tutc2}), (\ref{tutc3}), and (\ref{tutc4}), we plot the estimations
of allowed region in the $|\xi_{tu}|-|\xi_{tc}|$ plane in \autoref{tutc}. The strictest upper limit
$|\xi_{tc}|\lesssim3$ and $|\xi_{tu}|\lesssim70$ comes from (\ref{tutc3}), and the obvious behavior of the
correlation between $|\xi_{tc}|$ and $|\xi_{tu}|$ comes from (\ref{tutc1}). The boundary contains relative
errors of $\mathcal{O}(10\%)$ and it is not sensitive to $m_{\eta}$ for $m_{\eta}\sim(20-40)\textrm{GeV}$.
\subsection{Constraints on Lepton Flavor Violation}
\label{LFV}
In type III 2HDM \cite{type3} there exist direct $\ell_i\ell_j\phi$ vertices to be constrained. For the discovered
125 GeV Higgs boson, a straightforward calculation gives \cite{cmslfv}
\begin{equation}
\textrm{Br}(h\rightarrow\ell_i^{\pm}\ell_j^{\mp})=\frac{m_hm_im_j}{8\pi\Gamma_hv^2}\left(|c_{h,ij}|^2+|c_{h,ji}|^2\right).
\end{equation}
For $h\rightarrow\mu\tau$ process, direct searches by CMS \cite{cmslfv} and ATLAS \cite{atlaslfv} collaborations
gave $\textrm{Br}(h\rightarrow\mu\tau)<1.51\%$ and $\textrm{Br}(h\rightarrow\mu\tau)<1.85\%$ respectively, both
at $95\%$ C.L. \footnote{Especially for $h\rightarrow\mu\tau$ signal, the CMS result gave a $2.4\sigma$ hint
corresponding to the best-fit branching ratio $\textrm{Br}(h\rightarrow\mu\tau)=(0.84^{+0.39}_{-0.37})\%$ \cite{cmslfv}.}
In this scenario, $|c_{h,ij}|$ is suppressed to be $(0.1-0.3)|\xi_{ij}|$ for $m_{\eta}\sim(20-40)\textrm{GeV}$, assuming
$|c_{h,ij}|\sim|c_{h,ji}|$, we have the bound
\begin{equation}
\label{mutau}
|\xi_{\mu\tau}|\lesssim(5-16).
\end{equation}

\begin{figure}[h]
\caption{Feynman diagrams contributed to radiative LFV decays $\ell_i\rightarrow\ell_j\gamma$.}\label{LFVdiag}
\includegraphics[scale=0.9]{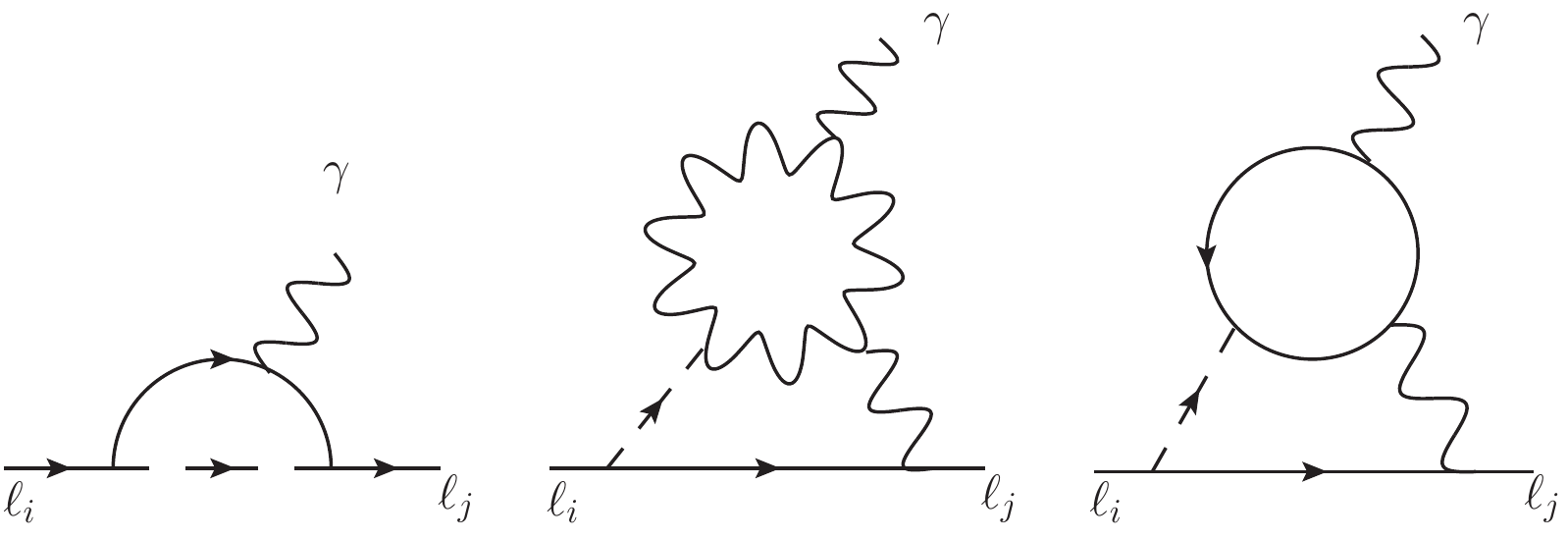}
\end{figure}

Another kind of strict constraints on $\ell_i\ell_j\phi$ vertices come from radiative LFV decays as $\tau\rightarrow\mu\gamma$ and
$\mu\rightarrow e\gamma$. For $\tau\rightarrow\mu(e)\gamma$, Belle and BaBar collaborations  gave the
$90\%$ C.L. upper limit as \cite{Bellelfv,BaBarlfv}
\begin{eqnarray}
&\textrm{Br}(\tau\rightarrow\mu\gamma)<4.5\times10^{-8},\quad\textrm{Br}(\tau\rightarrow e\gamma)<1.2\times10^{-7}\quad(\textrm{Belle});&\\
&\textrm{Br}(\tau\rightarrow\mu\gamma)<4.4\times10^{-8},\quad\textrm{Br}(\tau\rightarrow e\gamma)<3.3\times10^{-8}\quad(\textrm{BaBar}).&
\end{eqnarray}
While for $\mu\rightarrow e\gamma$, the MEG collaboration gave \cite{MEG}
\begin{equation}
\textrm{Br}(\mu\rightarrow e\gamma)<5.7\times10^{-13}
\end{equation}
at $90\%$ C.L. In SM, the branching ratios of $\ell_i\rightarrow\ell_j\gamma$ processes are estimated to be of
$\mathcal{O}(10^{-56}-10^{-54})$ \cite{PDG,smlfv,ourlfv} which are far below the experimental sensitivity. But
in 2HDM with LFV, it can be larger or even comparable to recent data. In this model, $\ell_i\rightarrow\ell_j\gamma$
process can be generated by Feynman diagrams in \autoref{LFVdiag}
and the branching ratios can be expressed as \cite{lfv1}
\begin{equation}
\frac{\textrm{Br}(\ell_i\rightarrow\ell_j\gamma)}{\textrm{Br}(\ell_i\rightarrow\ell_j\nu_i\bar{\nu}_j)}=
\frac{48\pi^3\alpha}{G_F^2}\left(|A_L|^2+|A_R|^2\right)
\end{equation}
where $A_{L(R)}$ are defined through \cite{lfv2,lfv3}
\begin{equation}
\mathcal{M}(\ell_i\rightarrow\ell_j\gamma)=em_i\bar{u}_j(p_j)\textrm{i}\sigma^{\mu\nu}q_{\nu}(A_LP_L+A_RP_R)u_i(p_i)\epsilon^*_{\mu}(q)
\end{equation}
in which $P_{L(R)}\equiv(1\mp\gamma^5)/2$ and $q$ is the momentum of photon. According to \autoref{LFVdiag}, there are
one-loop and two-loop contributions to these processes where the two-loop diagrams are called Barr-Zee type diagrams
\cite{bz} \footnote{This kind of two-loop diagrams were first used by Barr and Zee to calculate the electric dipole
moments for fermion in \cite{bz} which would also be discussed later.}. For $\tau\rightarrow\mu(e)\gamma$, The analytical
expression for left-handed (right-handed) amplitude should be \cite{ourlfv,lfv1,lfv2,lfv3,lfv4,lfv5}
\footnote{Notice that the analytical formulae for $\ell_i\rightarrow\ell_j\gamma$ decay process in these papers are not consistent with each other.
We checked the calculation during finishing our recent paper \cite{ourlfv} and confirmed the result by Omura et. al. \cite{lfv2,lfv3} is correct.}
\begin{eqnarray}
A_L^*(A_R)&=&A^*_{L,\textrm{one-loop}}(A_{R,\textrm{one-loop}})+A^*_{L,\textrm{two-loop}}(A_{R,\textrm{two-loop}})\nonumber\\
&=&\mathop{\sum}_{\phi}\frac{\sqrt{m_im_j}c_{\phi,ij}(c_{\phi,ji})}{16\pi^2v^2}\Bigg(\frac{m_i}{m^2_h}
\left(c_{\phi,i}\ln\left(\frac{m^2_h}{m^2_i}\right)-\frac{4}{3}|c_{\phi,i}|\cos(\alpha_{\phi,i})
-\frac{5}{3}\textrm{i}|c_{\phi,i}|\sin(\alpha_{\phi,i})\right)\nonumber\\
&&+\frac{c_{\phi,V}\alpha}{\pi m_i}\left(\left(3+\frac{m^2_{\phi}}{2m^2_W}\right)f\left(\frac{m^2_W}{m^2_{\phi}}\right)+
\left(\frac{23}{4}-\frac{m^2_{\phi}}{2m^2_W}\right)g\left(\frac{m^2_W}{m^2_{\phi}}\right)+\frac{3}{4}h\left(\frac{m^2_W}{m^2_{\phi}}\right)\right)\nonumber\\
\label{taudecay}
&&-\frac{8\alpha|c_{\phi,t}|}{3\pi m_i}\left(\cos(\alpha_{\phi,t})f\left(\frac{m^2_t}{m^2_{\phi}}\right)
+\textrm{i}\sin(\alpha_{\phi,t})g\left(\frac{m^2_t}{m^2_{\phi}}\right)\right)\Bigg)
\end{eqnarray}
where $i=\tau$, $j=e,\mu$, $\alpha_{\phi,f}\equiv\arg(c_{\phi,f})$ and the loop integration functions $f,g,h$ \cite{lfv1,lfv4}
are listed in (\ref{tau1})-(\ref{tau3}) in \autoref{loop}. Numerically the loop contributions with charge Higgs or $Z$ boson
inside are both small, thus we ignore them.  While for $\mu\rightarrow e\gamma$ decay which means $i=\mu$ and
$j=e$, The one-loop contribution should be changed to
\begin{equation}
A_{L,\textrm{one-loop}}^{*(\mu\rightarrow e\gamma)}=\sqrt{\frac{m_e}{m_{\mu}}}\mathop{\sum}_{\phi}\frac{m^2_{\tau}c_{\phi,\tau e}c_{\phi,\mu\tau}}{16\pi^2m_{\phi}^2v^2}\left(\ln\left(\frac{m^2_{\phi}}{m^2_{\tau}}\right)-\frac{3}{2}\right)
\end{equation}
because the loop with $\tau$ inside is expected to give larger contribution comparing with the $\mu$ case when adopting the Cheng-Sher ansatz \cite{CS}.
For $A_R$ we should take $c_{\phi,e\tau}c_{\phi,\tau\mu}$ instead of
$c_{\phi,\tau e}c_{\phi,\mu\tau}$ in $A^*_L$.

Numerically, we take the benchmark points as those in \autoref{bench}. For $m_{\eta}=20\textrm{GeV}$,
\begin{eqnarray}
\label{tmga1}
\textrm{Br}(\tau\rightarrow\mu\gamma)&\simeq&1.7\times10^{-10}\left(|\xi_{\tau\mu}|^2+|\xi_{\mu\tau}|^2\right)
\left|-5.7\xi_{\tau\tau}-5.4\xi_{tt}+1.2\textrm{i}\right|^2;\\
\label{tega1}
\textrm{Br}(\tau\rightarrow e\gamma)&\simeq&8.4\times10^{-13}\left(|\xi_{\tau e}|^2+|\xi_{e\tau}|^2\right)
\left|-5.7\xi_{\tau\tau}-5.4\xi_{tt}+1.2\textrm{i}\right|^2.
\end{eqnarray}
We used $\textrm{Br}(\tau\rightarrow e\nu_{\tau}\bar{\nu}_e)=17.8\%$ and $\textrm{Br}
(\tau\rightarrow\mu\nu_{\tau}\bar{\nu}_{\mu})=17.4\%$ \cite{PDG} in the calculations above.
For a typical case, $|\xi_{\mu(e)\tau}|\sim|\xi_{\tau\mu(e)}|$, $\xi_{tt}\sim0.6$ and $\xi_{\tau\tau}\sim1$, we have
$\textrm{Br}(\tau\rightarrow\mu\gamma)\sim3\times10^{-8}|\xi_{\mu\tau}|^2$, thus the upper limit for $|\xi_{\mu\tau}|$
should be around $1$. While for $\textrm{Br}(\tau\rightarrow e\gamma)\sim10^{-10}|\xi_{e\tau}|^2$,
$|\xi_{e\tau}|\sim\mathcal{O}(10)$ is still allowed. For $m_{\eta}=40\textrm{GeV}$,
\begin{eqnarray}
\label{tmga2}
\textrm{Br}(\tau\rightarrow\mu\gamma)&\simeq&1.7\times10^{-10}\left(|\xi_{\tau\mu}|^2+|\xi_{\mu\tau}|^2\right)
\left|-2\xi_{\tau\tau}-2.5\xi_{tt}-0.3+\textrm{i}\right|^2;\\
\label{tega2}
\textrm{Br}(\tau\rightarrow e\gamma)&\simeq&8.4\times10^{-13}\left(|\xi_{\tau e}|^2+|\xi_{e\tau}|^2\right)
\left|-2\xi_{\tau\tau}-2.5\xi_{tt}-0.3+\textrm{i}\right|^2.
\end{eqnarray}
Choosing the same parameters as above, $\textrm{Br}(\tau\rightarrow\mu\gamma)\sim5\times10^{-9}|\xi_{\mu\tau}|^2$ which
gives the upper limit of $|\xi_{\mu\tau}|$ to be around $3$. While for $\textrm{Br}(\tau\rightarrow e\gamma)
\sim2\times10^{-11}|\xi_{e\tau}|^2$, $|\xi_{e\tau}|\sim\mathcal{O}(10-10^2)$ are allowed. In the discussions
above, we assumed real $\xi_{tt(\tau\tau)}$. If $\xi_{tt(\tau\tau)}$ were complex, some accidental
cancelation would make larger $|\xi_{\mu\tau}|$ possible.

For $\tau\rightarrow\mu\gamma$ decay, it poses a stricter constraint than that from $h\rightarrow\mu\tau$ decay in (\ref{mutau})
with $m_{\eta}\sim(20-40)\textrm{GeV}$. Different from the cases discussed in \cite{ourlfv} in which the 125 GeV scalar
is the lightest one, in this scenario, the one-loop contribution from (20-40) GeV light scalar would be dominant or at least
comparable with the two-loop contributions. At the same time, $h\mu\tau$ vertex is suppressed by $s_{\theta}$ and
$t_{\beta}s_{\xi}$ to be of $\mathcal{O}(0.1)$. So that in this scenario, $\tau\rightarrow\mu\gamma$ decay gives dominant
constraint on the LFV vertex instead of $h\rightarrow\mu\tau$ decay. For $\tau\rightarrow e\gamma$ decay, $|\xi_{e\tau}|$ is
constrained to be less than $\mathcal{O}(10-10^2)$ which is still away from the expected magnitude by Cheng-Sher ansatz.

Numerically, for $\mu\rightarrow e\gamma$ decay, choosing typically $\xi_{ij}\sim\xi_{ji}$, we have
\begin{eqnarray}
\textrm{Br}(\mu\rightarrow e\gamma)&=&5.7\times10^{-9}|-\xi_{e\tau}\xi_{\mu\tau}+\xi_{e\mu}(-0.9\xi_{tt}+0.2\textrm{i})|^2,\quad(m_{\eta}=20\textrm{GeV});\\
\textrm{Br}(\mu\rightarrow e\gamma)&=&5.7\times10^{-11}|-3.6\xi_{e\tau}\xi_{\mu\tau}+\xi_{e\mu}(-7.7\xi_{tt}+1.6\textrm{i})|^2,\quad(m_{\eta}=40\textrm{GeV}).
\end{eqnarray}
Choosing $|\xi_{tt}|\sim0.6$ as usual, the three LFV couplings $\xi_{e\mu,e\tau,\mu\tau}$ are strongly correlated between
each other. The typical upper limit for $|\xi_{e\mu}|$ and $|\xi_{e\tau}\xi_{\mu\tau}|$ are both of $\mathcal{O}(10^{-2})$
for $m_{\eta}\sim(20-40)\textrm{GeV}$. For example, fixing $\xi_{e\tau}=0$ (or $\xi_{\mu\tau}=0$),
\begin{equation}
|\xi_{e\mu}|\lesssim(1.4-3.3)\times10^{-2};
\end{equation}
while fixing $\xi_{e\mu}=0$,
\begin{equation}
\label{etaumutau}
|\xi_{e\tau}\xi_{\mu\tau}|\lesssim(1.0-2.8)\times10^{-2}.
\end{equation}
\subsection{Constraints from Electric Dipole Moments}
The effective interaction for EDM of a fermion $f$ can be written as \cite{reviewEDM}
\begin{equation}
\mathcal{L}_{\textrm{EDM}}=-\frac{\textrm{i}}{2}d_f\bar{f}\sigma^{\mu\nu}\gamma^5fF_{\mu\nu}
\end{equation}
which violates both P and CP symmetries. In SM, the only origin of CP-violation is the
complex CKM matrix \cite{KM,CKM} thus the EDM for electron and neutron are generated at
four- and three-loop level respectively and they are estimated to be \cite{reviewEDM}
\begin{equation}
d_e\sim10^{-38}e\cdot\textrm{cm},\quad\textrm{and}\quad d_n\sim10^{-32}e\cdot\textrm{cm}.
\end{equation}
They are still far below the experimental upper limits \cite{eEDM,nEDM}
\begin{equation}
|d_e|<8.7\times10^{-29}e\cdot\textrm{cm},\quad\textrm{and}\quad |d_n|<2.9\times10^{-26}e\cdot\textrm{cm},
\end{equation}
both at $90\%$ C.L. In BSM with additional origins of CP-violation, the EDM for a fermion might be
generated at one- or two-loop level \footnote{Non-perturbation effects arising from $\theta$ term may also give significant contribution
to neutron EDM \cite{Strong}, but we don't include that in this paper.} thus they
can be quite larger than those in SM or even reach the sensitivity of recent data.

\begin{figure}[h]
\caption{Feynman diagrams contributed to EDM for a fermion $f$.}\label{EDMdiag}
\includegraphics[scale=0.72]{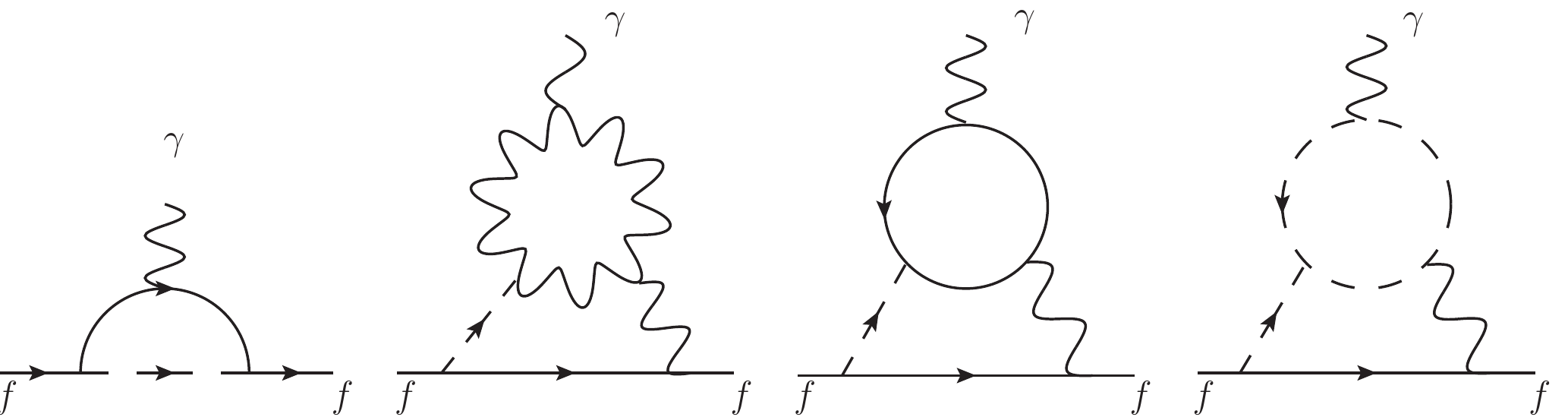}
\end{figure}
The EDM for a fermion $f$ can be generated from the Feynman diagrams in \autoref{EDMdiag} if there exist
CP-violation in $\phi f\bar{f}$ vertices. The two-loop diagrams are called Barr-Zee diagrams \cite{bz}.
If there is no CP-violation in flavor-changing vertices, the one loop contributions are proportional to
$(m_f/v)^3$ thus they are usually negligible for light fermions. The dominant contributions come from Barr-Zee
diagram as \cite{bz,bz1,bz2,bz3}
\begin{eqnarray}
\frac{d_f}{e}&=&\mathop{\sum}_{\phi}\frac{2\sqrt{2}\alpha_{\textrm{em}}G_FQ_fm_f|c_{\phi,f}|}{(4\pi)^3}\bigg(\sin\alpha_{\phi,f}\left(c_{\phi,V}
\mathcal{J}_1(m_W,m_{\phi})+g_{\phi,\pm}\mathcal{J}_0(m_{\pm},m_{\phi})\right)\nonumber\\
\label{bzEDM}
&&-\frac{8}{3}|c_{\phi,t}|\left(\sin\alpha_{\phi,t}\cos\alpha_{\phi,f}\mathcal{J}_{1/2}(m_t,m_{\phi})+
\cos\alpha_{\phi,t}\sin\alpha_{\phi,f}\mathcal{J}'_{1/2}(m_t,m_{\phi})\right)\bigg).
\end{eqnarray}
Here $Q_f$ is the electric charge for fermion $f$, $\alpha_{\phi,f}\equiv\arg(c_{\phi,f})$, and
the $\phi H^+H^-$ vertex $g_{\phi,\pm}\equiv(1/v)(\partial^3V/\partial\phi\partial
H^+\partial H^-)$ is defined in (\ref{sc}). The first term comes from
$W^{\pm}$ loop contribution (the second figure in \autoref{EDMdiag}); the second term comes from $H^{\pm}$ loop contribution
(the last figure in \autoref{EDMdiag}) \footnote{Numerically the charged Higgs contribution is small comparing with $W^{\pm}$
or top contributions as usual, but it may be comparable with experimental data especially for electron, so it's not negligible
like that in radiative LFV decay calculations.}; and the last two terms come from top loop contribution (the third figure in
\autoref{EDMdiag}). The loop functions $\mathcal{J}_i$ \cite{bz2} are all listed in \autoref{loop} in (\ref{edm1})-(\ref{edm4}).

\begin{figure}[h]
\caption{Constraints in $\alpha_{ee}-\alpha_{tt}$ plane by electron EDM. Fix $m_{\eta}=20\textrm{GeV}$, $|\xi_{tt}|=0.6$,
and $\xi_{ee}=1$. Yellow regions are allowed at $90\%$ C.L, the same till \autoref{EDMfig4}.}\label{EDMfig1}
\includegraphics[scale=0.57]{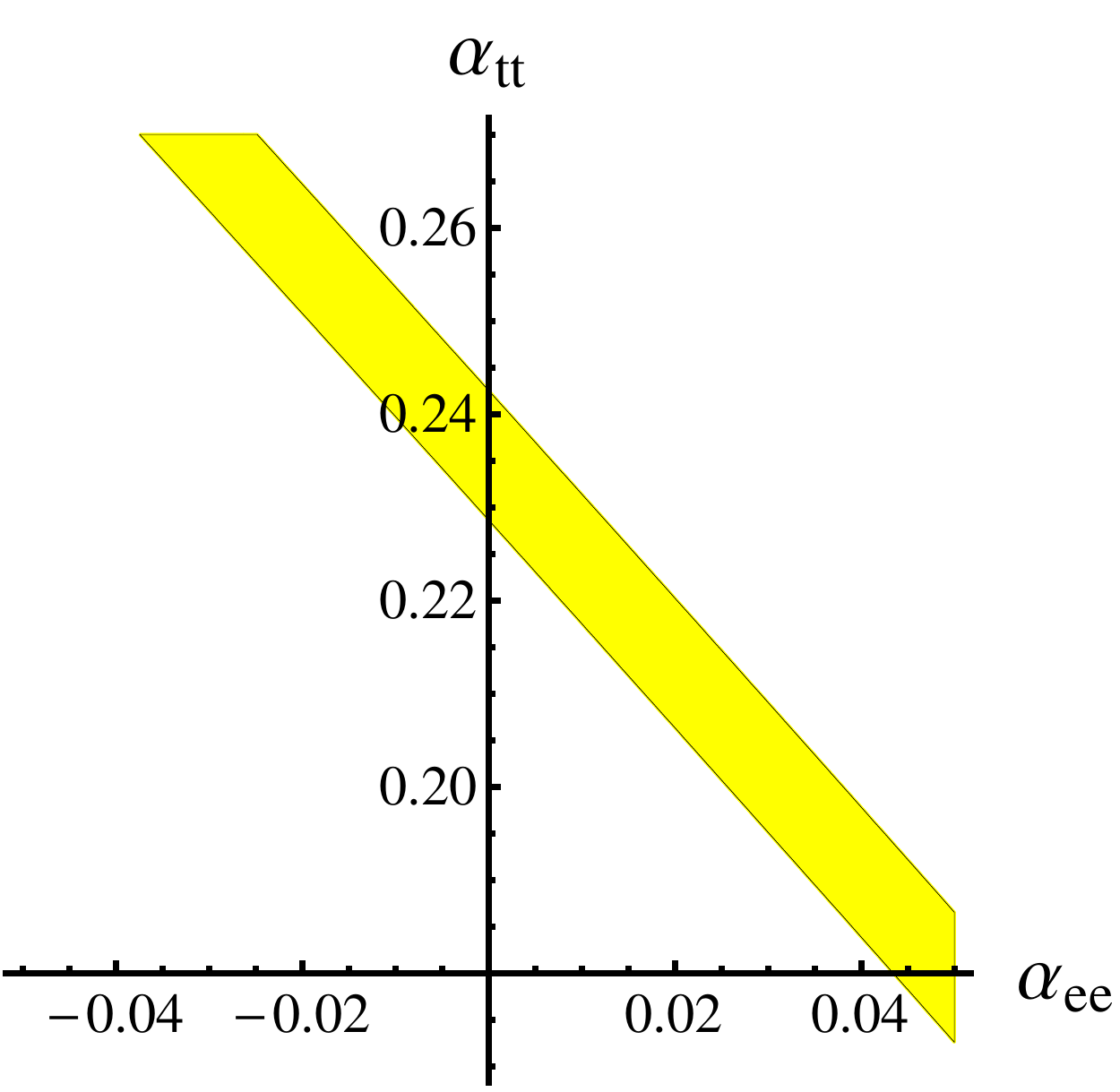}\quad\includegraphics[scale=0.57]{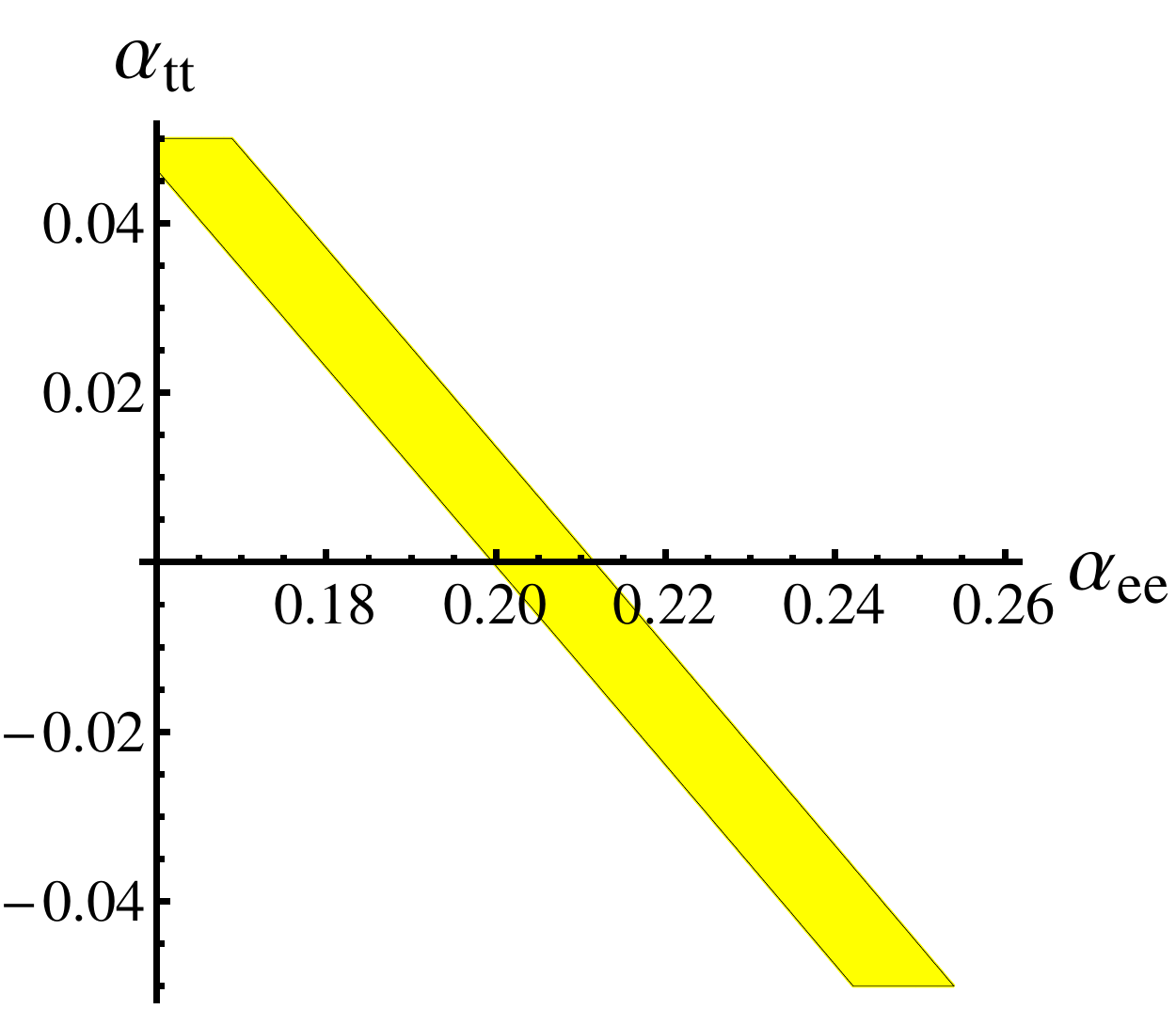}
\end{figure}

\begin{figure}[h]
\caption{Constraints in $\alpha_{ee}-\alpha_{tt}$ plane by electron EDM. Fix $m_{\eta}=20\textrm{GeV}$, $|\xi_{tt}|=0.6$,
and $\xi_{ee}=0.3$.}\label{EDMfig2}
\includegraphics[scale=0.57]{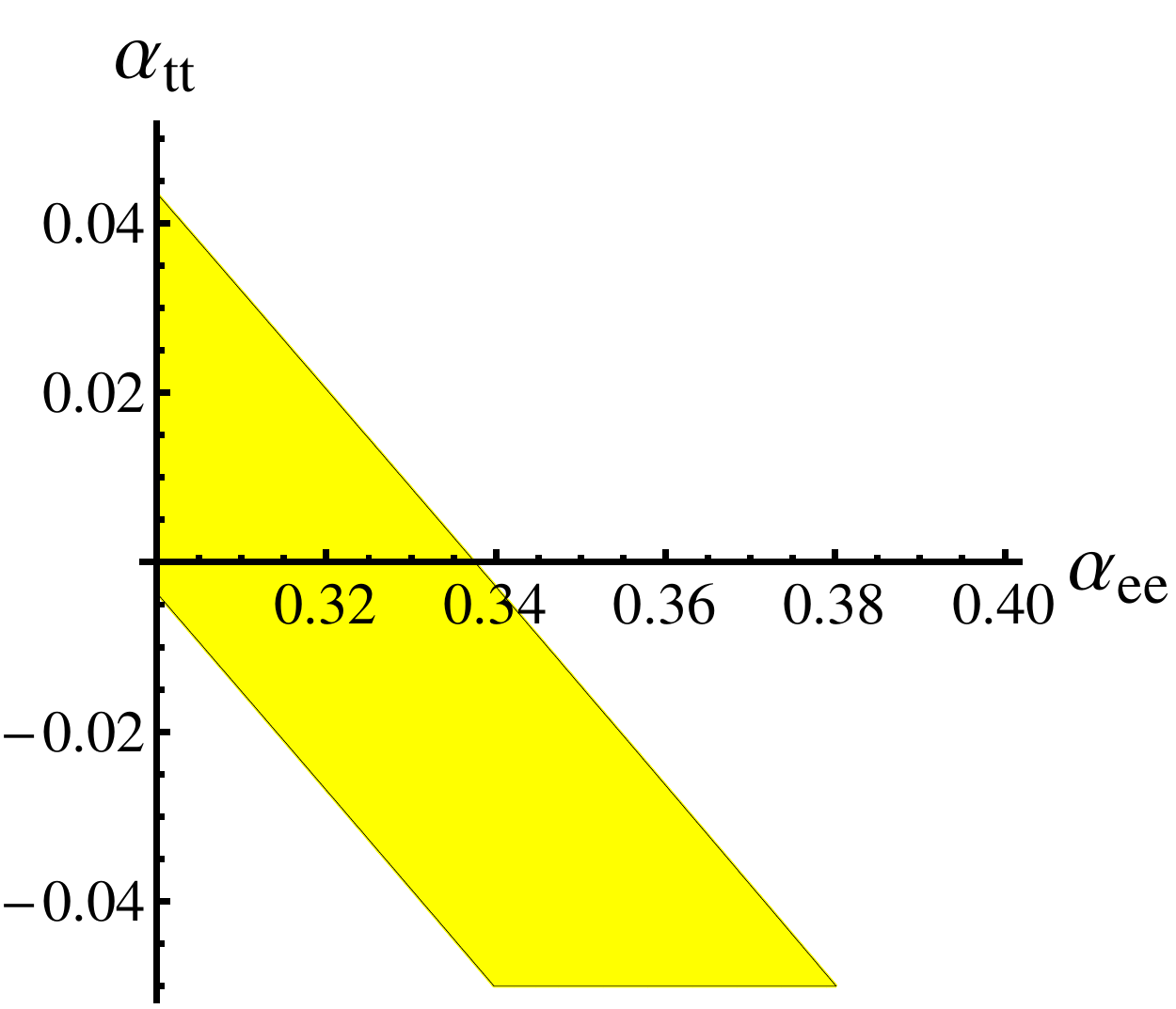}\quad\includegraphics[scale=0.57]{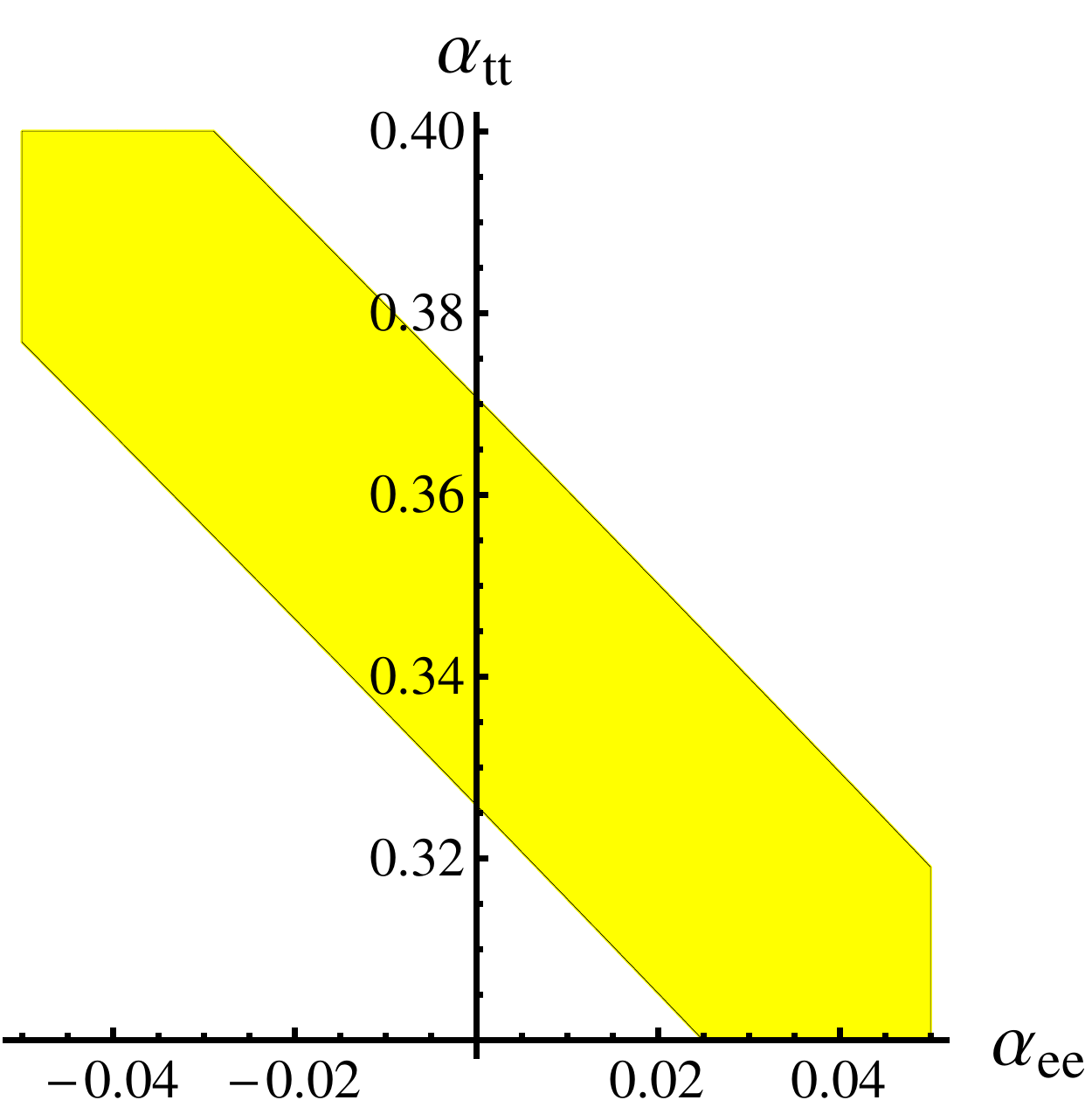}
\end{figure}

\begin{figure}[h]
\caption{Constraints in $\alpha_{ee}-\alpha_{tt}$ plane by electron EDM. Fix $m_{\eta}=40\textrm{GeV}$, $|\xi_{tt}|=0.6$,
and $\xi_{ee}=1$.}\label{EDMfig3}
\includegraphics[scale=0.57]{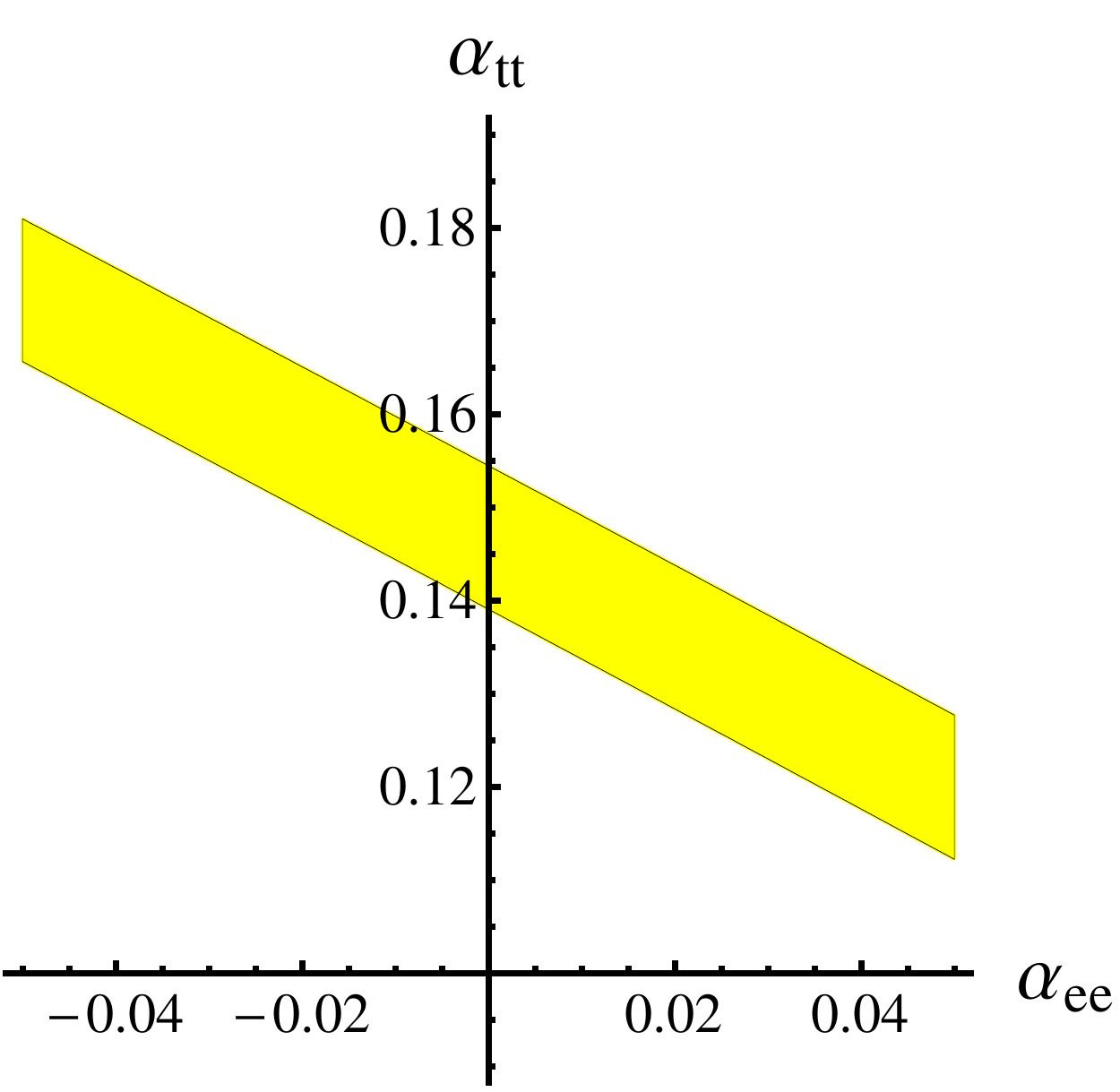}\quad\includegraphics[scale=0.57]{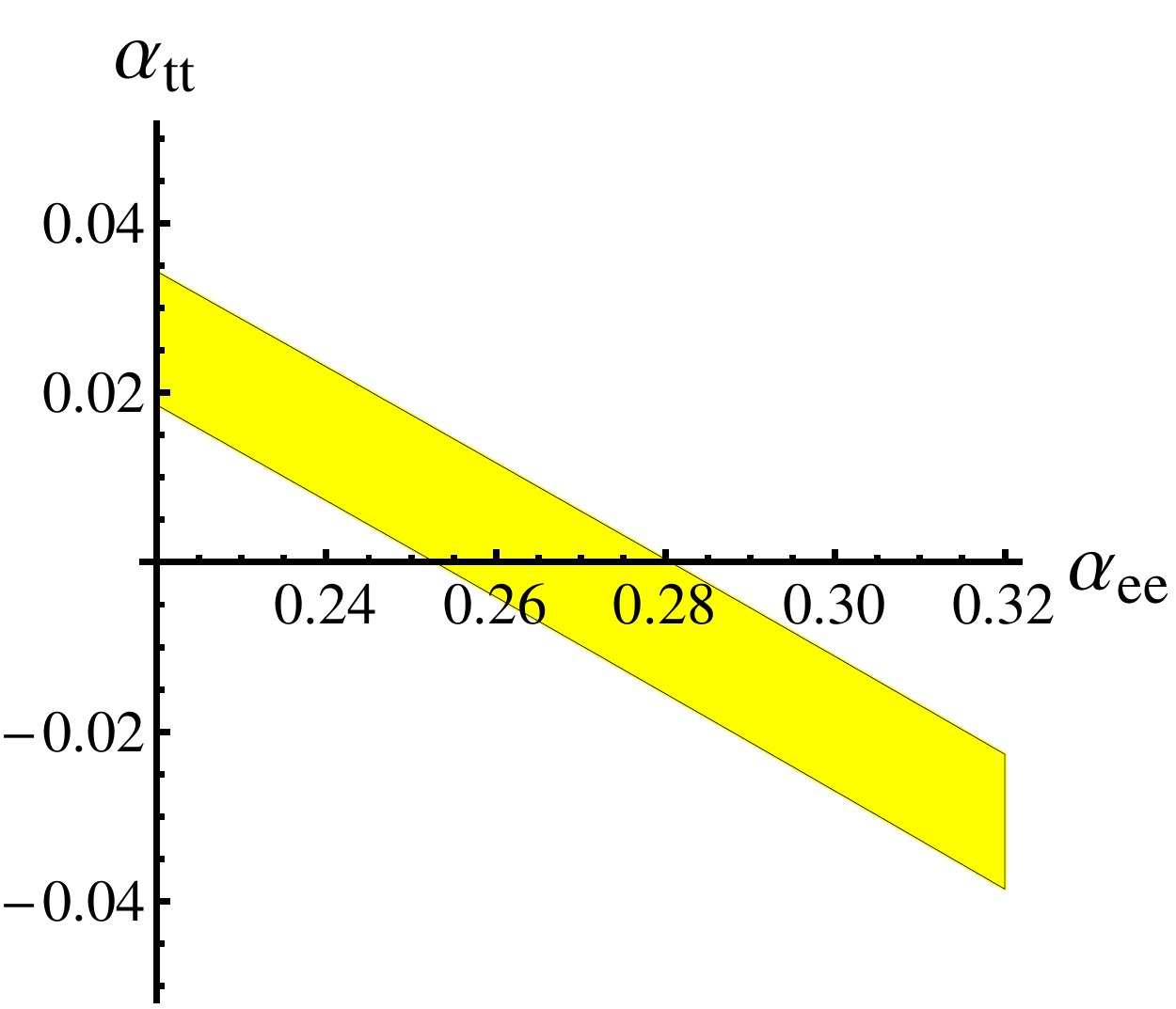}
\end{figure}

\begin{figure}[h]
\caption{Constraints in $\alpha_{ee}-\alpha_{tt}$ plane by electron EDM. Fix $m_{\eta}=40\textrm{GeV}$, $|\xi_{tt}|=0.6$,
and $\xi_{ee}=0.3$.}\label{EDMfig4}
\includegraphics[scale=0.57]{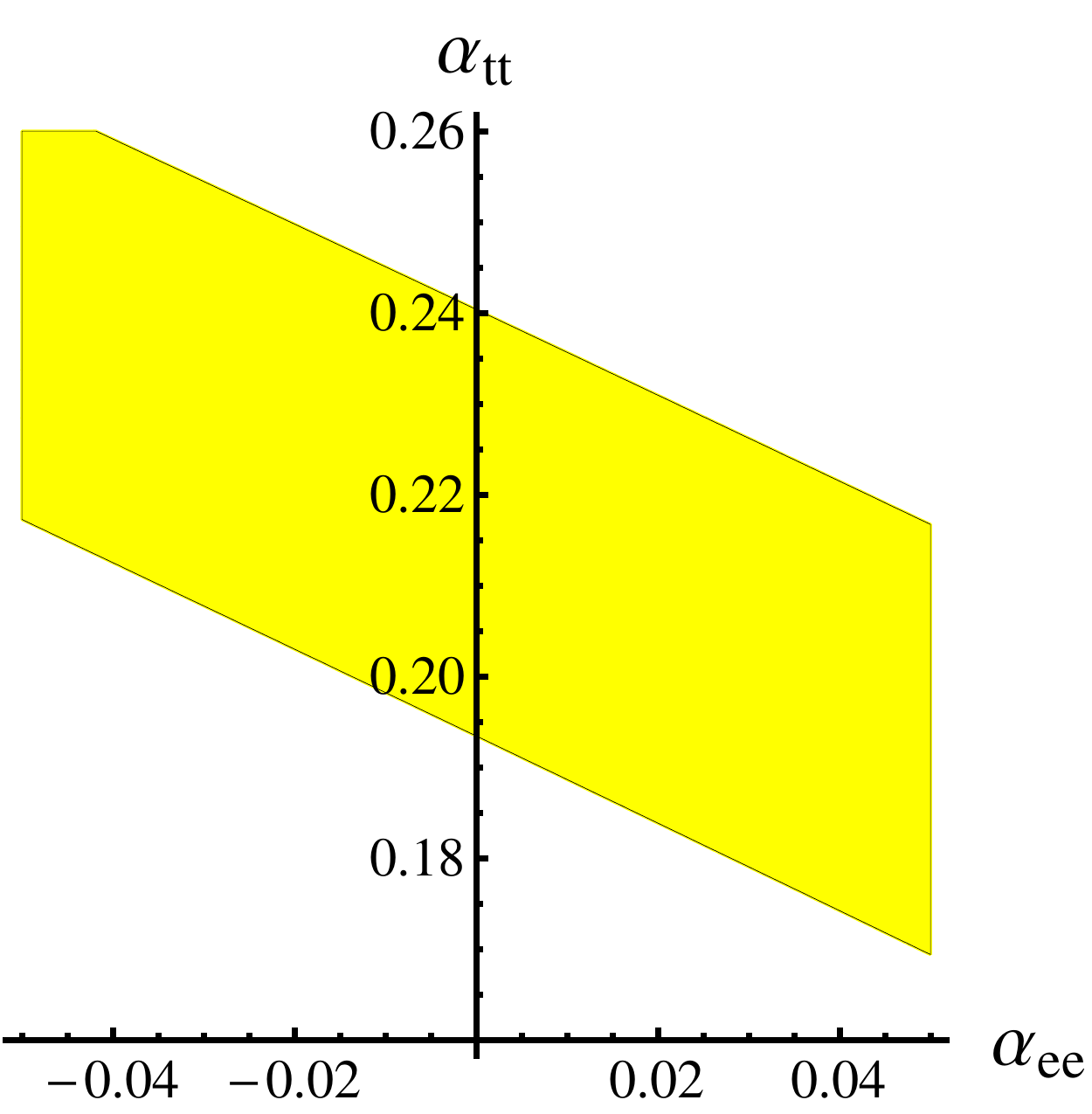}\quad\includegraphics[scale=0.57]{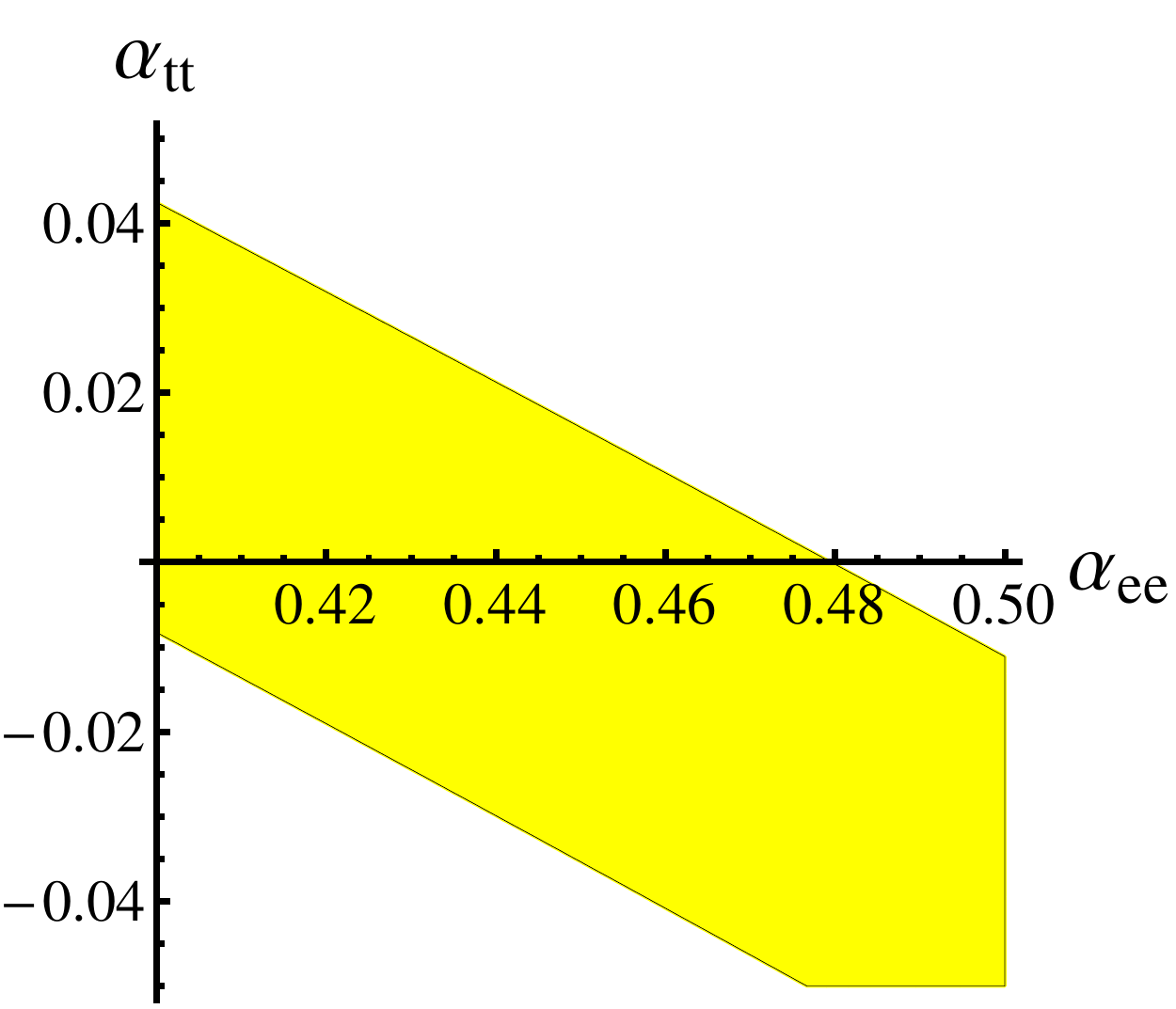}
\end{figure}

For an electron, (\ref{bzEDM}) can fully describe its EDM if we ignore the one-loop contributions. Numerically, we take the
benchmark points as those in \autoref{bench} and fix $|\xi_{tt}|=0.6$ as usual. Precision measurement by \cite{eEDM} requires
strong correlation among parameters to generate the cancelation between different contributions \cite{our,canc}.
Define $\alpha_{ij}\equiv\arg(\xi_{ij})$, we show some allowed regions at $90\%$ C.L. in \autoref{EDMfig1}-\autoref{EDMfig4} in
$\alpha_{ee}-\alpha_{tt}$ plane.

From the figures, We can see for fixing $|\xi_{ee,tt}|$, $\alpha_{tt}$ and $\alpha_{ee}$ have strong negative correlation.
In \autoref{EDMfig1} and \autoref{EDMfig2} we both choose $m_{\eta}=20\textrm{GeV}$. For $|\xi_{ee}|=1$, the allowed band
is very narrow that $\Delta\alpha\sim10^{-2}$; while for $|\xi_{ee}|=0.3$, the allowed band is wider that $\Delta\alpha
\sim(3-4)\times10^{-2}$. In \autoref{EDMfig3} and \autoref{EDMfig4} we both choose $m_{\eta}=40\textrm{GeV}$. The behaviors
are the same as the case $m_{\eta}=20\textrm{GeV}$, but the constraints are a bit weaker. For $|\xi_{ee}|=1$,
$\Delta\alpha\sim(1-2)\times10^{-2}$; while for $|\xi_{ee}|=0.3$, $\Delta\alpha\sim(5-7)\times10^{-2}$. The charged Higgs
loops give sub-dominant contributions, thus the final results are not sensitive to $\phi H^+H^-$ couplings. The location of
the allowed regions would shift a little bit for different choices of $\phi H^+H^-$ couplings.

\begin{figure}[h]
\caption{Constraints in $\alpha_{ee}-\alpha_{tt}$ plane by electron EDM. Fix $m_{\eta}=20\textrm{GeV}$, $|\xi_{tt}|=0.6$,
and $\xi_{ee}=1$. Yellow regions are allowed for the case without one-loop contribution. Green regions are for one-loop contribution
$\Delta d_e=+10^{-27}e\cdot\textrm{cm}$ while blue regions are for one-loop contribution $\Delta d_e=-10^{-27}e\cdot\textrm{cm}$.}\label{EDMfig5}
\includegraphics[scale=0.57]{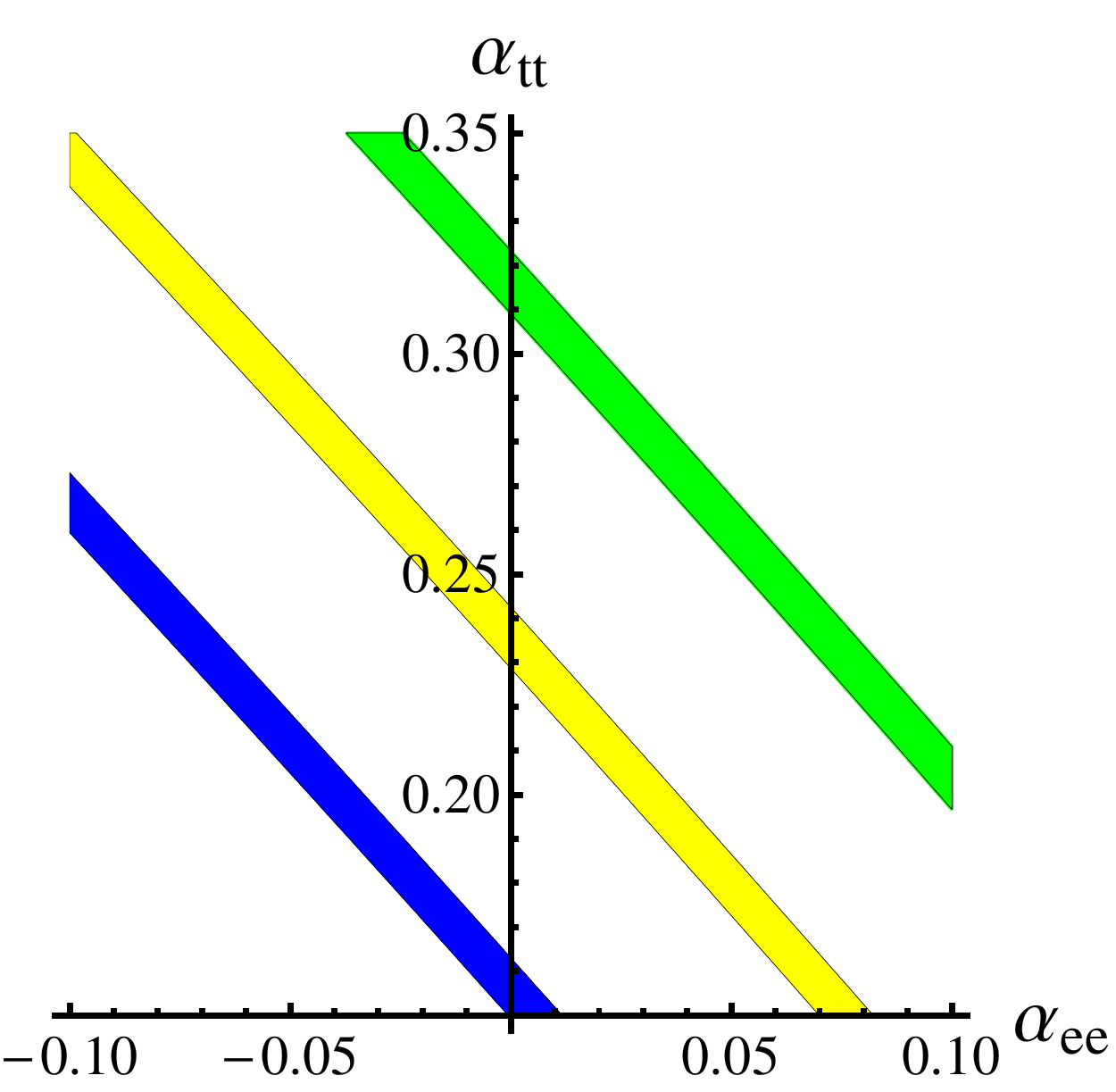}\quad\includegraphics[scale=0.57]{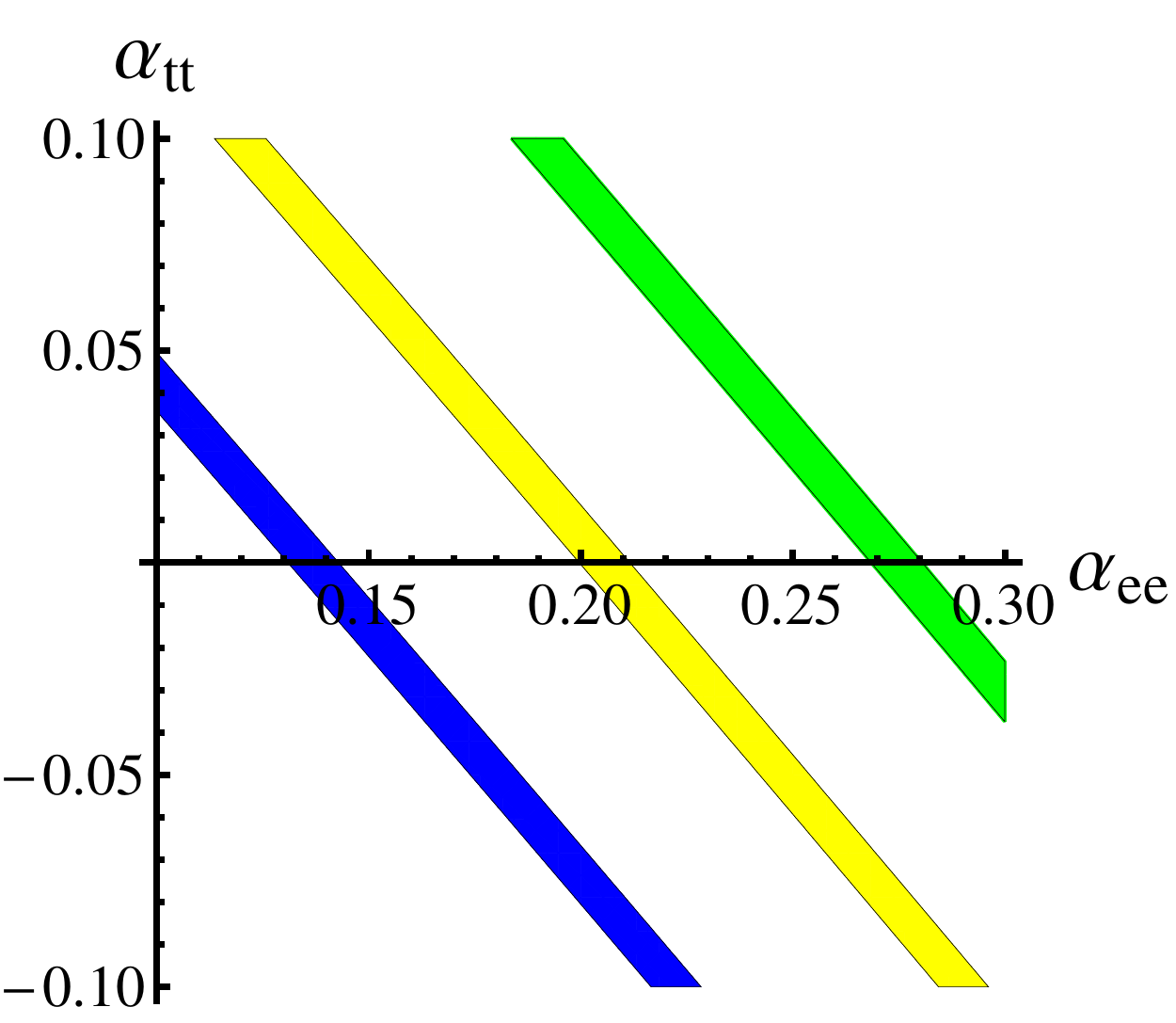}
\end{figure}

The one-loop contribution induced by flavor-diagonal interaction, showed as the first figure in \autoref{EDMdiag}, is estimated
for election as $|d_e|\sim(em_e^3/16\pi^2v^2m^2_{\phi})\ln(m^2_{\phi}/m^2_e)\sim10^{-32}e\cdot\textrm{cm}$ which is negligible small.
But the flavor-changing vertices should also generate CP-violation effects. If a $\tau$ runs in this loop, the one-loop
contribution for $d_e$ is \cite{lfv5,olEDM,olEDM2}
\begin{equation}
\Delta d_e=-\mathop{\sum}_{\phi}\frac{em_em_{\tau}^2|c_{\phi,e\tau}|^2\sin(2\alpha_{\phi,e\tau})}{16\pi^2v^2m^2_{\phi}}
\left(\ln\left(\frac{m^2_{\phi}}{m^2_{\tau}}\right)-\frac{3}{2}\right).
\end{equation}
For $|\xi_{e\tau}|\lesssim0.1$, one-loop contribution $|d_e|\lesssim10^{-28}e\cdot\textrm{cm}$ thus it is negligible
comparing with the recent experimental sensitivity \cite{eEDM}. While if $|\xi_{e\tau}|$ are larger, for example, we can take
$|\xi_{e\tau}|\sim\mathcal{O}(1)$ \footnote{Which means $|\xi_{\mu\tau}|\lesssim\mathcal{O}(10^{-2})$ according
to (\ref{etaumutau}).}, one-loop contribution to $|d_e|$ would reach $\mathcal{O}(10^{-27}-10^{-26})e\cdot\textrm{cm}$.
In this case, the allowed region would be modified a little bit. As an example, for the parameters in \autoref{EDMfig1},
we show the allowed region before and after adding the one-loop contribution $\Delta d_e=\pm10^{-27}e\cdot\textrm{cm}$ in \autoref{EDMfig5} .

The neutron EDM contains four types of contribution \cite{reviewEDM}, including quark EDM $d_q$, quark color EDM (CEDM) $\tilde{d}_q$,
Weinberg operator \cite{wen1,wen2} $w$, and strong CP term \cite{Strong} which would not be discussed in this paper. Thus \cite{reviewEDM,bz1}
\begin{equation}
\frac{d_n}{e}\simeq1.4\left(\frac{d_d}{e}-0.25\frac{d_u}{e}\right)+1.1\left(\tilde{d}_d+0.5\tilde{d}_u\right)+(22\textrm{MeV})w.
\end{equation}
Ignore the CP-violation effects in flavor-changing vertices now, the EDM for $u$ and $d$ quarks are just those in (\ref{bzEDM}) which
come from the Barr-Zee type contributions. And the CEDM from Barr-Zee diagrams are given by \cite{bz,bz1}
\begin{eqnarray}
\tilde{d}_q&=&-\mathop{\sum}_{\phi}\frac{2\sqrt{2}G_F\alpha_sm_q|c_{\phi,t}c_{\phi,q}|}{(4\pi)^3}\nonumber\\
&&\left(\sin\alpha_{\phi,t}\cos\alpha_{\phi,q}
\mathcal{J}_{1/2}(m_t,m_{\phi})+\cos\alpha_{\phi,t}\sin\alpha_{\phi,q}\mathcal{J}'_{1/2}(m_t,m_{\phi})\right)
\end{eqnarray}
where the loop functions are the same as those in (\ref{bzEDM}). The contribution from Weinberg operator is \cite{bz1,wen1,wen2}
\begin{equation}
\label{weinberg}
w=\mathop{\sum}_{\phi}\frac{\sqrt{2}G_Fg_s\alpha_s|c_{\phi,t}|^2}{4\cdot(4\pi)^3}
\sin\alpha_{\phi,t}\cos\alpha_{\phi,t}\mathcal{K}\left(\frac{m^2_t}{m^2_{\phi}}\right)
\end{equation}
where the loop function $\mathcal{K}$ is listed in (\ref{weinbergK}) in \autoref{loop}. Including also the running effects
for these operators (see the appendices in \cite{bz1}),
\begin{eqnarray}
\frac{d_n}{e}&\simeq&\frac{m_d(\mu_H)}{m_d(\mu_W)}\left(0.63\frac{d_d(\mu_W)}{e}+0.73\tilde{d}_d(\mu_W)\right)
+\frac{m_u(\mu_H)}{m_u(\mu_W)}\left(-0.16\frac{d_u(\mu_W)}{e}+0.19\tilde{d}_u(\mu_W)\right)\nonumber\\
&&+(8.8\textrm{MeV}+0.17m_d(\mu_H)+0.08m_u(\mu_H))w(\mu_W).
\end{eqnarray}
Here $\mu_H$ is the hadron scale and $\mu_W$ is the electro-weak scale, $\alpha_s(\mu_W)\approx0.11$ \cite{as},
and $m_d(\mu_H)\approx4.8\textrm{MeV}$, $m_u(\mu_H)\approx2.3\textrm{MeV}$ \cite{PDG}. $d_q(\mu_W)$, $\tilde{d}_q(\mu_W)$,
and $w(\mu_W)$ are all calculated at electro-weak scale.

\begin{figure}[h]
\caption{Allowed region on $\alpha_{uu}-\alpha_{dd}$ plane with constraint from neutron EDM. We fix $|\xi_{uu}|=|\xi_{uu}|=1$,
$|\xi_{tt}|=0.6$, and $\alpha_{tt}=0$. The left figure is for $m_{\eta}=20\textrm{GeV}$ and the right one is for $m_{\eta}=40\textrm{GeV}$.
All other benchmark points are the same as above. Yellow regions are allowed at $90\%$ C.L.}\label{nEDM1}
\includegraphics[scale=0.57]{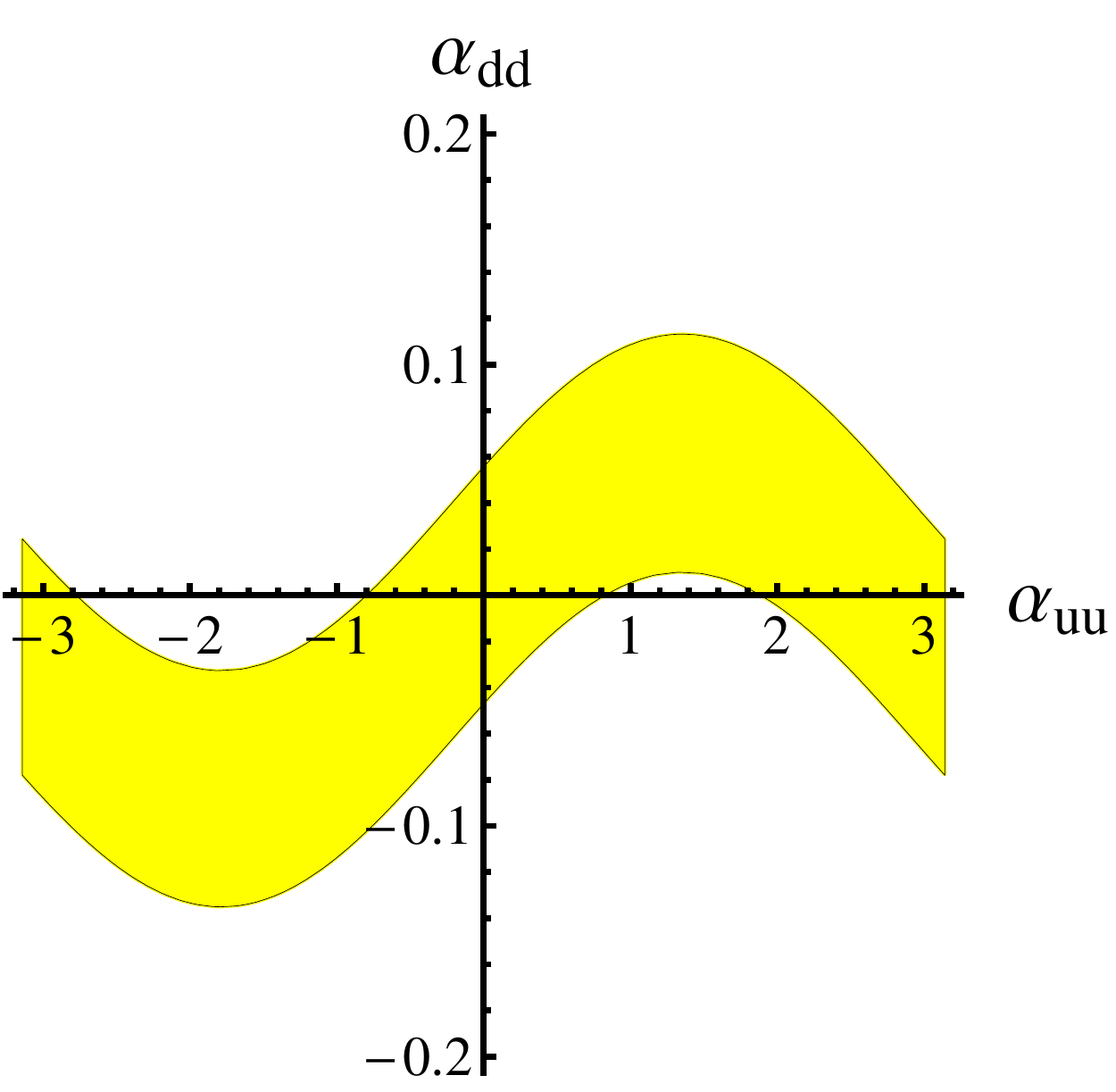}\quad\includegraphics[scale=0.57]{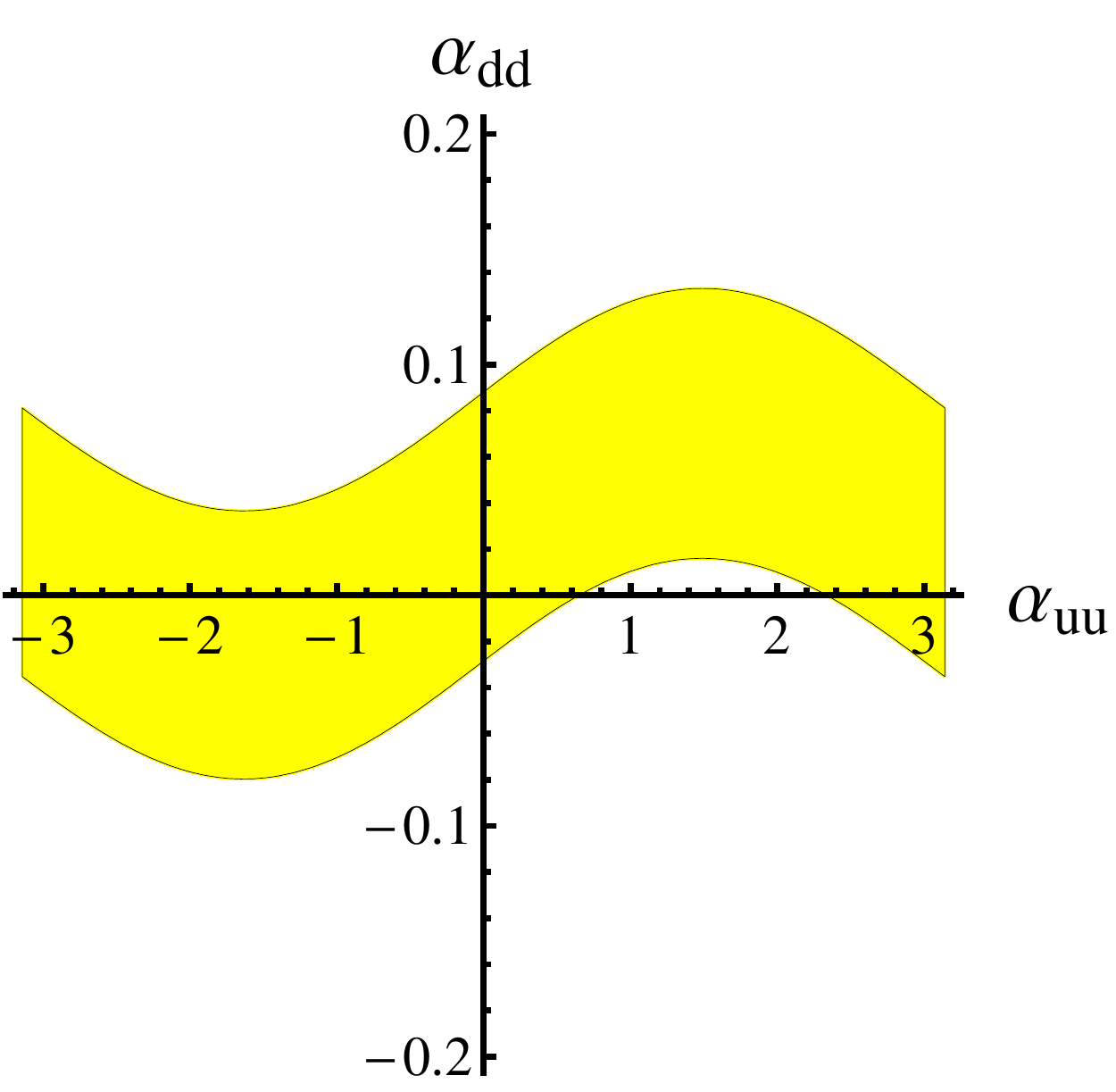}
\end{figure}

Numerically, we use the benchmark points the same as above. Fixing $|\xi_{uu}|=|\xi_{dd}|=1$, $|\xi_{tt}|=0.6$, and $\alpha_{tt}=0$.
For $m_{\eta}=20\textrm{GeV}$ and $m_{\eta}=40\textrm{GeV}$, we show the allowed regions in $\alpha_{uu}-\alpha_{dd}$ plane in
\autoref{nEDM1}. There exist cancelation between different contributions as well. From the figures, we can see $\alpha_{uu}$ is
almost free, and $\alpha_{dd}$ is constrained in a narrow band. For both cases, $\alpha_{uu}=\alpha_{dd}=0$ is inside the allowed
region, and $\Delta\alpha_{dd}\sim0.1$. The cancelation behavior is not sensitive to $m_{\eta}$. It is also a strict constraint
from neutron EDM, but not so strict as that from electron EDM.

\begin{figure}[h]
\caption{Allowed region by the constraint from neutron EDM. Benchmark points are the same as above in \autoref{nEDM1}.
Yellow regions are allowed for the case without one-loop contribution. Green regions are for one-loop contribution
$\Delta d_n=+10^{-25}e\cdot\textrm{cm}$ while blue regions are for one-loop contribution $\Delta d_n=-10^{-25}e\cdot\textrm{cm}$.}
\label{nEDM2}
\includegraphics[scale=0.57]{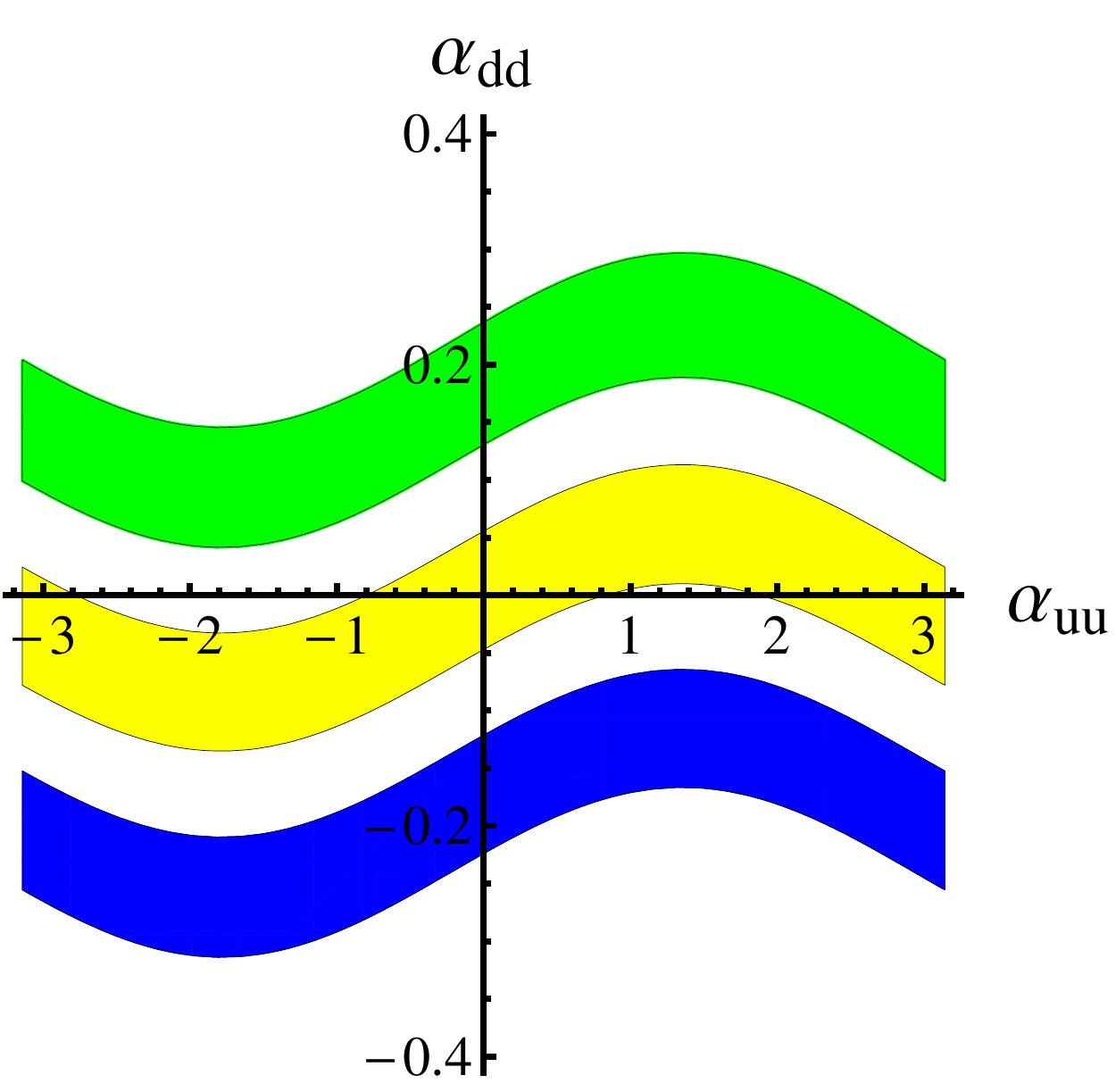}\quad\includegraphics[scale=0.57]{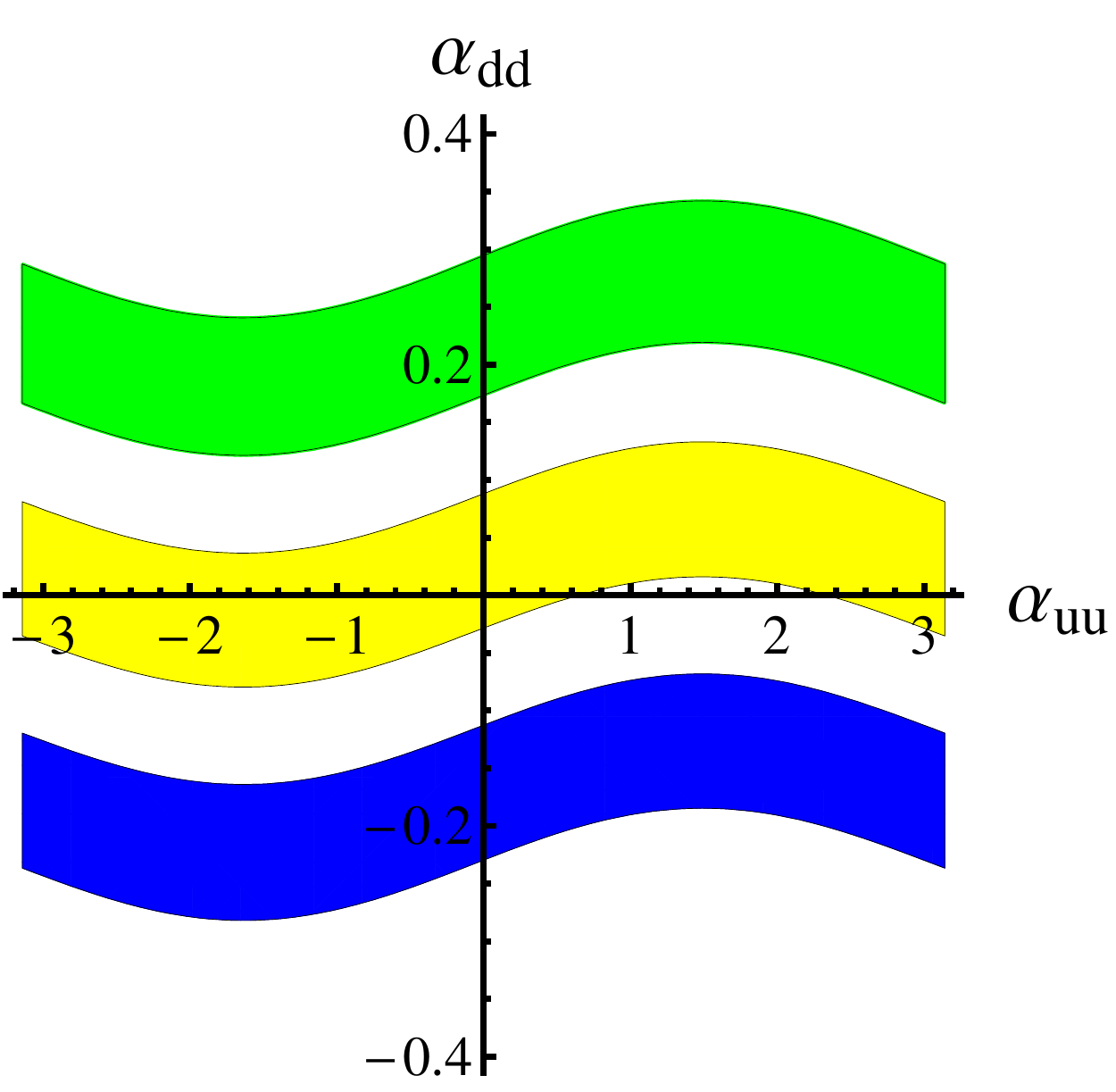}
\end{figure}

Next, consider the contributions from flavor-changing vertices. Strict constraints from meson mixing (see the text in \autoref{mixcons})
require that the contributions for $d_n$ from $bd\phi$, $sd\phi$, and $uc\phi$ vertices should be less than $\mathcal{O}(10^{-30})e\cdot\textrm{cm}$,
thus they can are ignorable. But CP-violation in $tu\phi$ vertex would give larger contribution to $d_u$ and $\tilde{d}_u$ \cite{lfv5,olEDM,olEDM2}
through the one-loop diagram as the left figure in \autoref{EDMdiag},
\begin{eqnarray}
\label{du1}
\frac{\Delta d_u}{e}&=&\mathop{\sum}_{\phi}\frac{m_u|c_{\phi,tu}|^2\sin(2\alpha_{\phi,tu})}{24\pi^2v^2}\mathcal{P}_1\left(\frac{m^2_t}{m^2_{\phi}}\right);\\
\label{du2}
\Delta\tilde{d}_u&=&\mathop{\sum}_{\phi}\frac{m_u|c_{\phi,tu}|^2\sin(2\alpha_{\phi,tu})}{16\pi^2v^2}\mathcal{P}_1\left(\frac{m^2_t}{m^2_{\phi}}\right).
\end{eqnarray}
The loop function $\mathcal{P}_1(x)$ is listed in (\ref{p1}) in \autoref{loop}. For $|\xi_{tu}|\sim1$, the additional contribution to
neutron EDM can reach $\Delta d_n\sim\mathcal{O}(10^{-26}-10^{-25})e\cdot\textrm{cm}$, which would change the cancelation behavior and
shift the allowed region a little bit. In \autoref{nEDM2}, we show the allowed region before and after adding the one-loop contribution $\Delta d_n=
\pm10^{-25}e\cdot\textrm{cm}$.
The case $|\Delta d_n|\gtrsim6\times10^{-25}e\cdot\textrm{cm}$ is excluded for this benchmark points choice because
enough cancelation between different contributions cannot be generated.
\subsection{Constraints from B Meson Rare Decays}
The leptonic decay $B^0_{(s)}\rightarrow\mu^+\mu^-$ was measured by CMS and LHCb collaborations and the results
\cite{rare1,rare2,rare3} are listed in \autoref{rareresult} together with their SM predictions \cite{SMrare1,SMrare2}.
\begin{table}[h]
\caption{Recent experimental and theoretical results for $B^0_{(s)}\rightarrow\mu^+\mu^-$ decay branching ratios.}\label{rareresult}
\begin{tabular}{|c|c|c|}
\hline
Result&$\textrm{Br}(B^0_s\rightarrow\mu^+\mu^-)$&$\textrm{Br}(B^0\rightarrow\mu^+\mu^-)$\\
\hline
CMS&$(2.8^{+1.1}_{-0.9})\times10^{-9}$&$(4.4^{+2.2}_{-1.9})\times10^{-10}$\\
\hline
LHCb&$(2.7^{+1.1}_{-0.9})\times10^{-9}$&$(3.3^{+2.4}_{-2.1})\times10^{-10}$\\
\hline
Combined&$(2.8^{+0.7}_{-0.6})\times10^{-9}$&$(3.9^{+1.6}_{-1.4})\times10^{-10}$\\
\hline
SM Prediction&$(3.65\pm0.23)\times10^{-9}$&$(1.06\pm0.09)\times10^{-10}$\\
\hline
\end{tabular}
\end{table}
Both measurements are almost consistent with SM predictions \footnote{The CMS result for
$\textrm{Br}(B^0\rightarrow\mu^+\mu^-)$ has a deviation from SM prediction at about $2\sigma$ level. The same
thing happens to the combined result for $\textrm{Br}(B^0\rightarrow\mu^+\mu^-)$.},  thus new physics contributions
would be limited.

The tree level contributions to $B^0_{(s)}\rightarrow\mu^+\mu^-$ is negligible \cite{our} due to the constraints
from B meson mixing. Here we consider the charged Higgs contribution only. In this scenario, $m_{\pm}\sim
v$ is favored as above. For $|\xi_{bb,dd,\ell\ell}|\sim\mathcal{O}(1)$,
the modified branching ratio for $B^0_{(s)}\rightarrow\mu^+\mu^-$ should be \cite{bmumu}
\begin{equation}
\label{blep}
\frac{\textrm{Br}(B^0_{(s)}\rightarrow\mu^+\mu^-)}{\textrm{Br}_{\textrm{SM}}(B^0_{(s)}\rightarrow\mu^+\mu^-)}
=\left(1-\frac{|\xi_{tt}|^2}{\eta}\frac{\mathcal{Y}_{\textrm{2HDM}}(m^2_t/m^2_W,m^2_{\pm}/m^2_W)}
{\mathcal{Y}_{\textrm{SM}}(m^2_t/m^2_W)}\right)^2
\end{equation}
where $\eta=0.987$ is the QCD and electro-weak correlation factor and the loop functions $\mathcal{Y}_i$ are listed
in (\ref{ysm})-(\ref{y2hdm}) in \autoref{loop}. Numerically, consider $B^0_s\rightarrow\mu^+\mu^-$, both CMS and LHCb
results give
\begin{equation}
|\xi_{tt}|\lesssim(0.7-0.8)
\end{equation}
at $95\%$ C.L. which is near the constraint by B meson mixing in (\ref{tt}). For $B^0_d\rightarrow\mu^+\mu^-$ decay,
these regions are also allowed at $95\%$ C.L. by both CMS and LHCb results \footnote{If considering the combined result,
$|\xi_{tt}|\lesssim(0.5-0.6)$ is still allowed by data due to $B^0_s\rightarrow\mu^+\mu^-$ which is a bit stricter than
that in (\ref{tt}). For $B^0_d\rightarrow\mu^+\mu^-$, we also need $|\xi_{tt}|\gtrsim(0.2-0.3)$ at $95\%$ C.L. because
the combined deviation between $\textrm{Br}(B^0\rightarrow\mu^+\mu^-)$ and its SM prediction is a bit larger than $2\sigma$.}.

\begin{figure}[h]
\caption{Constraints by $\textrm{Br}(\bar{B}\rightarrow X_s\gamma)$ fixing $m_{\pm}=200\textrm{GeV}$. In the left figure, we
take $|\xi_{tt}|=0.6$ and plot the allowed region in $|\xi_{bb}|-\alpha_{bt}$ plane. In the right figure, we take $\alpha_{bt}=0$
and plot the allowed region in $|\xi_{bb}|-|\xi_{tt}|$ plane. In both figures, green regions are allowed at $68\%$ C.L. and the yellow
regions are allowed at $95\%$ C.L.}\label{Brad1}
\includegraphics[scale=0.57]{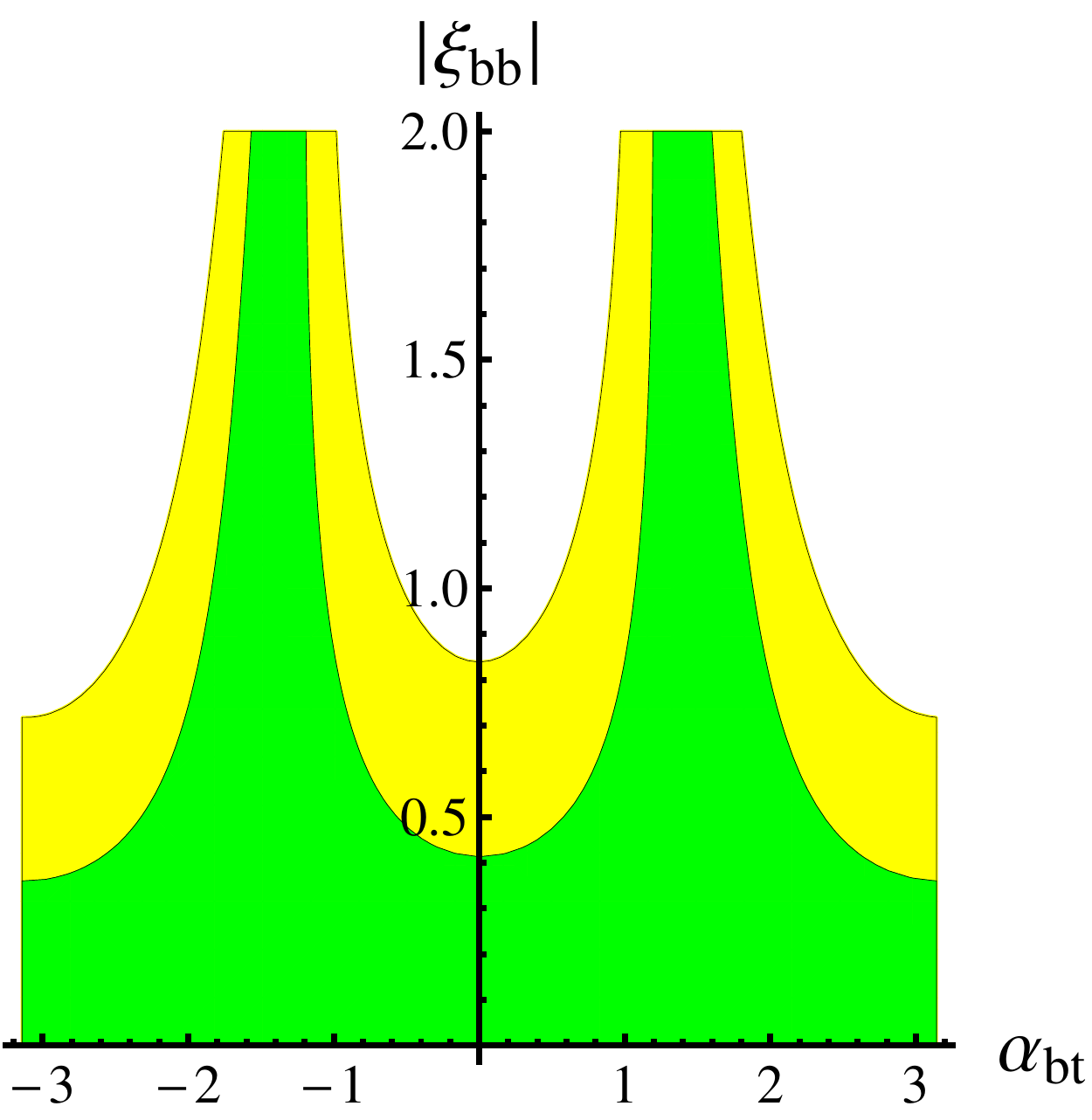}\quad\includegraphics[scale=0.57]{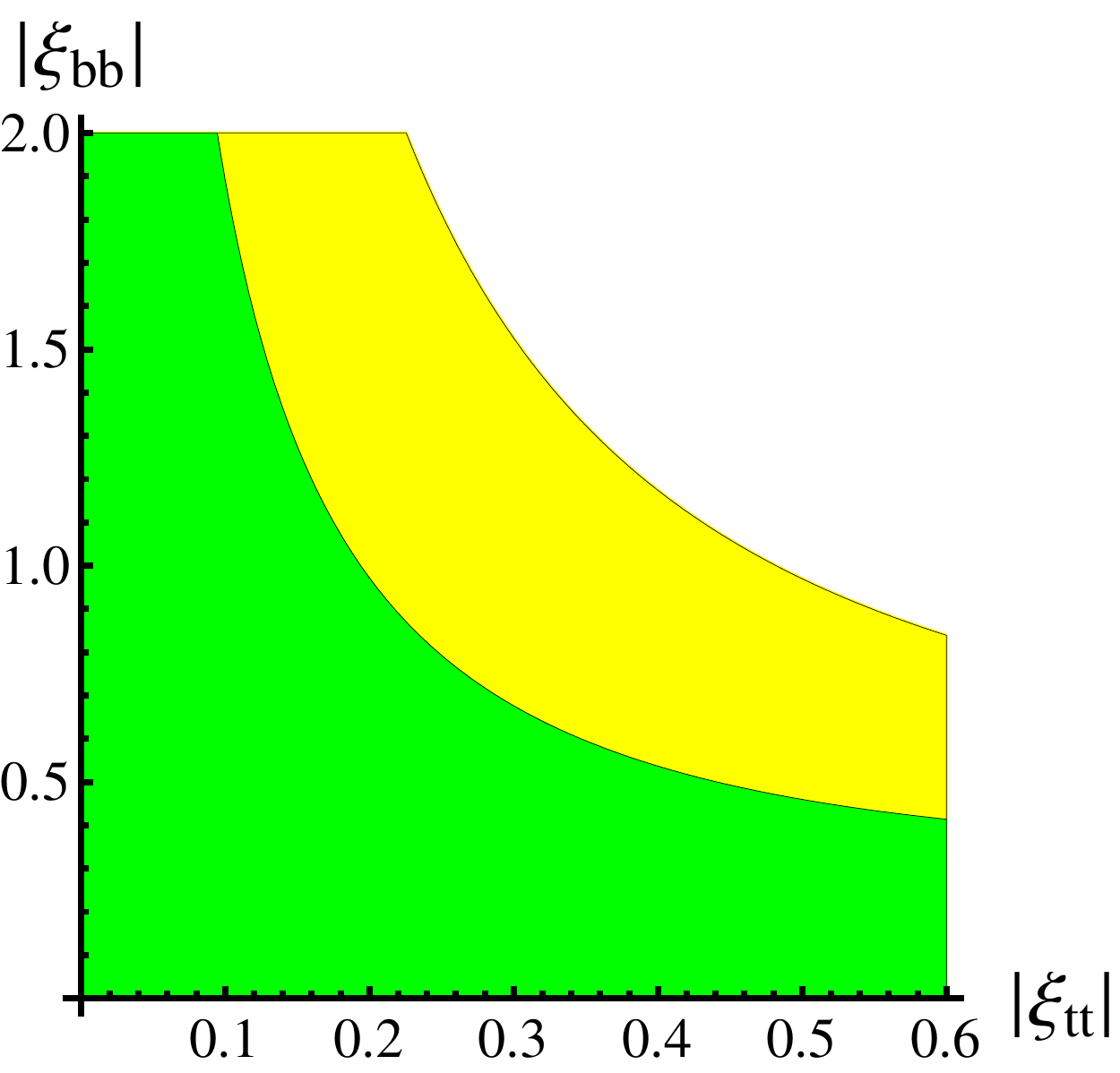}
\end{figure}

\begin{figure}[h]
\caption{Constraints by $\textrm{Br}(\bar{B}\rightarrow X_s\gamma)$ fixing $m_{\pm}=300\textrm{GeV}$. All other sets are the same
as those in \autoref{Brad1}.}\label{Brad2}
\includegraphics[scale=0.57]{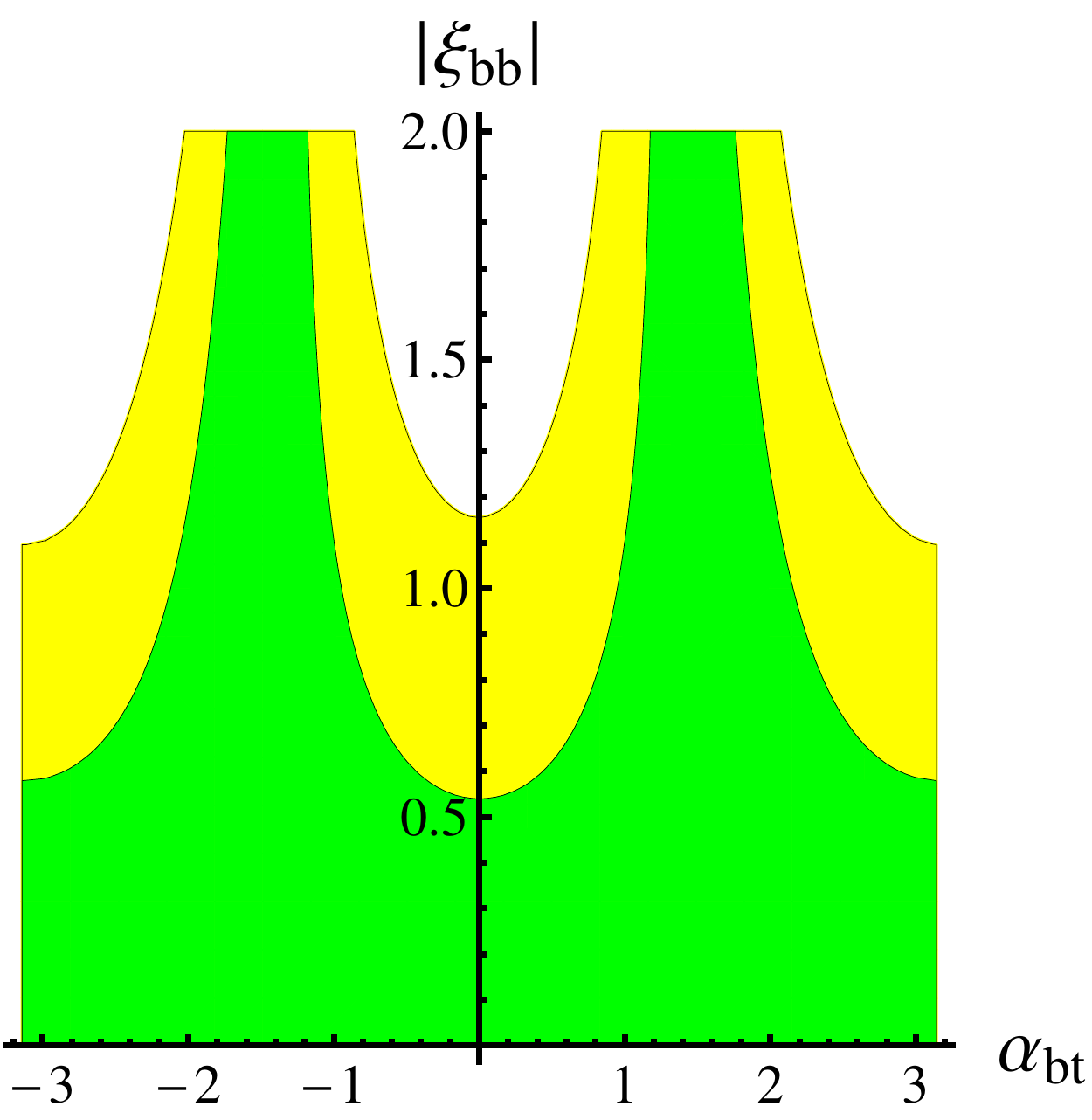}\quad\includegraphics[scale=0.57]{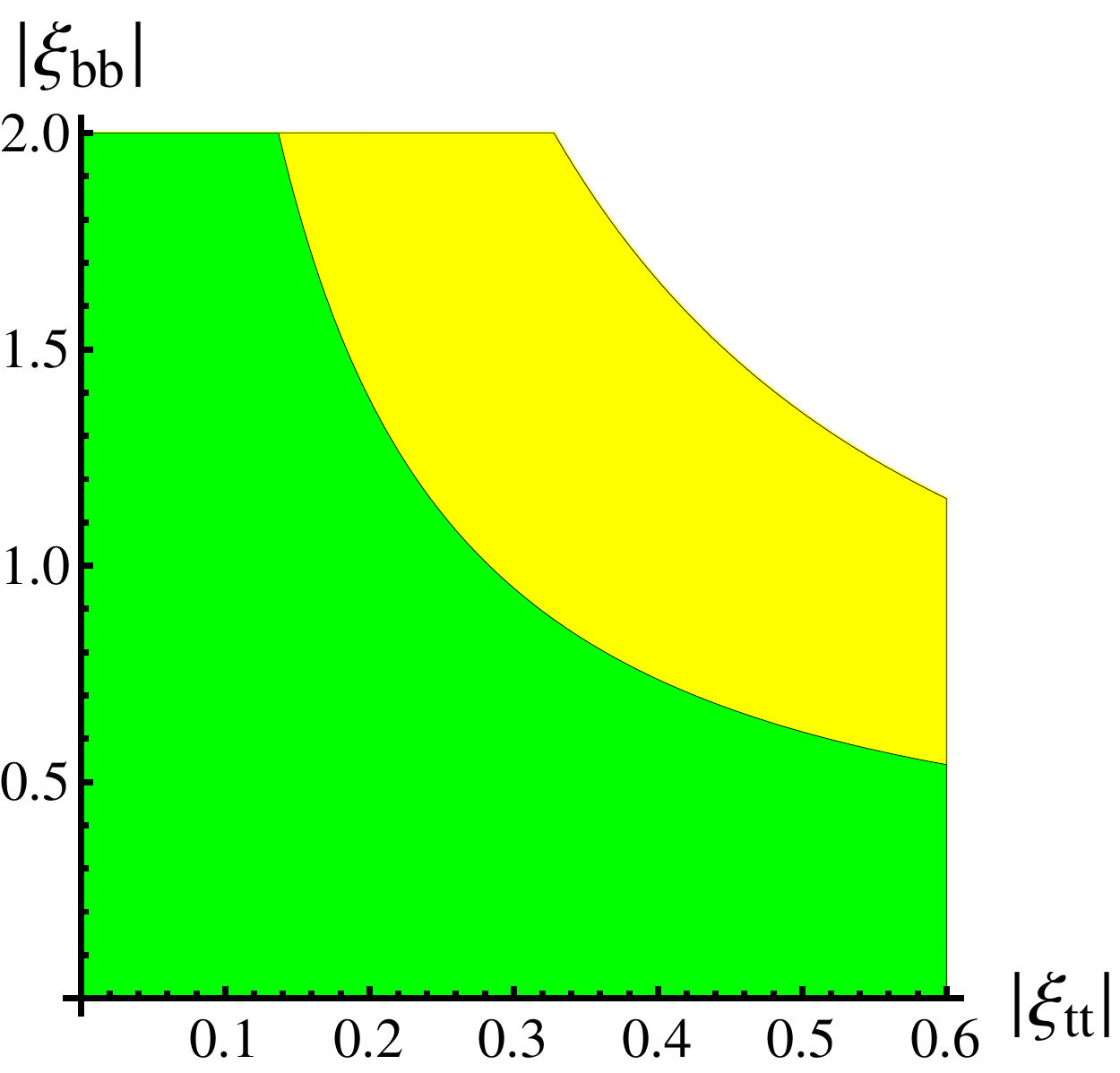}
\end{figure}

The world averaged value for B radiative decay branching ratio reads $\textrm{Br}_{\textrm{ave}}(\bar{B}\rightarrow X_s\gamma)
=(3.43\pm0.22)\times10^{-4}$ \cite{HFAG} which is consistent with its SM prediction $\textrm{Br}_{\textrm{SM}}(\bar{B}\rightarrow X_s\gamma)=(3.36\pm0.23)\times10^{-4}$ \cite{bsgammasm}. In 2HDM, according to (\ref{charyuk}), a charged Higgs boson can also run in
the loop instead of $W^{\pm}$ for the radiative decay process thus the branching ratio can be modified.
In type II 2HDM, the
charged Higgs mass is constrained to be larger than about $410$ GeV \cite{bsgamma2hdm}
\footnote{This value is different from the data in the text of \cite{bsgamma2hdm} because the
SM prediction result was updated in \cite{bsgammasm} recently.} at $95\%$ C.L. But in a general 2HDM,
a lighter charged Higgs boson may be allowed \cite{our}.
Different from leptonic decay, the radiative decay branching ratio is sensitive to not only $m_{\pm}$ and $\xi_{tt}$, but also
$\xi_{bb}$. For a general case, $\alpha_{bt}\equiv\arg(\xi_{bb}/\xi_{tt})$ is also
a free parameter. Based on \cite{bsgamma2hdm} and the mathematica code, we plot the constraints
on these parameters in \autoref{Brad1} and \autoref{Brad2}, fixing $m_{\pm}=200\textrm{GeV}$ and $m_{\pm}=300\textrm{GeV}$ respectively.
From the figures, we can see for $|\xi_{tt}|=0.6$, for most $\alpha_{bt}$ choice, we have $|\xi_{bb}|\lesssim1$; but for some $\alpha_{bt}$ choice,
a larger $|\xi_{bb}|$ is also allowed. While for fixed $\alpha_{bt}$, there is also larger allowed region in $|\xi_{bb}|-|\xi_{tt}|$ plane. The
constraint is not so strict as that for type II 2HDM because more parameters are free, just like the case in \cite{our}.
\section{Predictions and Future Tests for this Scenario}
\label{fut}
We have discussed all the constraints on the Lee model in an alternative scenario which is weakly-coupled. As shown above, it is still not
excluded by experimental results.
Comparing with the scenario in \cite{our}, the particle spectrum are the same. But in this scenario, all the scalars are required to have their
mass around electro-weak scale or lighter. Especially, the lightest scalar is required to have its mass $m_{\eta}\sim\mathcal{O}(10\textrm{GeV})$
which means new physics is hidden in the scale lower than electro-weak scale. That's different from the scenario in \cite{our} in which
new physics would appear at $\mathcal{O}(\textrm{TeV})$ or higher scale.

In this scenario, the couplings of the 125 GeV Higgs boson is SM-like, but other particles are not decoupled thus they would face future tests at colliders.
A lighter scalar can also appear through Higgs decay channels $h\rightarrow Z\eta,\eta\eta$ which are worth to search. Different scalars
may also be produced associated with each other or with heavy quark (pair). $h\rightarrow Z\eta,\eta\eta$ rare decays would also be constrained
by Higgs signal strengths which would be measured precisely in the future. Experiments on flavor changing processes and EDM measurements would
also help to confirm or exclude this scenario indirectly.
\subsection{Direct Searches for Extra Scalars at Future Colliders}
The key prediction of this scenario (weakly-coupled Lee model) is a light particle $\eta$ with its mass of $\mathcal{O}(10\textrm{GeV})$. It should
be a CP-mixing state with pseudoscalar component dominant. Its low mass is correlated with the smallness of CP-violation. At LHC, it certainly
can be produced through gluon fusion or $b\bar{b}$ fusion with large cross section, but such a light particle would be hidden below the huge QCD
background \cite{SLH3}, thus it is difficult to be discovered. At LHC with $\sqrt{s}=(13-14)\textrm{TeV}$, $\eta$ can also be produced in associated with
top quark pair with a cross section of $\mathcal{O}(0.1)\textrm{pb}$ \cite{lighthig}. According to \cite{lighthig}, at LHC with $\sqrt{s}=14\textrm{TeV}$
and $3\textrm{ab}^{-1}$ luminosity, for $m_{\eta}\sim(20-40)\textrm{GeV}$, the constraint $|\xi_{tt}|\lesssim(0.34-0.54)$ at $95\%$ C.L. would be achieved assuming no positive results. It would be stricter than all the recent constraints obtained from indirect processes. On the other hand, for $m_{\eta}\sim(30-40)\textrm{GeV}$, the benchmark case $|\xi_{tt}|=0.6$ would be discovered at more than $5\sigma$.

$\eta$ can also appear as the decay final state of other scalars, such as $h,H\rightarrow\eta\eta,Z\eta$, etc. We will study the cascade
decay channels in details in the future. LHeC \cite{lhec1,lhec2} would
be a better collider in searching for the exotic Higgs decays \cite{lheczc}. At future $e^+e^-$ colliders \cite{TLEP,CLIC,CEPC,ILC},
$\eta$ is also possible to be discovered through Higgs rare decay processes, such as $e^+e^-\rightarrow Zh(\rightarrow\eta\eta)$. At the Higgs factories with $\sqrt{s}\sim(240-250)\textrm{GeV}$ like CEPC \cite{CEPC} or TLEP \cite{TLEP}, this process can be discovered at $5\sigma$ with $5\textrm{ab}^{-1}$
luminosity if $\textrm{Br}(h\rightarrow\eta\eta)>10^{-3}$ \cite{CEPC}.
$\eta$ can also be produced in associated with $Z$ or $h$ at CEPC or TLEP. With a roughly estimation comparing with LEP
results \cite{LEP1,LEP2,LEP3,LEP4} we used in \autoref{directsearch}, using $\mathcal{O}(10^2-10^3)
\textrm{fb}^{-1}$ luminosity, the sensitivity to $c_{\eta,V}$ and $c_{h\eta}(=c_{H,V})$ would improve at least an order. At $e^+e^-$ colliders
with $\sqrt{s}>m_{\eta}+m_H$, it is possible to produce $\eta$ and $H$ through $e^+e^-\rightarrow\eta H$ \footnote{If $m_H\sim200\textrm{GeV}$,
Higgs factory mentioned above is also allowed for this process.}. It's worth noting that under weak-coupling assumption, $m_H$ should be around
the electro-weak scale, and $c_{\eta H}=c_h\sim1$ would never be suppressed, thus this is also a key process to confirm or exclude this scenario
at future $e^+e^-$ colliders.

For the heavy Higgs boson $H$, it is required to have a mass around $v$ thus it is possible to be discovered at LHC \cite{cmsfuthig}. For
$m_H\sim(200-300)\textrm{GeV}$, choose $|\xi_{tt}|=0.6$ and $c_{H,V}=0.3$ we take above, the cross section $\sigma(pp\rightarrow
H\rightarrow ZZ)\sim(120-200)\textrm{fb}$ according to \cite{futheavyth} at future LHC with $\sqrt{s}=14\textrm{TeV}$. It is larger
than the $5\sigma$ discovery threshold $(50-100)\textrm{fb}$ using $3\textrm{ab}^{-1}$ luminosity \cite{cmsfuthig}, thus it would be
easily discovered. While if no signal evidence were found, according to \cite{cmsfuthig}, the $95\%$ C.L. limit for $\sigma(pp\rightarrow
H\rightarrow ZZ)$ would be $(20-40)\textrm{fb}$ for the mass region $m_H\sim(200-300)\textrm{GeV}$. Since the dominant production channel
for $H$ is gluon fusion, this result means the future LHC would be able to set the constraint
\begin{equation}
|\xi_{tt}|c_{H,V}\lesssim0.08
\end{equation}
at $95\%$ C.L. if no evidence for this channel were found.

Through the oblique parameter constraints, the charged Higgs mass is around $v$ in this scenario. It must face the direct searches at LHC
or $e^+e^-$ colliders. At LHC, it can be produced through $gb\rightarrow tH^-$ associated production \cite{bgth} which was used to search
for the charged Higgs boson in \cite{chargh}. For a light charged higgs with $m_{\pm}\sim(200-300)\textrm{GeV}$, for $|\xi_{tt}|\sim0.6$ and $|\xi_{bb}|\sim
\mathcal{O}(1)$, it would be discovered at LHC with $\sqrt{s}=(13-14)\textrm{TeV}$ and $300\textrm{fb}^{-1}$ luminosity; and the polarization
of top quark would also be useful to test the chiral structure in $tbH^-$ vertex \cite{pol1,pol2,pol3}. At $e^+e^-$ colliders with
$\sqrt{s}\gtrsim500\textrm{GeV}$, we can discover the charged Higgs boson through $e^+e^-\rightarrow H^+H^-$ process \cite{ILC2,eehh}. This
process would not be suppressed as well, thus it is useful to confirm or exclude this scenario. In \autoref{futtest1} and \autoref{futtest2} 
we summarize the mentioned channel above which would be useful to test this scenario in the future 
\cite{lighthig,CEPC,cmsfuthig,futheavyth,bgth,eehh,ILC2,pol1,pol2,pol3,added}.
\begin{table}[h]
\caption{Examples for main processes which would be useful to test this scenario at future $pp$ collider. ``*"
means we will study this process in details in the future. In this table, all masses are chosen as: $m_{\eta}=40\textrm{GeV}$,
$m_{h}=125\textrm{GeV}$, and $m_H=m_{\pm}=300\textrm{GeV}$ as an example. The benchmark points listed here for collider or model
parameters are possible choices but not the only choice for the corresponding processes.}\label{futtest1}
\begin{tabular}{|c|c|c|c|c|c|c|}
\hline
Collider & Process & \begin{tabular}{c}$\sqrt{s}$\\(TeV)\end{tabular} & \begin{tabular}{c}Couplings and/or\\Branching Ratios\end{tabular} & \begin{tabular}{c}Cross Section\\(pb)\end{tabular} & Implications\\
\hline
LHC&$pp\rightarrow t\bar{t}\eta$&14&$|\xi_{tt}|=0.6$&$0.18$&
\begin{tabular}{c}Over $5\sigma$ discovery\\with $3\textrm{ab}^{-1}$ luminosity\\assuming $\textrm{Br}_{\eta\rightarrow b\bar{b}}=1$.\end{tabular}\\
\hline
LHC&$pp\rightarrow H(ZZ)$&14&\begin{tabular}{c}$|\xi_{tt}|=0.6$\\$\textrm{Br}_{H\rightarrow ZZ}=3\%$\end{tabular}&$0.12$&
\begin{tabular}{c}Over $5\sigma$ discovery\\ with $3\textrm{ab}^{-1}$ luminosity.\end{tabular}\\
\hline
LHC&$pp\rightarrow H(Z\eta,\eta\eta)$&14&\begin{tabular}{c}$c_{H\eta}=c_{h,V}=0.95$,\\$|\xi_{tt}|=0.6$,$g_{H\eta\eta}\sim1$\end{tabular}&
$4\times\textrm{Br}_{H\rightarrow Z\eta,\eta\eta}$&*To be studied.\\
\hline
LHC&$pp(bg)\rightarrow tH^-(\bar{t}b)$&14&\begin{tabular}{c}$|\xi_{tt}|=0.6$,$|\xi_{bb}|\lesssim1$\\$\textrm{Br}_{H^-\rightarrow \bar{t}b}=1$\end{tabular}&
$0.6$&\begin{tabular}{c}$5\sigma$ discovery with\\$\mathcal{O}(10^2)\textrm{fb}^{-1}$ luminosity.\end{tabular}\\
\hline
\end{tabular}
\end{table}
\begin{table}[h]
\caption{Examples for main processes which would be useful to test this scenario at future $e^+e^-$ colliders. ``*"
means we will study this process in details in the future. In this table, all masses are chosen as: $m_{\eta}=40\textrm{GeV}$,
$m_{h}=125\textrm{GeV}$, and $m_H=m_{\pm}=300\textrm{GeV}$ as an example. The benchmark points listed here for collider or model
parameters are possible choices but not the only choice for the corresponding processes.}\label{futtest2}
\begin{tabular}{|c|c|c|c|c|c|c|}
\hline
Collider & Process & \begin{tabular}{c}$\sqrt{s}$\\(TeV)\end{tabular} & \begin{tabular}{c}Couplings and/or\\Branching Ratios\end{tabular} & \begin{tabular}{c}Cross Section\\(pb)\end{tabular} & Implications\\
\hline
CEPC&$e^+e^-\rightarrow Z\eta$&0.25&$c_{\eta,V}=0.1$&$4.4\times10^{-3}$&\begin{tabular}{c}*Sensitivity to $c_{\eta,V}$\\ would reach $\mathcal{O}(10^{-2})$\\with $5\textrm{ab}^{-1}$ luminosity.\end{tabular}\\
\hline
CEPC&$e^+e^-\rightarrow h\eta$&0.25&$c_{h\eta}=c_{H,V}=0.3$&$7.3\times10^{-3}$&\begin{tabular}{c}*Sensitivity to $c_{H,V}$\\ would reach $\mathcal{O}(10^{-2})$\\with $5\textrm{ab}^{-1}$ luminosity.\end{tabular}\\
\hline
CEPC&$e^+e^-\rightarrow Zh(\eta\eta)$&0.25&\begin{tabular}{c}$c_{h,V}=0.95$\\$\textrm{Br}_{\eta\rightarrow b\bar{b}}=1$\end{tabular}&$0.19\times\textrm{Br}_{h\rightarrow\eta\eta}$&\begin{tabular}{c}$5\sigma$ discovery with\\ $5\textrm{ab}^{-1}$ luminosity\\
if $\textrm{Br}_{h\rightarrow\eta\eta}>10^{-3}$\end{tabular}\\
\hline
ILC&$e^+e^-\rightarrow H\eta$&0.5&$c_{h,V}=0.95$&$1\times10^{-2}$&*To be studied.\\
\hline
ILC&$e^+e^-\rightarrow H^+H^-$&0.8&$\textrm{Br}_{H^-\rightarrow \bar{t}b}=1$&$1.4\times10^{-2}$&\begin{tabular}{c}Cross section can be\\
measured to $9\%$ with\\$1\textrm{ab}^{-1}$ luminosity.\end{tabular}\\
\hline
\end{tabular}
\end{table}

If all the three neutral scalars and their couplings to $VV$ were discovered in the future, the associated productions for any two scalars
are important to confirm CP-violation in Higgs sector as well. Since in a general model, if no CP-violation exists in scalar sector, all
the three discovered neutral scalars should be CP-even \footnote{This case cannot appear in 2HDM, there must be additional scalar degree
of freedoms, such as another Higgs doublet.} thus there would be no direct $h_ih_jZ$ vertices. The $e^+e^-\rightarrow h_ih_j$ process would
be loop induced in this case thus the cross section would be highly suppressed. If the cross sections showed there exists tree level
$h_ih_jZ$ vertices \footnote{For example, if the cross sections satisfied the relations in (\ref{rel}).}, the scalars must contain different
CP components thus we would be able to confirm CP-violation in scalar sector \cite{added} \footnote{Notice this is a model-independent method
to confirm CP-violation in scalar sector. But it cannot be used to exclude CP-violation in scalar sector, because in some models, there are no 
direct $h_ih_jZ$ vertices even CP-violation exists in scalar sector.}.
\subsection{Future Measurements on 125 GeV Higgs Boson}
In this scenario, the couplings between 125 GeV Higgs boson $h$ and SM particles should be SM-like. Exotic decay channels $h\rightarrow\eta\eta$
or $Z\eta$ make the total width of $h$ larger, which would also affect other decay channels of $h$. In the future, LHC can measure the signal strengths
$h\rightarrow\gamma\gamma,ZZ^*,WW^*,b\bar{b},\tau^+\tau^-$ to the precision $(11-14)\%$ with $300\textrm{fb}^{-1}$
luminosity, and $(7-8)\%$ with $3\textrm{ab}^{-1}$ luminosity \cite{strfut1,strfut2}.

In this scenario, the modification of Higgs couplings to fermions or gauge bosons should be at percent level, for some cases it can reach
$10\%$. Under this assumption, if the future signal strengths were all consistent with SM prediction, we perform a global-fit and estimate that
at least $\Gamma_{\textrm{exo}}\lesssim(0.4-0.6)\textrm{MeV}$ (or equivalently $\textrm{Br}_{\textrm{exo}}\lesssim(10-15)\%$) would still
be allowed in this scenario. The direct measurements at future LHC cannot reach the sensitivity to test the modification of Higgs signal
strengths from those in SM.

At future $e^+e^-$ colliders, such as the Higgs factories CEPC \cite{CEPC} or ILC \cite{ILC,ILC2} with $\sqrt{s}\sim(240-250)\textrm{GeV}$,
to the luminosity of $\mathcal{O}(\textrm{ab}^{-1})$, all the channels mentioned above together with Higgs total width can be measured to
percent level or even better. For $c_{h,V}\sim0.95$, $\Delta\sigma_{Zh}/\sigma_{Zh}\sim10\%$ which can be measured with $\mathcal{O}(0.1
\textrm{ab}^{-1})$ luminosity. The precision measurements on $h\rightarrow b\bar{b},\tau^+\tau^-$ are also helpful to distinguish this scenario
from SM. If no deviations were found, $|\xi_{bb,\tau\tau}|$ would be constrained to $\mathcal{O}(1)$. For $h\rightarrow gg$ decay, it is
sensitive to both $|c_{h,t}|$ and $\alpha_{h,t}$. The exotic decays $h\rightarrow\eta\eta,Z\eta$ would also be discovered or further constrained
at Higgs factory.

CP-violation is a main feather in Lee model, for example, the $hf\bar{f}$ couplings would contain CP-violation. For $\tau$ lepton and top quark,
the decay distribution would include its polarization information thus it is possible to test the CP-violation effects in $ht\bar{t}$ and
$h\tau^+\tau^-$ vertices \cite{ILC2}. At LHC with $\sqrt{s}=13\textrm{TeV}$ and $3\textrm{ab}^{-1}$ luminosity, using $h\rightarrow\tau^+\tau^-$
decay mode, it is possible to measure $\alpha_{h,\tau}$ to the sensitivity $\Delta\alpha_{h,\tau}\sim4^{\circ}$ with the help of final states
distribution in $\tau$ decay \cite{htautau1}; while at $e^+e^-$
collider, with $\sqrt{s}=250\textrm{GeV}$ and $1\textrm{ab}^{-1}$ luminosity, this sensitivity would reach $\Delta\alpha_{h,\tau}\sim2.8^{\circ}$
\cite{htautau2}; which are both enough to test this scenario. For $ht\bar{t}$ coupling, we can use $e^+e^-\rightarrow t\bar{t}h$ associated
production to test $\alpha_{h,t}$ \cite{olEDM2,ILC2,htt}, with $\sqrt{s}>2m_t+m_h\sim470\textrm{GeV}$.

\subsection{EDM for Third Generation Fermions}
As mentioned above, the polarization of a $\tau$ lepton or top quark can affect on the distribution of its decay final states. With this property,
it may be possible to test their anomalous electro-weak couplings including EDM. For a heavy fermion such as $\tau,b,t$, if one-loop contribution
to CP-violation (see \autoref{EDMdiag}) exists, the Barr-Zee type contribution would be ignorable. The one-loop contribution reads
\cite{olEDM,olEDM2}
\begin{equation}
\label{df}
d_f=\frac{Q_fm_f^3}{16\pi^2v^2}\mathop{\sum}_{\phi}\frac{|c_{\phi,f}|^2\sin(2\alpha_{\phi,f})}{m^2_{\phi}}\mathcal{P}_2\left(\frac{m_f^2}{m^2_{\phi}}\right)
\end{equation}
where the loop function $\mathcal{P}_2(x)$ is listed in (\ref{p2}) in \autoref{loop}.

For a $\tau$ lepton,
\begin{equation}
|d_{\tau}|\lesssim\frac{m^3_{\tau}|\xi_{\tau\tau}|^2}{16\pi^2v^2m^2_{\eta}}
\left(\ln\left(\frac{m^2_{\eta}}{m^2_{\tau}}\right)-\frac{3}{2}\right)\sim10^{-22}|\xi_{\tau\tau}|^2e\cdot\textrm{cm}.
\end{equation}
If $|\xi_{\tau\tau}|\sim1$, it is still far away from the future sensitivity of $\tau$ EDM, around $\mathcal{O}(10^{-19}e\cdot\textrm{cm})$,
given by SuperB \cite{supb1,supb2} with $\sqrt{s}=m_{\Upsilon(4S)}$ and $(50-75)\textrm{ab}^{-1}$ luminosity or CEPC \cite{CEPC} with
$\sqrt{s}=240\textrm{GeV}$ and $5\textrm{ab}^{-1}$ luminosity.
But for a top quark, it can be larger due to its large mass. With the benchmark points in \autoref{bench}, for $|\xi_{tt}|=0.6$ and $m_{\eta}
\sim (20-40)\textrm{GeV}$, $|d_t|$ can reach $\mathcal{O}(10^{-19}-10^{-18})e\cdot\textrm{cm}$ which would be possibly tested at future $e^+e^-$ colliders with
$\mathcal{O}(\textrm{ab}^{-1})$ luminosity \cite{olEDM2,tedm}.
\subsection{Future Tests in Flavor Physics}
At future SuperB with $(50-75)\textrm{ab}^{-1}$ luminosity \cite{supb1,supb2} and LHCb with $50\textrm{fb}^{-1}$ luminosity
\cite{suplhcb} experiments, for $B^0_{(s)}-\bar{B}^0_{(s)}$ mixing, the sensitivity to
$\Delta_{B(B_s)}$ in (\ref{mixpar}) would reach $(3-7)\times10^{-2}$ given by \cite{offdiag2}. With these sensitivity, if no deviations
in B meson mixing were found, it would require $|\xi_{tt}|\lesssim(0.2-0.3)$ at $95\%$ C.L. While the benchmark point we choose in the
text above, $|\xi_{tt}|\sim0.6$, would lead to at least a $5\sigma$ deviation from SM prediction in B meson mixing results.

Another important indirect constraint on $\xi_{tt}$ comes from the leptonic decay of B meson. Future measurements on
$\textrm{Br}(B^0_s\rightarrow\mu^+\mu^-)$ would reach $12\%$ with $3\textrm{ab}^{-1}$ luminosity by CMS \cite{cmsfutB}
and $4\%$ with $50\textrm{fb}^{-1}$ by LHCb \cite{suplhcb}.
If no deviation from SM were found, it would require $|\xi_{tt}|\lesssim0.4$ at $95\%$ C.L. If $|\xi_{tt}|=0.6$, the
LHCb result would be larger than the SM prediction at $3\sigma$ level.

At LHC with $\sqrt{s}=13\textrm{TeV}$ and $300\textrm{fb}^{-1}$ luminosity, if no LFV signal were found, it would require $\textrm{Br}(h\rightarrow\mu^{\pm}\tau^{\mp})<7.7\times10^{-4}$ at $95\%$ C.L. \cite{futhmutau}, or equivalently $|\xi_{\mu\tau}|\lesssim
(1.1-3.5)$ which is still not strict. To $3\textrm{ab}^{-1}$, the upper limit for $|\xi_{\mu\tau}|$ would reduce to $(0.6-2.0)$.
At SuperB factory with $75\textrm{ab}^{-1}$ luminosity, $\textrm{Br}(\tau\rightarrow\mu\gamma)$ can be constrained to
less than $2.4\times10^{-9}$ at $90\%$ C.L., or be discovered at $3\sigma$ level if it is larger than $5.4\times10^{-9}$
\cite{supb2}.

\begin{figure}[h]
\caption{Distributions of $\textrm{Br}(\tau\rightarrow\mu\gamma)$ in $\alpha_{tt}-\alpha_{\tau\tau}$ plane with the benchmark points in \autoref{LFV}.
The left figure is for $m_{\eta}=20\textrm{GeV}$ and the right figure is for $m_{\eta}=40\textrm{GeV}$. In both figures, $|\xi_{\mu\tau}|=|\xi_{\tau\mu}|=
|\xi_{\tau\tau}|=1$ and $\xi_{tt}=0.6$. The green regions are for $\textrm{Br}(\tau\rightarrow\mu\gamma)\leq2.4\times10^{-9}$; the yellow regions are for $2.4\times10^{-9}<\textrm{Br}(\tau\rightarrow\mu\gamma)\leq5.4\times10^{-9}$; the blue regions are for $5.4\times10^{-9}<\textrm{Br}(\tau\rightarrow\mu\gamma)
\leq9\times10^{-9}$; and the orange regions are for $\textrm{Br}(\tau\rightarrow\mu\gamma)>9\times10^{-9}$.}\label{taumuga}
\includegraphics[scale=0.57]{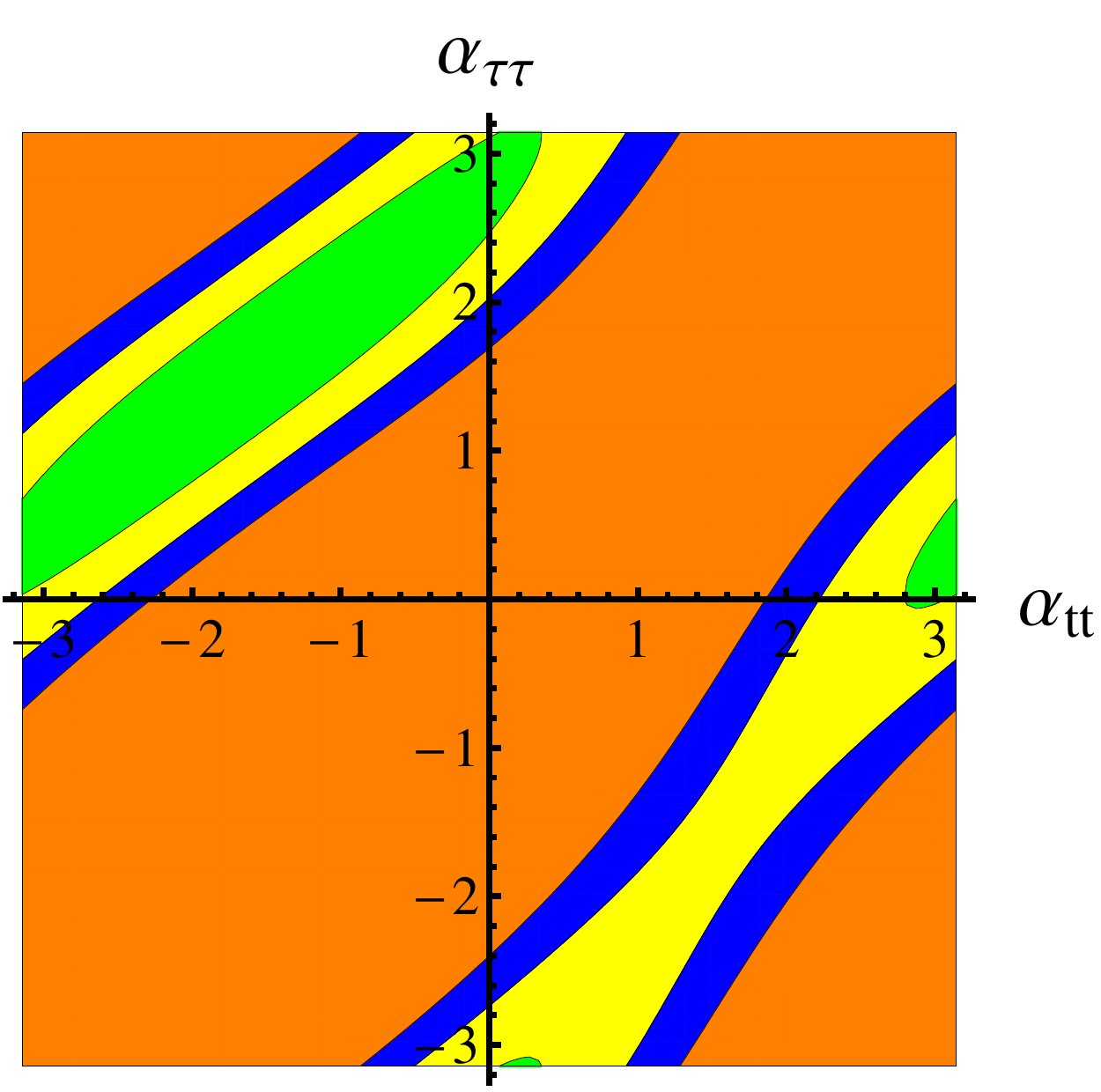}\quad\includegraphics[scale=0.57]{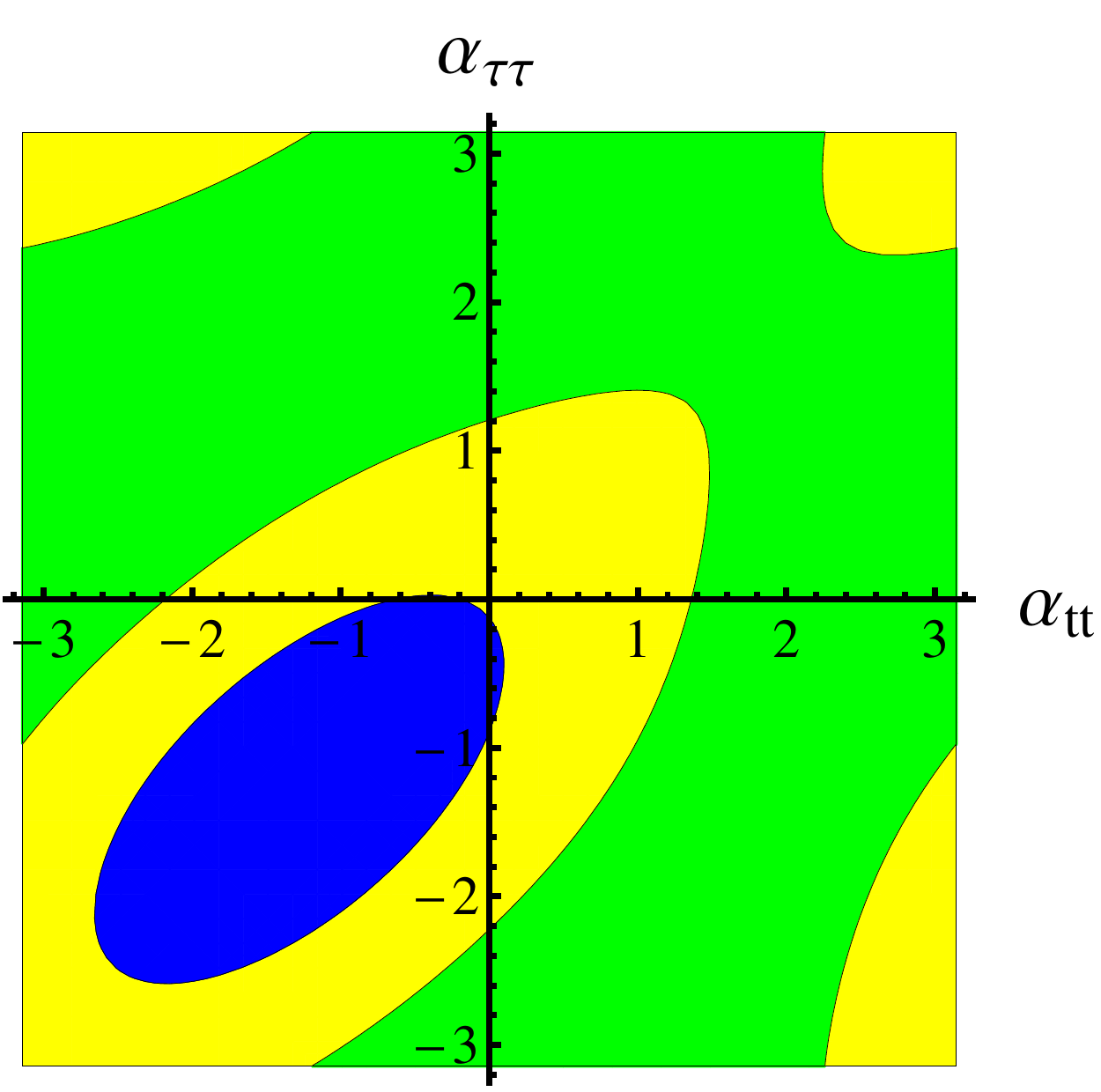}
\end{figure}

According to (\ref{tmga1}) and (\ref{tmga2}) taking the benchmark points in \autoref{bench}, fix $|\xi_{\mu\tau}|=|\xi_{\tau\mu}|=|\xi_{\tau\tau}|=1$
and $\xi_{tt}=0.6$, we plot the $\textrm{Br}(\tau\rightarrow\mu\gamma)$ distributions in $\alpha_{tt}-\alpha_{\tau\tau}$ plane in \autoref{taumuga}, for
$m_{\eta}=20\textrm{GeV}$ (left) and $m_{\eta}=40\textrm{GeV}$ (right).
If no evidence for $\tau\rightarrow\mu\gamma$ were found, the parameters would be constrained to be in green regions. While if the parameters were in
blue (orange) regions, $\tau\rightarrow\mu\gamma$ would be discovered at $3(5)\sigma$ level at SuperB factory with $75\textrm{ab}^{-1}$ luminosity.
Fixing $|\xi_{\mu\tau}|=|\xi_{\tau\mu}|=1$ and leaving other parameters free, if no evidence were found at SuperB, $|\xi_{\tau\tau}|$ would be required
less than $1.2$ for the $m_{\eta}=20\textrm{GeV}$ case or less than $2.6$ for the $m_{\eta}=40\textrm{GeV}$ case.

At SuperB factory, the dominant background for $\tau\rightarrow\mu\gamma$ should be $e^+e^-\rightarrow\tau^+\tau^-\gamma$ \cite{supb1} which would be
suppressed at a collider with $\sqrt{s}$ not far above $2m_{\tau}$, such as Super tau-charm factory \cite{stauchar}. At Super tau-charm factory with
$10\textrm{ab}^{-1}$ luminosity, the sensitivity of $\textrm{Br}(\tau\rightarrow\mu\gamma)$ would reach around $2\times10^{-10}$ \cite{stcsensi}, which
can give further constraints. Future MEG experiments on $\textrm{Br}(\mu\rightarrow e\gamma)$ would reach the sensitivity $6\times10^{-14}$ in three
years \cite{MEGfut}, which can give stricter constraints for all the three LFV couplings $\xi_{e\mu,e\tau,\mu\tau}$.
\section{Conclusions and Discussions}
\label{conc}
In this paper, based on weakly-coupled spontaneous CP-violation 2HDM (named Lee model), using the correlation between the lightest scalar and smallness
of CP-violation through small $t_{\beta}s_{\xi}$ which was proposed in our recent paper \cite{our}, we predicted that a light CP-mixing scalar with
its mass of $\mathcal{O}(10\textrm{GeV})$ should exist. It is pseudoscalar dominant with only about $\mathcal{O}(0.1)$ scalar component. In this scenario,
other scalars' masses are all around the electro-weak scale $v$. It's attractive because there should be new physics hidden at $\mathcal{O}(10\textrm{GeV})$
scale which is below the electro-weak scale, different from the scenario we discussed in \cite{our} in which the Higgs sector is strong-coupled and new
physics appear at $\mathcal{O}(\textrm{TeV})$ or higher scale.

We discussed all experimental constraints, at both high and low energy, for two typical lightest scalar ($\eta$) masses, $m_{\eta}=20\textrm{GeV}$
($h\rightarrow Z\eta$ decay allowed) and $m_{\eta}=40\textrm{GeV}$ ($h\rightarrow Z\eta$ decay forbidden). For these $\eta$ masses, $c_{\eta,V}\sim0.1$
is required theoretically and it is also allowed by data. The 125 GeV Higgs boson $h$ has SM-like couplings to fermions and gauge bosons. With a global-fit
to higgs signal strengths, branching ratio for exotic decay channels are constrained to less than about $30\%$, which leads to strict constraints on
$h\eta\eta$ (and $hZ\eta$ if $m_{\eta}<34\textrm{GeV}$) couplings. The constraints from oblique parameters require $m_{\pm}\sim m_H\sim v$ under the
weak-coupling assumption. The typical benchmark points listed in \autoref{bench} are chosen according to these constraints.

In Lee model, there is no additional discrete symmetry except CP, thus there may exist flavor-changing interactions at tree level. We adopted the
Cheng-Sher ansatz to parameterize the flavor-changing effects. High energy processes including top flavor-changing interactions cannot give strict
constraints. The most strict constraint from high energy experiments comes from an indirect test, top quark widths limit from $t\bar{t}$ pair production,
which requires $|\xi_{tc}|\lesssim1$. A more strict constraint comes from $D^0-\bar{D}^0$ mixing which requires $|\xi_{tc}\xi_{tu}|\lesssim6$. All other
$|\xi_{ij}|$ in quark sector are constrained to be less than around $\mathcal{O}(10^{-2})$, through meson mixing measurements. In lepton sector, indirect
tests (especially radiative LFV decays) require $|\xi_{\mu\tau}|\lesssim\mathcal{O}(1)$, while upper limit on $|\xi_{e\mu}|$ or $|\xi_{e\tau}\xi_{\mu\tau}|$
are of $\mathcal{O}(10^{-2})$. EDM tests also favor $|\xi_{e\tau,ut}|\lesssim\mathcal{O}(1)$. These constraints are usually stricter than those in \cite{our}
as we discussed, that's because in this scenario, a lighter scalar would give more significant contribution to the flavor-changing processes.

B meson mixing and B leptonic decay processes are all sensitive to $\xi_{tt}$. With these data, $|\xi_{tt}|\lesssim0.6$ is favored at $95\%$ C.L.
which is the reason why in most part of the text we choose $|\xi_{tt}|=0.6$ as a benchmark point. The B radiative decay process is sensitive to both
$\xi_{tt}$ and $\xi_{bb}$. With the assumption $m_{\pm}\sim(200-300)\textrm{GeV}$ and $|\xi_{tt}|=0.6$, $\xi_{bb}\lesssim\mathcal{O}(1)$ is allowed by data.
That is a difference between this scenario and type II 2HDM in which charged Higgs should be heavier than around 410 GeV at $95\%$ C.L.

The EDM constraints are also strict just like the scenario we discussed in \cite{our}. For both electron and neutron EDM, we need large cancelation
between different contributions, as shown in \autoref{EDMfig1}-\autoref{nEDM2}. In each of the figures, the two shown parameters are constrained in
a narrow band which means a strong correlation between them.

We also discussed the future tests for this scenario. For the lightest scalar $\eta$, the dominant ways to discover it at LHC are associated
production $pp\rightarrow t\bar{t}\eta$ and cascade decay $pp\rightarrow h,H\rightarrow\eta\eta,Z\eta$. While since the heaviest neutral scalar is
also required to have its mass around $v$, it can also be searched through $Z\eta$ or $VV$ final states. At LHeC or $e^+e^-$ colliders, the
exotic decays $h\rightarrow\eta\eta,Z\eta$ would be tested. Especially at Higgs factory, with $\mathcal{O}(10-10^2)\textrm{fb}^{-1}$ luminosity,
$c_{\eta,V}\sim0.1$ can be discovered at $(3-5)\sigma$ through re-scaling the LEP constraints. If nothing were found, constraints on $c_{\eta,V}$
would improve an order which also implies $m_{\eta}\sim\mathcal{O}(\textrm{GeV})$, thus this scenario is disfavored. The mass of 
charged Higgs boson is predicted to be around
$v$ which is possible to be discovered at future LHC or $e^+e^-$ colliders with $\sqrt{s}\gtrsim500\textrm{GeV}$, using $\mathcal{O}(0.1-1)
\textrm{ab}^{-1}$ luminosity. Note that the $e^+e^-\rightarrow H^+H^-$ process cannot be suppressed with fixed $m_{\pm}$, if nothing were found,
this scenario would be excluded. 

If all the three scalars are discovered and they all have direct vertices to massive gauge boson pairs, the $Z$-
mediated Higgs associated pair production via $e^+e^-\rightarrow h_ih_j$ would be a key observable to confirm CP-violation in scalar sector.
It can be used to distinguish Lee model and models in which the scalar sector contains more CP-even degrees of freedom but no CP-violation.

Indirect tests on B meson mixing and B leptonic decay can be used to test a nonzero $\xi_{tt}$ or give a stricter limit on $|\xi_{tt}|$. For
the case $|\xi_{tt}|=0.6$ we used in this paper, there would appear $(3-5)\sigma$ deviations in these experiments. If nothing anomaly were
found, $|\xi_{tt}|$ would be pushed to less than about $(0.2-0.3)$. Radiative LFV decays would also confirm a nonzero LFV vertex or push them
to a smaller number, depending on the results positive or negative.

In this attractive scenario, all new physics would appear below or at electro-weak scale which behaves different from most
models in which new physics appear at or above $\mathcal{O}(\textrm{TeV})$ scale. It means this scenario is testable. The roughly
estimation showed it is able to discover or exclude this scenario, especially for $\eta$ who is hidden at $\mathcal{O}(10\textrm{GeV})$
scale. It is also possible to distinguish whether CP-violation in Higgs sector exists if all neutral scalars were found. Thus it is
worth further studying in details, especially at future $e^+e^-$ colliders.
\section*{Acknowledgement}
We thank Chen Zhang for helpful discussions. This work was supported in part by the Natural Science Foundation of China (Grants No. 11135003 and No. 11375014).

\numberwithin{equation}{section}
\appendix
\section{Spectrum and Couplings}
\label{coup}
Expansion for the neutral scalar mass matrix
\begin{equation}
\left(\begin{array}{ccc}(\lambda_4-\lambda_7)s^2_{\xi}&
-((\lambda_4-\lambda_7)s_{\beta}c_{\xi}+\lambda_2c_{\beta})s_{\xi}&
-((\lambda_4-\lambda_7)c_{\beta}c_{\xi}+\lambda_5s_{\beta})s_{\xi}\\
&&\\
&\begin{array}{c}4\lambda_1c^2_{\beta}+\lambda_2s_{2\beta}c_{\xi}+(\lambda_4-\lambda_7)s^2_{\beta}c^2_{\xi}\end{array}&
\begin{array}{c}((\lambda_3+\lambda_7)+(\lambda_4-\lambda_7)c^2_{\xi}/2)s_{2\beta}\\
+\lambda_2c^2_{\beta}c_{\xi}+\lambda_5s^2_{\beta}c_{\xi}\end{array}\\
&&\\
&&\begin{array}{c}(\lambda_4-\lambda_7)c^2_{\beta}c^2_{\xi}\\+\lambda_5s_{2\beta}c_{\xi}+4\lambda_6s^2_{\beta}\end{array}\end{array}\right)
\end{equation}
is $\tilde{m}=\tilde{m}_0+(t_{\beta}s_{\xi})\tilde{m}_1+(t_{\beta}s_{\xi})^2\tilde{m}_2+\ldots$ To the first order, we have
\begin{equation}
m_{\eta}=0,\quad\quad m_{h,H}=\frac{v}{2}\sqrt{(4\lambda_1+\lambda_4-\lambda_7)\mp
\left((4\lambda_1-(\lambda_4-\lambda_7))c_{2\theta}+2\lambda_2s_{2\theta}\right)}
\end{equation}
where $\theta=(1/2)\arctan(2\lambda_2/(4\lambda_1-\lambda_4+\lambda_7))$ is the mixing angle. The scalar fields
\begin{equation}
\eta_0=I_2,\quad\quad\left(\begin{array}{c}h\\H\end{array}\right)_0=\left(\begin{array}{cc}c_{\theta}&s_{\theta}\\
-s_{\theta}&c_{\theta}\end{array}\right)\left(\begin{array}{c}R_1\\R_2\end{array}\right).
\end{equation}
Calculation by perturbation method to the leading order of $(t_{\beta}s_{\xi})$ gives
\begin{eqnarray}
\eta&=&I_2-(t_{\beta}s_{\xi})\left(\frac{(\tilde{m}_1)_{12}}{(\tilde{m}_0)_{22}}(c_{\theta}R_1+s_{\theta}R_2)
+\frac{(\tilde{m}_1)_{13}}{(\tilde{m}_0)_{33}}(c_{\theta}R_2-s_{\theta}R_1)\right)-t_{\beta}c_{\xi}I_1;\\
h&=&c_{\theta'}R_1+s_{\theta'}R_2+\frac{(\tilde{m}_1)_{12}}{(\tilde{m}_0)_{22}}(t_{\beta}s_{\xi})I_2;\\
H&=&-s_{\theta'}R_1+c_{\theta'}R_2+(t_{\beta}s_{\xi})\left(I_1+\frac{(\tilde{m}_1)_{13}}{(\tilde{m}_0)_{33}}I_2\right);\\
m_{\eta}&=&\frac{vt_{\beta}s_{\xi}}{\sqrt{2}}\sqrt{(\tilde{m}_2)_{11}-\frac{(\tilde{m}_1)^2_{12}}{(\tilde{m}_0)_{22}}-
\frac{(\tilde{m}_1)^2_{13}}{(\tilde{m}_0)_{33}}}.
\end{eqnarray}
Here
\begin{equation}
\theta'=\theta+\frac{(t_{\beta}s_{\xi})(\tilde{m}_1)_{23}}{(\tilde{m}_0)_{22}-(\tilde{m}_0)_{33}}.
\end{equation}

The scalar self-interactions
\begin{equation}
\mathcal{L}=-\sum\left(\frac{1}{S_{ijk}}g_{ijk}vh_ih_jh_k+\frac{1}{S_{ijkl}}g_{ijk}h_ih_jh_kh_l\right)
\end{equation}
where the symmetric factor $S\equiv\prod(n_i!)$ in which $n_i$ denotes the appearance time for $h_i$ in the lagrangian.
The couplings can be obtained directly from
\begin{equation}
\label{sc}
g_{ijk}=\left.\frac{1}{v}\frac{\partial^3V}{\partial h_i\partial h_j\partial h_k}\right|_{\textrm{all }h_i=0},\quad\quad
g_{ijkl}=\left.\frac{\partial^4V}{\partial h_i\partial h_j\partial h_k \partial h_l}\right|_{\textrm{all }h_i=0}.
\end{equation}
As an example, the $h\eta\eta$ vertex is given by
\begin{eqnarray}
g_{h\eta\eta}&=&\frac{\partial^3V}{\partial h\partial\eta^2}\nonumber\\
&=&(\lambda_3+\lambda_7)c_{\theta'}+\frac{1}{2}\lambda_5s_{\theta'}-\frac{t_{\beta}c_{\xi}}{2}(\lambda_2c_{\theta'}+(\lambda_4-\lambda_7)s_{\theta'}).
\end{eqnarray}
For $m_{\eta}<m_h/2$, The strict constraints from $h\rightarrow\eta\eta$ rare decay showed that
\begin{equation}
\lambda_3+\lambda_7+\frac{\lambda_5t_{\theta'}}{2}\simeq\frac{t_{\beta}c_{\xi}}{2}(\lambda_2+(\lambda_4-\lambda_7)t_{\theta'})
\end{equation}
which means $(\tilde{m}_1)_{12}\sim\mathcal{O}(\beta)$ is ignorable in the formula above.

For the $h_iVV$ and $h_ih_jZ$ couplings, the effective interaction should be written as
\begin{eqnarray}
\mathcal{L}_{h_iVV}&=&c_{i,V}h_i\left(\frac{2m^2_W}{v}W^{+\mu}W^-_{\mu}+\frac{m^2_Z}{v}Z^{\mu}Z_{\mu}\right);\\
\mathcal{L}_{h_ih_jZ}&=&\frac{c_{ij}g}{2c_W}Z^{\mu}\left(h_i\partial_{\mu}h_j-h_j\partial_{\mu}h_i\right).
\end{eqnarray}
With a straightforward calculation, we have
\begin{equation}
\label{rel}
c_{\eta,V}=c_{hH},\quad\quad c_{h,V}=c_{H\eta},\quad\quad c_{H,V}=c_{\eta h}
\end{equation}
thus $\sum c^2_{i,V}=\sum c^2_{ij}=1$. In this scenario, to the leading order of $t_{\beta}s_{\xi}$,
\begin{eqnarray}
c_{\eta,V}&=&t_{\beta}s_{\xi}\left(1+s_{\theta}\frac{(\tilde{m}_1)_{13}}{(\tilde{m}_0)_{33}}-c_{\theta}\frac{(\tilde{m}_1)_{12}}{(\tilde{m}_0)_{22}}\right);\\
c_{h,V}&=&c_{\theta'}+t_{\beta}c_{\xi}s_{\theta'};\\
c_{H,V}&=&-s_{\theta'}+t_{\beta}c_{\xi}c_{\theta'}.
\end{eqnarray}
For the case $m_{\eta}<m_h-m_Z$, strict constraints from $h\rightarrow Z\eta$ rare decay showed that
$c_{H,V}\ll1$ thus $t_{\theta'}\simeq t_{\beta}c_{\xi}$.

The Yukawa interactions
\begin{equation}
\mathcal{L}_Y=-\mathop{\sum}_{\phi}\left(\mathop{\sum}_f\frac{c_{\phi,f}m_f}{v}\bar{f}_Lf_R\phi+\mathop{\sum}_{i\neq j}\frac{c_{\phi,ij}\sqrt{m_im_j}}{v}\bar{f}_{Li}f_{Rj}\phi\right)+\textrm{h.c.}
\end{equation}
where $\phi$ denotes any scalar and $f$ denotes any fermion. The factors for diagonal terms can be generated directly as
\begin{eqnarray}
c_{\eta,f}&=&\pm\textrm{i}\xi_{ff}\left(1+t_{\beta}s_{\xi}\left(\frac{(\tilde{m}_1)_{12}}{(\tilde{m}_0)_{22}}c_{\theta'}
-\frac{(\tilde{m}_1)_{13}}{(\tilde{m}_0)_{33}}s_{\theta'}\right)\right)\nonumber\\
&&+t_{\beta}s_{\xi}\left(1-c_{\theta'}\left(\xi_{ff}\frac{(\tilde{m}_1)_{12}}{(\tilde{m}_0)_{22}}+\frac{(\tilde{m}_1)_{13}}{(\tilde{m}_0)_{33}}\right)
-s_{\theta'}\left(\xi_{ff}\frac{(\tilde{m}_1)_{12}}{(\tilde{m}_0)_{22}}-\frac{(\tilde{m}_1)_{13}}{(\tilde{m}_0)_{33}}\right)\right);\\
\label{chf}
c_{h,f}&=&c_{\theta'}+\xi_{ff}s_{\theta'}+t_{\beta}c_{\xi}(s_{\theta'}-\xi_{ff}c_{\theta'})\pm\textrm{i}t_{\beta}s_{\xi}\xi_{ff}
\left(\frac{(\tilde{m}_1)_{12}}{(\tilde{m}_0)_{22}}-c_{\theta'}\right);\\
c_{H,f}&=&-s_{\theta'}+\xi_{ff}c_{\theta'}+t_{\beta}c_{\xi}(c_{\theta'}+\xi_{ff}s_{\theta'})\pm\textrm{i}t_{\beta}s_{\xi}\xi_{ff}
\left(\frac{(\tilde{m}_1)_{13}}{(\tilde{m}_0)_{33}}+s_{\theta'}\right).
\end{eqnarray}
While the factors for off-diagonal term are
\begin{eqnarray}
c_{\eta,ij}&=&\pm\textrm{i}\xi_{ij}\left(1+t_{\beta}s_{\xi}\left(\frac{(\tilde{m}_1)_{12}}{(\tilde{m}_0)_{22}}c_{\theta'}
-\frac{(\tilde{m}_1)_{13}}{(\tilde{m}_0)_{33}}s_{\theta'}\right)\right)\nonumber\\
\label{etaij}
&&-t_{\beta}s_{\xi}\xi_{ij}\left(c_{\theta'}
\frac{(\tilde{m}_1)_{13}}{(\tilde{m}_0)_{33}}+s_{\theta'}\frac{(\tilde{m}_1)_{12}}{(\tilde{m}_0)_{22}}\right);\\
\label{hij}
c_{h,ij}&=&\xi_{ij}\left(s_{\theta'}-t_{\beta}c_{\xi}c_{\theta'}\pm\textrm{i}t_{\beta}s_{\xi}
\left(\frac{(\tilde{m}_1)_{12}}{(\tilde{m}_0)_{22}}-c_{\theta'}\right)\right);\\
\label{Hij}
c_{H,ij}&=&\xi_{ij}\left(c_{\theta'}+t_{\beta}c_{\xi}s_{\theta'}\pm\textrm{i}t_{\beta}s_{\xi}
\left(\frac{(\tilde{m}_1)_{13}}{(\tilde{m}_0)_{33}}+s_{\theta'}\right)\right).
\end{eqnarray}
In each of the six formula, when ``$\pm$" appears, ``$+$" stands for down type fermions and ``$-$" stands for up type fermions.
\section{Useful Analytical Loop Integrations}
\label{loop}
The loop integration functions for $h\rightarrow\gamma\gamma(gg)$ decay width in (\ref{gg}) and (\ref{gaga}) are
\begin{eqnarray}
\label{gggaga1}
\mathcal{A}_0(x)&=&\frac{x-f(x)}{x^2},\\
\mathcal{A}_{1/2}(x)&=&-\frac{x+(x-1)f(x)}{x^2},\\
\mathcal{B}_{1/2}(x)&=&-\frac{2f(x)}{x},\\
\mathcal{A}_1(x)&=&\frac{2x^2+3x+3(2x-1)f(x)}{x^2}
\end{eqnarray}
where
\begin{equation}
\label{gggaga5}
f(x)=\left\{\begin{array}{ll}\arcsin^2(\sqrt{x}),&(\textrm{for }x\leq1);\\
-\frac{1}{4}\left(\ln\frac{1+\sqrt{1-x^{-1}}}{1-\sqrt{1-x^{-1}}}-\textrm{i}\pi\right),&(\textrm{for }x>1).\end{array}\right.
\end{equation}
The difference between $\mathcal{A}_{1/2}$ and $\mathcal{B}_{1/2}$ comes from the different
tensor structures for the scalar and pseudoscalar components.

The loop integration functions for oblique parameters in (\ref{DS}) and (\ref{DT}) are
\begin{eqnarray}
\label{ST1}
F(x,y)&=&\frac{x+y}{2}-\frac{xy}{x-y}\ln\left(\frac{x}{y}\right);\\
G(x,y)&=&-\frac{16}{3}+5(x+y)-2(x-y)^2+3\left(\frac{x^2+y^2}{x-y}+y^2-x^2+\frac{(x-y)^3}{3}\right)\ln\left(\frac{x}{y}\right)\nonumber\\
&&+(1-2(x+y)+(x-y)^2)f(x+y-1,1-2(x+y)+(x-y)^2);\\
H(x)&=&-\frac{79}{3}+9x-2x^2+\left(-10+18x-6x^2+x^3-9\frac{x+1}{x-1}\right)\ln x\nonumber\\
&&+(12-4x+x^2)f(x,x^2-4x);
\end{eqnarray}
where
\begin{equation}
\label{ST4}
f(x,y)=\left\{\begin{array}{l}\sqrt{y}\ln\left|\frac{x-\sqrt{y}}{x+\sqrt{y}}\right|,\quad\quad y\geq0;\\
2\sqrt{-y}\arctan\left(\frac{\sqrt{-y}}{x}\right),\quad\quad y<0.\end{array}\right.
\end{equation}

The loop integration functions for meson mixing in (\ref{Dbox}) and (\ref{Bbox}) are
\begin{eqnarray}
\label{box1}
\mathcal{F}_0(x)&=&\frac{x(1-x^2+2x\ln x)}{(1-x)^3};\\
\mathcal{F}_1(x,y,z)&=&\frac{2y}{1-z}\left(\frac{(z-4)\ln y}{(1-y)^2}+\frac{3z\ln x}{(1-x)^2}-
\frac{(1-z)(4-x)}{(1-y)(1-x)}\right);\\
\label{box3}
\mathcal{F}_2(x)&=&1+\frac{9}{1-x}-\frac{6}{(1-x)^2}-\frac{6x^2\ln x}{(1-x)^3}.
\end{eqnarray}

The loop integration functions for two-loop radiative LFV $\tau$ decay in (\ref{taudecay}) are
\begin{eqnarray}
\label{tau1}
f(z)&=&\frac{z}{2}\int^1_0dx\frac{1-2x(1-x)}{x(1-x)-z}\ln\left(\frac{x(1-x)}{z}\right);\\
g(z)&=&\frac{z}{2}\int^1_0dx\frac{1}{x(1-x)-z}\ln\left(\frac{x(1-x)}{z}\right);\\
\label{tau3}
h(z)&=&-\frac{z}{2}\int^1_0dx\frac{1}{x(1-x)-z}\left(1-\frac{z}{x(1-x)-z}\ln\left(\frac{x(1-x)}{z}\right)\right).
\end{eqnarray}
For $z<1/4$, the integrations are defined as their Cauchy principle value.

The loop integration functions for two-loop Barr-Zee type contribution in calculating the EDM
for a fermion $f$ in (\ref{bzEDM}) are
\begin{eqnarray}
\label{edm1}
\mathcal{J}_0(m_{\pm},m_{\phi})&=&\frac{v^2}{2m^2_{\phi}}\left(\mathcal{I}\left(\frac{m_{\pm}^2}{m_{\phi}^2}\right)
-\mathcal{I}'\left(\frac{m_{\pm}^2}{m_{\phi}^2}\right)\right);\\
\mathcal{J}_{1/2}(m_t,m_{\phi})&=&\frac{m^2_t}{m^2_{\phi}}\mathcal{I}\left(\frac{m_t^2}{m_{\phi}^2}\right);\\
\mathcal{J}'_{1/2}(m_t,m_{\phi})&=&\frac{m^2_t}{m^2_{\phi}}\mathcal{I}'\left(\frac{m_t^2}{m_{\phi}^2}\right);\\
\label{edm4}
\mathcal{J}_1(m_W,m_{\phi})&=&\frac{m^2_W}{m^2_{\phi}}\left(\left(5-\frac{m^2_{\phi}}{2m^2_W}\right)\mathcal{I}\left(\frac{m_W^2}{m_{\phi}^2}\right)
+\left(3+\frac{m^2_{\phi}}{2m^2_W}\right)\mathcal{I}'\left(\frac{m_W^2}{m_{\phi}^2}\right)\right);
\end{eqnarray}
where
\begin{eqnarray}
\mathcal{I}(z)&=&\int^1_0dx\frac{1}{x(1-x)-z}\ln\left(\frac{x(1-x)}{z}\right);\\
\mathcal{I}'(z)&=&\int^1_0dx\frac{1-2x(1-x)}{x(1-x)-z}\ln\left(\frac{x(1-x)}{z}\right).
\end{eqnarray}
For $z<1/4$, the integrations are defined as their Cauchy principle value as above. The loop function for Weinberg operator
in (\ref{weinberg}) is
\begin{equation}
\label{weinbergK}
\mathcal{K}(x)=4x^2\int_0^1du\int_0^1dv\frac{(uv)^3(1-v)}{(xv(1-uv)+(1-u)(1-v))^2}.
\end{equation}
The loop functions for one-loop contribution to fermion EDM in (\ref{du1})-(\ref{du2}) and (\ref{df}) are
\begin{eqnarray}
\label{p1}
\mathcal{P}_1(x)&=&\frac{x}{(x-1)^2}\left(\frac{x-3}{2}+\frac{\ln x}{x-1}\right);\\
\label{p2}
\mathcal{P}_2(x)&=&\int_0^1dz\frac{z^2}{1-z+xz^2}.
\end{eqnarray}

The loop integration functions for B meson leptonic decays in (\ref{blep}) are
\begin{eqnarray}
\label{ysm}
\mathcal{Y}_{\textrm{SM}}(x)&=&\frac{x}{8}\left(\frac{x-4}{x-1}+\frac{3x\ln x}{(x-1)^2}\right);\\
\label{y2hdm}
\mathcal{Y}_{\textrm{2HDM}}(x)&=&\frac{x^2}{8}\left(\frac{1}{y-x}+\frac{y}{(y-x)^2}\ln\left(\frac{x}{y}\right)\right).
\end{eqnarray}
\section{Formalism for Meson Mixing}
\label{mix}
Begin with the Schr$\ddot{\textrm{o}}$dinger equation
\begin{equation}
\textrm{i}\frac{\partial}{\partial t}|\psi\rangle=\mathcal{H}|\psi\rangle=\left(\mathbf{m}-\frac{\textrm{i}}{2}\mathbf{\Gamma}\right)|\psi\rangle
\end{equation}
where $|\psi\rangle=(|M^0\rangle,|\bar{M}^0\rangle)^T$ with normalization condition $\langle M^0|M^0\rangle=\langle\bar{M}^0|\bar{M}^0\rangle=2m_M$
in position space, and $\mathbf{m}$, $\mathbf{\Gamma}$ are $2\times2$ matrixes. The hamiltonian can be written as
\begin{equation}
\mathcal{H}=\mathcal{H}_0+\mathcal{H}_{\Delta F=1}+\mathcal{H}_{\Delta F=2}.
\end{equation}
The matrix element
\begin{equation}
\left(\mathbf{m}-\frac{\textrm{i}}{2}\mathbf{\Gamma}\right)_{ij}=m_M\delta_{ij}+\frac{1}{2m_M}\langle\psi_i|\mathcal{H}_{\Delta F=2}|\psi_j\rangle
+\frac{1}{2m_M}\int d\Pi_f\frac{\langle\psi_i|\mathcal{H}_{\Delta F=1}|f\rangle\langle f|\mathcal{H}_{\Delta F=1}|\psi_j\rangle}{m_M-E_f+\textrm{i}\epsilon}
\end{equation}
where the states $|\psi_{i,j}\rangle$ mean $|M^0\rangle$ or $|\bar{M}^0\rangle$, and $|f\rangle$ denotes a mediated state. The second and third
terms correspond to short- and long-distance contributions respectively, and from the third term,
\begin{equation}
\label{deltm}
\mathbf{\Gamma}_{ij}=\frac{1}{2m_M}\int d\Pi_f\langle\psi_i|\mathcal{H}_{\Delta F=1}|f\rangle\langle f|\mathcal{H}_{\Delta F=1}|\psi_j\rangle
2\pi\delta(E_f-m_M).
\end{equation}
The masses and widths for the mass eigenstates are
\begin{eqnarray}
m_{H(L)}&=&m_M\pm\textrm{Re}\left(\sqrt{\left(\mathbf{m}_{12}-\frac{\textrm{i}}{2}\mathbf{\Gamma}_{12}\right)
\left(\mathbf{m}_{12}^*-\frac{\textrm{i}}{2}\mathbf{\Gamma}_{12}^*\right)}\right);\\
\Gamma_{H(L)}&=&\Gamma_M\mp\textrm{Im}\left(\sqrt{\left(\mathbf{m}_{12}-\frac{\textrm{i}}{2}\mathbf{\Gamma}_{12}\right)
\left(\mathbf{m}_{12}^*-\frac{\textrm{i}}{2}\mathbf{\Gamma}_{12}^*\right)}\right);
\end{eqnarray}
where H (L) denotes the heavy (light) mass eigenstate
\begin{equation}
|M_{H(L)}\rangle=p|M^0\rangle\mp q|\bar{M}^0\rangle.
\end{equation}
$p$ and $q$ are determined through
\begin{equation}
|p|^2+|q|^2=1,\quad\textrm{and}\quad\left(\frac{p}{q}\right)^2=
\frac{\mathbf{m}_{12}-\textrm{i}\mathbf{\Gamma}_{12}/2}{\mathbf{m}_{12}^*-\textrm{i}\mathbf{\Gamma}_{12}^*/2}.
\end{equation}
In $K^0-\bar{K}^0$ system, $\mathbf{m}_{12}$ is almost real and $\mathbf{\Gamma}_{12}\sim\mathbf{m}_{12}$, thus
$\Delta m_K\approx2\textrm{Re}\mathbf{m}_{12}$; while in $B^0_{(s)}-\bar{B}^0_{(s)}$ system, $|\mathbf{\Gamma}_{12}|
\ll|\mathbf{m}_{12}|$, thus $\Delta m_B\approx2|\mathbf{m}_{12}|$.

Transform to momentum space, take $\mathcal{H}$ as the hamiltonian density and change the normalization condition to
$\langle M^0|M^0\rangle=\langle\bar{M}^0|\bar{M}^0\rangle=2m_M\delta^{(3)}(\mathbf{p})$. With the matrix elements
\begin{eqnarray}
\langle0|\bar{f}_i\gamma^{\mu}\gamma^5f_j|M^0(p)\rangle=-\textrm{i}f_Mp^{\mu},\quad
\langle0|\bar{f}_i\gamma^5f_j|M^0(p)\rangle=\textrm{i}\frac{m^2_Mf_M}{m_i+m_j};
\end{eqnarray}
the useful $\Delta F=2$ matrix elements
\begin{eqnarray}
&&\langle\bar{M}_0|\bar{f}_{Li}\gamma^{\mu}f_{Lj}\bar{f}_{Li}\gamma_{\mu}f_{Lj}|M_0\rangle=
\langle\bar{M}_0|\bar{f}_{Ri}\gamma^{\mu}f_{Rj}\bar{f}_{Ri}\gamma_{\mu}f_{Rj}|M_0\rangle=\frac{2}{3}f_M^2m^2_M;\\
&&\langle\bar{M}_0|\bar{f}_{Li}\gamma^{\mu}f_{Lj}\bar{f}_{Ri}\gamma_{\mu}f_{Rj}|M_0\rangle=-\frac{5}{6}f_M^2m^2_M;\\
&&\langle\bar{M}_0|\bar{f}_{Li}f_{Rj}\bar{f}_{Li}f_{Rj}|M_0\rangle=
\langle\bar{M}_0|\bar{f}_{Ri}f_{Lj}\bar{f}_{Ri}f_{Lj}|M_0\rangle=-\frac{5}{12}f_M^2m^2_M;\\
&&\langle\bar{M}_0|\bar{f}_{Li}f_{Rj}\bar{f}_{Ri}f_{Lj}|M_0\rangle=\frac{7}{12}f_M^2m^2_M;
\end{eqnarray}
where the bag parameters are all taken as 1 for simplify.

\clearpage\end{CJK*}

\end{document}